# Radiation-driven winds of hot luminous stars
# XVI. Expanding atmospheres of massive and very massive stars and the evolution of dense stellar clusters

A. W. A. Pauldrach[1], D. Vanbeveren[2,3], and T. L. Hoffmann[1]

[1] Institut für Astronomie und Astrophysik der Universität München, Scheinerstraße 1, 81679 München, Germany
e-mail: uh10107@usm.lmu.de, e-mail: hoffmann@usm.lmu.de
[2] Astrophysical Institute, Vrije Universiteit Brussel, Pleinlaan 2, 1050 Brussels, Belgium
e-mail: dvbevere@vub.ac.be
[3] Leuven Engineering College, Groep T, Association KU Leuven, Vessaliusstraat 13, 3000 Leuven, Belgium

**ABSTRACT**

*Context.* Starbursts play an essential role in the evolution of galaxies. In these environments, massive stars, with their short lifetimes, are of particular importance. The stellar winds of massive stars significantly influence not only on their surroundings, but the associated mass loss also profoundly affects the evolution of the stars themselves. The evolution of the dense cores of massive starburst clusters is also affected by dynamical processes induced by $N$-body interactions, in addition to the evolution of each star, and the formation of very massive stars with masses up to several thousand solar masses may be decisive for the evolution of the cluster. The interpretation of the corresponding observations relies mainly on the theoretical modeling of such starbursts, which is a major challenge.
*Aims.* The primary objective is to introduce an advanced diagnostic method of O-type stellar atmospheres with winds, including an assessment of the accuracy of the determinations of abundances, stellar and wind parameters. Moreover, observational results are interpreted in the framework of our stationary, one-dimensional theory of line driven winds. Possible effects caused by non-homogeneous time dependent structures are also discussed.
*Methods.* We combine consistent models of expanding atmospheres with stellar evolutionary calculations of massive and very massive (up to several 1000 solar masses) single stars with regard to the evolution of dense stellar clusters. Essential in this context are accurate dynamic parameters of the winds of very massive stars. Because the atmospheric mass outflow has substantial influence on the radiation field and the atomic occupation numbers, and the radiation field and the occupation numbers in turn directly influence the radiative acceleration and thus the strength and velocity of the outflow, the determination of the hydrodynamic structures requires a highly consistent treatment of the statistical equilibrium and the hydrodynamic and radiative processes in the expanding atmospheres.
*Results.* We present computed mass loss rates, terminal wind velocities, and spectral energy distributions of massive and very massive stars of different metallicities, calculated from atmospheric models with an improved level of consistency. These computations have important implications for (i) the primordial chemical enrichment of Population III very massive stars; (ii) the age determination of globular clusters; and (iii) the formation of intermediate mass black holes in dense stellar clusters with respect to the importance of stellar wind mass loss for the evolution of their progenitor stars.
*Conclusions.* Stellar evolutionary calculations, using the mass loss rates of very massive stars obtained in the present paper, show that very massive stars with a low metallicity lose only a very small amount of their mass; thus it is unlikely that very massive population III stars cause a significant helium enrichment of the interstellar medium. Solar-metallicity stars have higher mass-loss rates, but these are not so high to exclude very massive stars of solar metallicity, formed by dynamical processes in dense clusters, from ending their life massive enough to form intermediate-mass black holes.

## 1. Introduction

One of the standard paradigms for structure formation at early times is that very massive population III stars initiated the development of galaxies and the reionization of the universe. In the recent years it has become evident that very massive stars (VMS) – stars with an initial mass of $M \geq 100\,M_\odot$ – play a key role in the development of galaxies even today.

Numerical simulations indicate that the first – metal-free – stars (population III stars) were distributed according to an initial mass function (IMF) more skewed towards massive and very massive stars than the present IMF (e.g., Bromm et al. 2001, Abel & Wandelt 2002, Bromm & Larson 2004, Yoshida et al. 2004). These VMS may explain the observed metallicities ($Z$) of population II stars, and they may have contributed significantly to the reionization of the intergalactic cosmic matter (for extensive reviews, see Carr et al. 1984, Bromm & Larson 2004). Moreover, the ionizing properties of VMS at very low metallicity may explain the extreme Ly$\alpha$-emitting galaxies at high redshift (Kudritzki et al. 2000, Malhotra & Rhoads 2002).

But even today, very massive stars may be key players in the development of massive star clusters especially in starburst galaxies. The significance of these starbursts is not only that they heat the intergalactic medium and enrich it with metals via the combined action of stellar winds and supernovae, but also that young dense clusters are believed to harbor intermediate-mass black holes with masses in the range of several 100 to several 1000 solar masses and thus could provide an essential clue to un-derstanding the formation of supermassive black holes in galactic centers.

Star clusters are also an important actor in a key unsolved issue, that of massive star formation. The high radiation pressure by hot stars strongly counteracts accretion during the short pre-main-sequence phase of such objects, hence preventing the growth of objects to more than some 20 solar masses. The fact that all known young high mass stars are found within dense stellar clusters suggests an intriguing alternative scenario to direct star formation, namely that massive stars originate from coalescence of smaller objects.

Young massive starbursts may be very dense, so dense that their early evolution is significantly affected by stellar dynamics. In addition to their influence as a result of their own evolution (energy and momentum input into their environment), very massive stars play a dominant role in the dynamics of these starbursts despite their short lifetimes. Attempts to include these dynamics to follow the early evolution of massive starbursts have been presented by Portegies Zwart et al. (1999), Ebisuzaki et al. (2001), Portegies Zwart & McMillan (2002), Gürkan et al. (2004), and Freitag et al. (2006), in which the authors investigate the conditions necessary to initiate a runaway collision (runaway merger), which may lead to the formation of a very massive object at the center of the cluster, and the possible formation of an intermediate mass black hole.

Mass segregation as a result of stellar dynamics in a dense young cluster, associated with core collapse and the formation of a runaway stellar collision process, was promoted by Portegies Zwart et al. (2004) to explain ultra-luminous X-ray sources (ULX), point sources with X-ray luminosities up to $10^{42}$ erg s$^{-1}$. Such an ULX is associated with the cluster MGG-11, a young dense star cluster with solar-type metallicity about 200 pc from the center of the starburst galaxy M82, the parameters of which have been studied by McCrady et al. (2003). From the natural assumption that the X-rays are due to Eddington-limited mass accretion onto a black hole (BH), it is straightforward to show that the mass of the BH must be at least 1000 $M_\odot$.

Two very recent studies further demonstrate the probable existence of VMS in the local Universe: First, Gal-Yam et al. (2009) discovered a new optical transient (SNF20070406-008) and classified it as a Type Ic supernova. The observed light curve fits the theoretical one calculated from pair-instability supernova with a helium core mass around 100 $M_\odot$. Therefore, the progenitor must have been a VMS. Second, using HST and VLT spectroscopy and high resolution near-IR photometry, Crowther et al. (2010) concluded that the LMC cluster R136 hosts several stars whose initial masses were significantly larger than 100 $M_\odot$, perhaps up to ∼ 300 $M_\odot$.

We note here that the actual observation of massive-star formation in ultradense H II regions has only become possible recently, due to the remarkable progress in infrared and millimeter astronomy at high angular resolution. But work is also in progress to determine abundances in H II regions in galaxies of different metallicities and to draw conclusions about the spectral energy distributions of the irradiating stellar fluxes and thus about the upper mass range of the stellar content of the clusters (high-quality far-infrared spectra taken with the Spitzer Space Telescope are meanwhile available and are interpreted, e.g, by Rubin et al. 2007).

One of the most important factors influencing starburst evolution is the fact that massive and very massive stars lose a substantial amount of their mass on a timescale which is short compared to the typical dynamical timescale of starbursts. This is due to the high luminosities of those stars, and the associated radiation pressure which acts as a driving mechanism for a stellar outflow. The rate of mass loss clearly has strong influence on the further evolution of the stars themselves.

The evolution of VMS has been the subject of a few studies in the last decade. Marigo et al. (2003) considered population III VMS (with a metallicity of $Z/Z_\odot = 10^{-4}$) with initial masses in the range 120 to 1000 $M_\odot$. Belkus et al. (2007) studied the evolution of VMS with an initial mass up to 1000 $M_\odot$ and with $0.001 \leq Z/Z_\odot \leq 0.04$. However, the authors of both papers use the analytic formalism presented by Kudritzki (2002) to calculate the effect of stellar wind mass loss during core hydrogen burning (CHB), a formalism which we show in this paper greatly overestimates the actual mass loss rates of VMS.

Evolutionary models for LMC stars with an initial mass between 85 and 500 $M_\odot$ were presented by Crowther et al. (2010). They adopted the OB-type stellar wind mass loss prescription of Vink et al. (2001), but in part also applying the recipe beyond the range of parameters for which it had been developed (see below). For core hydrogen burning stars with the properties of late-type WN stars, they continued the computations by applying the stellar wind mass loss formalism of Nugis & Lamers (2000).

Yungelson et al. (2008) evolved an extended grid of VMS with solar-type metallicity but with an "ad-hoc" parameterized stellar wind mass loss formalism. They calibrated their formalism by comparing their rate of a 60 $M_\odot$ star to the one predicted by the Vink et al. (2001) formalism.

However, all these investigations suffered from the fact that they had to extrapolate existing predictions of mass loss rates into the regime of stellar parameters they were exploring. The stellar wind mass loss rates of Kudritzki (2002) were computed for O-type stars with a luminosity $\log(L/L_\odot) \leq 7.03$ (corresponding to stars with an initial mass $\leq 300\, M_\odot$ – this restriction also applies to a recent consideration of predicted mass loss rates by Vink et al. 2011). To calculate the evolution of VMS with an initial mass up to 1000 $M_\odot$, Marigo et al. (2003) and Belkus et al. (2007) had to extrapolate these rates. The Vink et al. (2001) formalism holds for hot stars with an initial mass < 100 $M_\odot$ so that the evolutionary models of Crowther et al. (2010) and those of Yungelson et al. (2008) rely on extrapolated mass loss rates as well. The formalism of Nugis & Lamers (2000) holds for WR stars with a luminosity $\log(L/L_\odot) \leq 6$, so that here also Crowther et al. (2010) had to extrapolate.

But extrapolation of mass loss rates into the regime of very massive stars is not necessary as a matter of principle, because these values can be calculated directly, provided one is able to correctly describe the physics of the expanding atmospheres of those stars. Furthermore, the simulations of the atmospheres can be verified on basis of a comparison of observed and predicted spectral energy distributions and characteristic spectral line features.

Not only do all massive stars show direct spectroscopic evidence of winds throughout their lifetime, but hot star winds are also able to modify the ionizing radiation of the stars dramatically (cf. Pauldrach 1987) and contribute significantly to the dynamics and energetics of the ambient interstellar medium via their output of momenta and energies. But modelling of hot star atmospheres is complicated by the fact that the outflow dominates the physics of the atmospheres of hot stars, in particular regarding the density stratification and the radiative transfer which are drastically modified through the presence of the macroscopic velocity field. The associated mass loss influences the evolution of massive stars in terms of evolutionary time-scales as well as surface abundances and stellar luminosities.

One of the most important points of stellar winds, however, is that they offer complete and completely independent quantitative spectroscopic studies of the most luminous stellar objects (cf. Pauldrach 1987, Pauldrach et al. 1990, Pauldrach et al. 1994, Pauldrach 2003), and as their broad stellar wind lines can be identified in the spectra of galaxies even at high redshift (Steidel et al. 1996), stellar wind lines can be used to determine metallicities of dense stellar clusters and starbursting galaxies (Pettini et al. 2000) and they can yield important information about the population of stars (Leitherer et al. 1999, Leitherer et al. 2010).

One of the objectives of the present paper thus is to present the advanced diagnostic method of stellar wind spectra, including an assessment of the accuracy of the determination of the parameters involved (Sect. 2 and Sect. 3). As a prerequisite for the evolutionary calculations of very massive stars we present consistent computations of stellar wind mass loss rates of VMS with different metallicities and initial masses up to 3000 $M_\odot$, by using detailed hydrodynamic atmospheric models that include a full description of non-LTE[1] line blocking[2] and blanketing[3] and the radiative force (Sect. 4 and Sect. 5).

As massive starbursts contain several $10^5$ stars in a relatively small volume, the evolution of a starburst is significantly affected by dynamical processes induced by $N$-body interactions, where the subtle interplay between stellar dynamics and stellar physics leaves a formidable modelling problem. The paper will finish with a discussion of the influence of realistically determined stellar wind properties on the evolution of dense stellar clusters by means of $N$-body simulations (Sect. 6).

As a direct result the importance of stellar wind mass loss on the formation of intermediate mass black holes will be demonstrated (Sect. 6.2). Sect. 7 gives conclusions and an outlook on future work.

## 2. The tool to determine stellar and wind parameters of massive stars

### 2.1. The wind parameters

Stationary and spherically symmetric winds of hot stars can be characterized by two essential parameters, the mass loss rate $\dot{M}$ and the terminal velocity $v_\infty$. As the winds are initiated and primarily continuously accelerated by the absorption of stellar photons in spectral lines, $v_\infty$ (which is reached at approximately 100 stellar radii) corresponds to the maximum velocity of the stellar wind. The mass loss rate $\dot{M}$ is linked to the velocity field $v(r)$ and the density structure $\rho(r)$ at any radial coordinate $r$ in the wind by the equation of continuity

$$\dot{M} = 4\pi r^2 \rho(r) v(r). \tag{1}$$

These parameters result from consistent simulations of the entire atmospheres of hot stars and they are determined from a comparison of observed and consistently calculated synthetic spectra. This is however only possible on the basis of realistic models.

In the following we describe the diagnostic method which yields complete sets of the stellar parameters and the corresponding wind parameters of massive stars.

### 2.2. The strategy for the determination of the required parameters

To determine complete sets of the stellar parameters (abundances (from H to Zn), the effective temperature $T_{\rm eff}$, the luminosity $L_*$ or the stellar radius $R_*$, the mass $M_*$ or the surface gravity $\log g$, the mass loss rate $\dot{M}$, and the terminal wind velocity $v_\infty$) just from observed UV-spectra is in principle straightforward and simple, because, given the right tools, these parameters are direct observables (see below). All that is required are stellar atmospheric models which include *the hydrodynamic effects of the winds, the radiative transfer, the energy balance, and the non-LTE effects* (deviations from local thermodynamic equilibrium) *of the statistical equilibrium of all important ions* in a realistic way.

The reliability of the results to be obtained depends of course on the degree of sophistication of the underlying model and the fact that UV-spectra are used for the determination of the basic stellar quantities, because these spectra contain many more spectral lines of different elements and ions than the number of parameters to be determined. Thus, due to the hundreds of wind-contaminated spectral lines which the UV-spectra contain from a large number of different ionization stages, *the system is highly overdetermined and thus allows deducing the basic stellar parameters exclusively from a comparison of observed and synthetic spectra*. The observational properties of hot stars can therefore be well explained by the outlined theory if the method used is adequately worked out and the required physics is treated in a detailed and consistent way. Regarding in particular the latter point, a formidable effort in terms of non-LTE multi-line atmospheric modelling is required in order to reach the stage where the consistently calculated synthetic spectra are in sufficiently good agreement with the observed ones (cf. Pauldrach 1987, Pauldrach et al. 1990, Pauldrach et al. 1993, Pauldrach et al. 1994, Taresch et al. 1997, Haser et al. 1998, Pauldrach et al. 1998, Pauldrach et al. 2001, Pauldrach 2003, Pauldrach et al. 2004, Pauldrach 2005).

### 2.3. The method

The diagnostic tool to determine the physical properties of hot stars via quantitative UV spectroscopy requires the construction of detailed atmospheric models. In this section we describe the status of our continuing work to construct realistic models for expanding atmospheres.

Essential steps in working out the theory of radiation-driven winds go back a long way, beginning with a paper by Milne (1926). The next fundamental step was due to Sobolev (1957), who developed the basic ideas of radiative transfer in expanding atmospheres. Radiation pressure as a driving mechanism for stellar outflow was rediscovered by Lucy & Solomon (1970) who developed the first attempt to the solution of the theory. The pioneering step in the formulation of the theory in a quasi consistent manner was performed by Castor et al. (1975). Although these approaches were only qualitative (due to many simplifications), development of the theory was continued due to the promising results obtained by these authors.

---

[1] The term non-LTE refers to detailed modelling of the statistical equilibrium of the occupation numbers of the atomic levels, obtained from actually solving the system of rate equations, without assuming the approximation of local thermodynamic equilibrium (LTE, where the level populations would follow a Boltzmann distribution at the local temperature and density).

[2] The effect of line blocking refers to an attenuation of the radiative flux in the EUV and UV spectral ranges due to the combined opacity of a huge number of metal lines present in hot stars in these frequency ranges. It drastically influences the ionization and excitation and the momentum transfer of the radiation field through radiative absorption and scattering processes.

[3] As a consequence of line blocking, only a small fraction of the radiation is re-emitted and scattered in the outward direction, whereas most of the energy is radiated back to the surface of the star producing there a backwarming. Due to the corresponding increase of the temperature, more radiation is emitted at lower energies, an effect known as line blanketing.

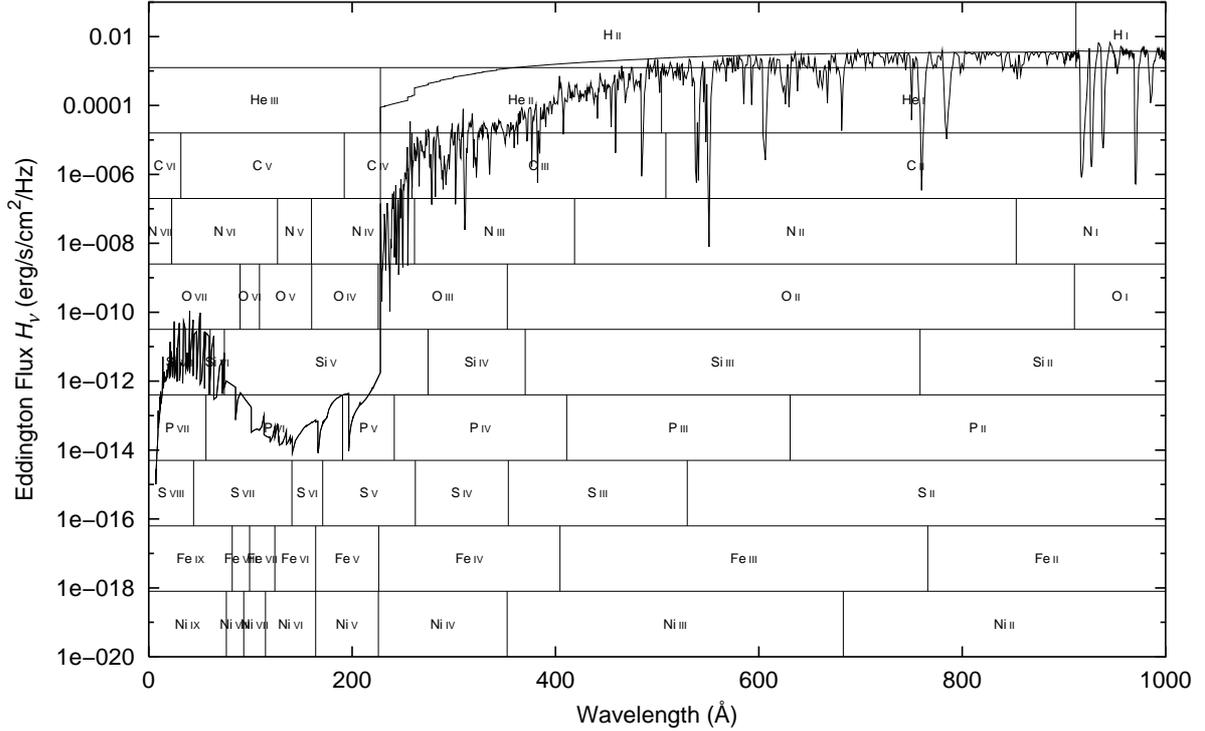

**Fig. 1.** Spectral energy distribution shortward of the Lyman ionization threshold of a $\zeta$ Puppis model atmosphere (see Sect. 3.1.3). The ionization balance depends almost entirely on the ionizing flux – the small vertical bars indicate the ionization thresholds for all important ions –, and this influence can be traced by the spectral lines in the observable part of the UV spectrum[4].

The general concept. The basis of our approach in constructing detailed atmospheric models for hot stars is the concept of *homogeneous, stationary, and spherically symmetric radiation driven atmospheres*. Although the approximations inherent in this seem to be quite restrictive, it has already been shown that the time-averaged mean of the observed UV spectral features can be described correctly by such a method (cf. Pauldrach et al. 1994). But even on basis of these approximations the method is not simple, since modelling the atmospheres of hot stars involves the replication of a tightly interwoven mesh of physical processes: the equations of radiation hydrodynamics including the energy equation, the statistical equilibrium for all important ions with detailed atomic physics, and the radiative transfer equation at all transition frequencies have to be solved simultaneously. For a detailed description of the treatment of non-LTE, radiation transfer, and line blocking and blanketing in our models we refer to Pauldrach et al. (2001); here we will just give an overview of the physics to be treated.

The principal features of the method are:

- *The hydrodynamic equations* are solved. Here the crucial term is the radiative acceleration with minor contributions from continuous absorption and major contributions from scattering and line absorption (including the effects of line-overlap and multiple scattering (cf. Fig. 2)). As the theoretical concept for calculating the radiative acceleration consistently with the non-LTE model is in the main focus of the present approach to the theory, we have described this issue in detail in Appendix A. We note that the consistent treatment of the hydrodynamics is a crucial point, because the hydrodynamics affects the non-LTE model via the density structure and the velocity field, and the radiative transfer with respect to Doppler-shifted spectral lines, but in turn is controlled by the line force determined by the occupation numbers and the radiative transfer of the non-LTE model.

- *The occupation numbers* are determined by the *rate equations* containing collisional and radiative transition rates for all important ions (in total 149 ionization stages of the 26 most abundant elements (H to Zn, apart from Li, Be, B, and Sc) have been considered – a detailed description of the atomic models used is to be found in Sect. 3 and Table 1 of Pauldrach et al. 2001 and in Sect. 2 of Pauldrach et al. 1994 where several Tables and Figures illustrating and explaining the overall procedure are shown). Low-temperature dielectronic recombination is included and Auger ionization due to K-shell absorption (essential for C, N, O, Ne, Mg, Si, and S) of soft X-ray radiation arising from shock-heated matter is taken into account using detailed atomic models for all important ions. Note that the hydrodynamical equations are coupled directly with the rate equations. The velocity field enters into the radiative rates via the Doppler shift, while the density is important for the collisional rates and the total number density.

---

[4] Note that any change of the abundances also produces a change of the EUV and hence of the visible UV spectrum, and that a change of the EUV can even be produced by elements which have no lines in the visible UV.

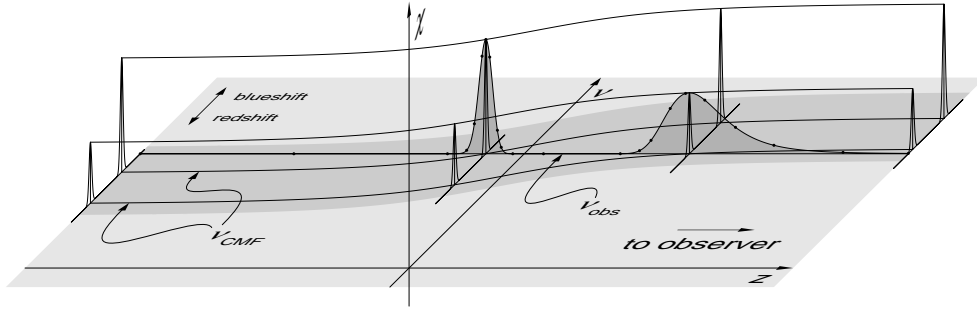

**Fig. 2.** Diagram illustrating the basic relationship of the rest-frame frequencies of spectral lines ($\nu_{\text{CMF}}$) to observer's frame frequency ($\nu_{\text{obs}}$) for one particular (non-core) $p$-ray in the spherically symmetric geometry ($p, z$ geometry). Shown are two spectral lines of different opacity $\chi$ which get shifted across the observer's frame frequency by the velocity field in the wind. The dots represent the stepping points of the adaptive microgrid used in solving the transfer equation in the radiative line transfer[5].

- *The spherical transfer equation* which yields the radiation field at every depth point, including the thermalized layers where the diffusion approximation is used as inner boundary, is solved for the total opacities and source functions of all important ions. Hence, the influence of the spectral lines – the strong EUV *line blocking* including the effects of line-overlap (cf. Fig. 2) – which affects the ionizing flux that determines the ionization and excitation of levels, is naturally taken into account. This is also the case for the effect of Stark-broadening, which is essential for the diagnostic use of certain spectral lines. *Stark-broadening* has therefore, as a new feature, been included in our procedure (cf. Sect. 3.1.3). Moreover, the *shock source functions* produced by radiative cooling zones which originate from a simulation of shock heated matter, together with K-shell absorption, are also included in the radiative transfer (the shock source function is incorporated on the basis of an approximate calculation of the volume emission coefficient of the X-ray plasma in dependence of the velocity-dependent post-shock temperatures and a filling factor).

- *The temperature structure* is determined by the microscopic *energy equation* which, in principle, states that the luminosity must be conserved in the atmosphere. *Line blanketing* effects which reflect the influence of line blocking on the temperature structure are taken into account.

The iterative solution of the total system of equations yields the hydrodynamic structure of the wind (i.e., the *mass-loss rate* and the *velocity structure*) together with *synthetic spectra* and *ionizing fluxes*.

The question whether $\dot{M}$, $v_\infty$, and the spectral energy distribution from our current models of hot stars are already realistic enough to be used for diagnostic issues requires of course an ultimate test, which can only be provided by a comparison of the synthetic and observed UV spectra of individual hot stars. Such a comparison is indeed an ultimate test, since it involves hundreds of spectral signatures of various ionization stages with different ionization thresholds, and these cover a large frequency range. Because almost all of the ionization thresholds lie within the spectral range shortward of the hydrogen Lyman edge (cf. Fig. 1), and as *the ionization balances of all elements depend sensitively on the ionizing radiation throughout the entire wind*, the ionization balance can be traced reliably through the strength and structure of the wind lines formed throughout the atmosphere. And the quality of the ionization balance ascertains the quality of the spectral energy distribution, and it also ascertains the quality of the mass-loss rate and the terminal velocity. Regarding the latter point it is important to realize that the accuracy of the calculation of the radiative acceleration is of the same quality as that of the synthetic spectrum, since the radiative acceleration is calculated analogously and in parallel to the synthetic spectrum (cf. Fig. 2 and Appendix A). This means that the velocity field $v(r)$ and the mass loss rate $\dot{M}$, which are just functions of the basic stellar parameters and the radiative acceleration, are as realistic as the synthetic spectrum is. Hence, the only reliable and remaining step to test the quality of the cur-

---

[5] The method employed is an integral formulation of the transfer equation using an adaptive stepping technique on every $p$-ray in which the radiation transfer in each micro-interval is solved as a weighted sum on the microgrid:

$$I(\tau_0(p,z)) = I(\tau_n)e^{-(\tau_n-\tau_0)} + \sum_{i=0}^{n-1}\left(e^{-(\tau_i-\tau_0)}\int_{\tau_i}^{\tau_{i+1}} S(\tau)e^{-(\tau-\tau_i)}\,d\tau(p,z)\right)$$

where $I$ is the specific intensity, $S$ is the source function and $\tau$ is the optical depth (increasing from $\tau_0$ on the right to $\tau_n$ on the left in the figure – cf. Pauldrach et al. 2001).
To accurately account for the variation of the line opacities and emissivities due to the Doppler shift, all line profile functions are evaluated correctly for the current microgrid-$(z, p)$-coordinate on the ray, thus effectively resolving individual line profiles.
Based on that, application of the Sobolev technique gives for the radiative line acceleration:

$$g_{\text{lines}}(r) = \frac{2\pi}{c}\frac{1}{\rho(r)}\sum_{\text{lines}}\chi_{\text{line}}(r)\int_{-1}^{+1} I_{\nu_0}(r,\mu)\frac{1-e^{-\tau_s(r,\mu)}}{\tau_s(r,\mu)}\mu\,d\mu$$

where

$$\tau_s(r,\mu) = \chi_{\text{line}}(r)\frac{c}{\nu_0}\left[(1-\mu^2)\frac{v(r)}{r} + \mu^2\frac{dv(r)}{dr}\right]^{-1}$$

is the Sobolev optical depth, and $\nu_0$ is the frequency at the center of each line (cf. Appendix A) – thus, the effects of line-overlap and multiple scattering are naturally included ($\chi_{\text{line}}(r)$ is the local line absorption coefficient, $\mu$ is the cosine of the angle between the ray direction and the outward normal on the spherical surface element, and $c$ is the speed of light).
Note that a comparison of the line acceleration of strong and weak lines evaluated with the comoving frame method and the Sobolev technique without consideration of the continuum acceleration is presented in Fig. 5 of Pauldrach et al. (1986), and with the comoving frame method and the Sobolev-with-continuum technique with consideration of the complete continuum acceleration in Fig. 3 of Puls & Hummer (1988), showing the excellent agreement of the two methods.
Finally, the hydrodynamics is solved by iterating the complete continuum acceleration $g_{\text{cont}}(r)$ (which includes in our case the force of Thomson scattering and of the continuum opacities $\chi_\nu^{\text{cont}}(r)$ – free-free and bound-free – of all important ions (cf. Pauldrach et al. 2001)) together with the line acceleration $g_{\text{lines}}(r)$ – obtained from the spherical non-LTE model – and the density $\rho(r)$, the velocity $v(r)$, and the temperature structure $T(r)$ (cf. Appendix A).

rent models is by virtue of their direct product: *the UV spectra of O stars*.

However, the discussion of the stellar parameters determined on basis of this method obviously requires a detailed strategy of the diagnostic steps beforehand, since we do not know whether the models are yet sophisticated enough so that the results of the analysis method applied appear to be reliable. This strategy will be investigated in the following.

## 3. Consistent analyses of massive O Stars

The objectives of a detailed comparison of synthetic and observed UV spectra of O stars are at least twofold. First, it has to be verified that the higher level of consistency in the description of the theoretical concept of the hydrodynamics with respect to line blocking and blanketing effects, and the corresponding modifications to the models, leads to changes in the line spectra with much better agreement to the observed spectra than the previous, less elaborated and less consistent models. Second, it has to be shown that the stellar parameters, the wind parameters, and the abundances can indeed be determined diagnostically via a comparison of observed and calculated high-resolution spectra covering the observable UV region alone.

### 3.1. The case of ζ Puppis

By means of one of the most luminous O supergiants in the Galaxy we will investigate in the next subsections thoroughly whether the kind of models described lead to consistently obtained realistic results. With respect to this objective we will examine in a first step whether our improvements to the WM-basic code already lead to improvements regarding the wind dynamics and UV spectra calculated on the basis of stellar parameters which are obtained from photospheric diagnostics.

#### 3.1.1. Consistent wind dynamics and UV spectra with stellar parameters from photospheric diagnostics

In general, the diagnostics of O stars requires an estimate of the stellar parameters. These are usually obtained by means of model atmospheres which are applied to photospheric H and He lines in order to determine $T_{\text{eff}}$ and $\log g$ from a fit of the spectral lines (cf. Kudritzki 1980, Puls et al. 1996, Puls et al. 2006). Using the distance modulus the stellar radius is then obtained from the model atmosphere flux and de-reddened photometry which, together with $T_{\text{eff}}$, then gives the luminosity. Finally, the stellar mass is determined from the radius and the surface gravity. Thus, *the mass and the luminosity is as accurate as the distance d is.* But, since the distance is highly uncertain, the mass and the luminosity of an O star are quantities which commonly should be regarded as only roughly known values. Unfortunately this is also the case for the chemical composition, which is, apart from a few special O star cases (see below), in general assumed to be solar.

Thus the situation regarding the mass, the luminosity, and the abundances is very uncomfortable. This is however not the case for the effective temperature and the surface gravity, which are – together with the terminal velocity – the only parameters that are independent of the distance. Table 1 lists the values which are currently regarded as a standard parameter set for ζ Puppis. (Note that the value of $\dot{M}$ is not independent of the distance because it is not a directly observable quantity; rather, the quantity derived from observed Hα profiles via model atmospheres is

**Table 1.** Standard stellar parameters (cf. Puls et al. 2006) and wind parameters of ζ Puppis.

| parameter | | standard value |
|---|---|---|
| $T_{\text{eff}}$ | (K) | 39000 |
| $\log g$ | (cgs) | 3.6 |
| $R_*$ | ($R_\odot$) | 18.6 |
| $Y_{\text{He}}$ | ($Y_{\text{He},\odot}$) | 2 |
| $v_\infty^{\text{P}}$ | (km/s) | 2250 |
| $\dot{M}^{\text{P}}$ | ($10^{-6}\ M_\odot$/yr) | 4.2 |
| $\dot{M}^{\text{C}}$ | ($10^{-6}\ M_\odot$/yr) | 3.5 |
| $\dot{M}^{\text{R}}$ | ($10^{-6}\ M_\odot$/yr) | 3.9 |

Note that, like stellar radii determined from the measured flux, the so-called "observed" mass loss rates depend sensitively on the assumed distance (see text).
P from Puls et al. (2006).
C from Cohen et al. (2010); the sensitive dependence of the mass loss rate on the assumed abundances has been extensively discussed by the authors.
R from Pauldrach et al. (1994); the value given is based on observed radio flux values (Abbott et al. 1980 and Abbott et al. 1985) and obtained by using the formula by Wright & Barlow (1975) for bound-free and free-free emission.

$Q \sim \dot{M}(R_* v_\infty)^{-3/2}$, relating mass loss rate, terminal velocity, and stellar radius. Thus, the Hα mass loss rates ($\dot{M}^{\text{P}}$ in Table 1) scale as $\dot{M} \sim d^{3/2}$.)

Based on these parameters and solar abundances we have calculated a consistent atmospheric model (model D$^+$). The dynamical parameters obtained are presented in Table 2, the computed synthetic spectrum is shown in the upper panel of Fig. 3.

As is observed, our model yields a mass loss rate of $\dot{M} = 3.8 \cdot 10^{-6}\ M_\odot$/yr, and this consistently calculated value lies just in between the mass loss rate $\dot{M}^{\text{P}}$ obtained from a fit of the Hα line profile and the mass loss rate $\dot{M}^{\text{C}}$ obtained from a fit of spectral lines in the X-ray frequency range and extremely close to the radio mass loss rate $\dot{M}^{\text{R}} = 3.9 \cdot 10^{-6}\ M_\odot$/yr (cf. Table 1).

**Table 2.** Consistently calculated wind parameters based on standard stellar parameters of ζ Puppis (cf. Fig. 3).

| parameter | | model D$^+$ |
|---|---|---|
| $T_{\text{eff}}$ | (K) | 39000 |
| $\log g$ | (cgs) | 3.55 |
| $R_*$ | ($R_\odot$) | 19.0 |
| $Y_{\text{He}}$ | ($Y_{\text{He},\odot}$) | 2 |
| $v_\infty$ | (km/s) | 2100 |
| $\dot{M}$ | ($10^{-6}\ M_\odot$/yr) | 3.8 |

This can be regarded as agreement between theory and observations, and this is certainly an encouraging result, since it shows that the radiative force has been computed properly in the frame of our procedure. Nevertheless, with respect to our statement "*the radiative acceleration is as realistic as the synthetic spectrum is, since the radiative acceleration is calculated analogous and in parallel to the synthetic spectrum*" our claim is not settled with this result, since the model does by no means reproduce the observed UV spectrum (cf. Fig. 3) in required detail![6]

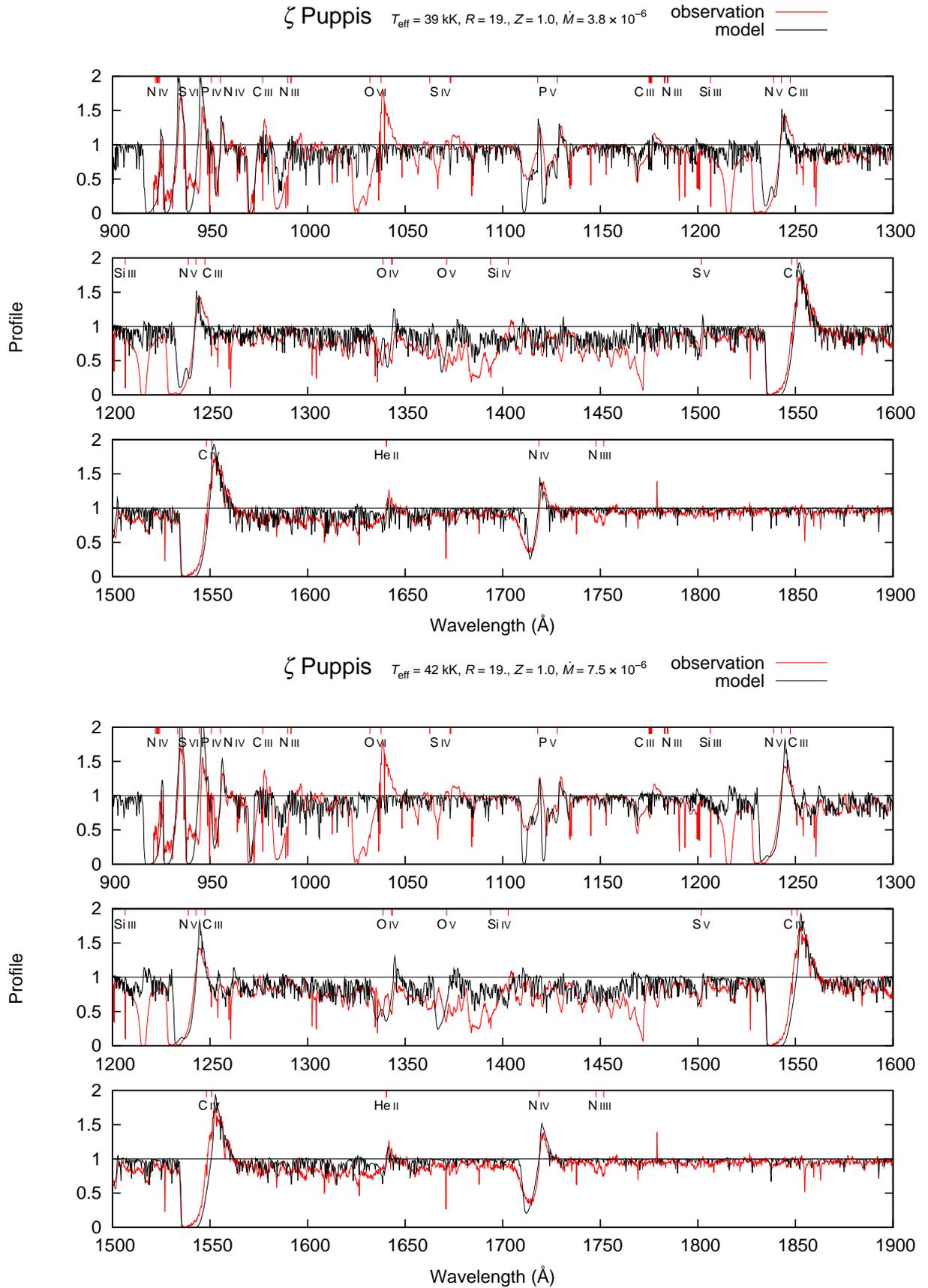

**Fig. 3.** Calculated and observed UV spectra for ζ Puppis. The observed spectrum shows the Copernicus (Morton & Underhill 1977) and IUE (Walborn et al. 1985) high-resolution observations (thin line), and the calculated spectrum (thick line) represents the standard model with respect to the stellar parameters (model D$^+$ – upper panel). The calculated spectrum in the lower panel represents also the standard model, however, here a value at the upper end has been chosen for $T_{\text{eff}}$ (model D$^-$).

Although the model does not reproduce the observed UV spectrum in required detail, an important conclusion can be drawn from its output, and this conclusion concerns the effective temperature $T_{\text{eff}}$. This will be done in the next step.

### 3.1.2. The effective temperature $T_{\text{eff}}$ of ζ Puppis determined from UV diagnostics

Although a preliminary inspection of the visual and/or UV spectrum of the star to be analyzed already gives an estimate of $T_{\text{eff}}$, the effective temperature is much more accurately determined from the ionization balance in the wind, which is reflected in the strengths of spectral lines of successive ionization stages of a number of elements. In particular we have found that the Fe IV/Fe V/Fe VI, the N III/N IV, the O IV/O V, and the C III/C IV ionization balances are well suited for this purpose. Whilst the prominent line features of the latter three pairs of ionization stages are immediately recognized in the UV spectra of O stars (cf. Fig. 3), recognition of the spectral signatures of Fe IV, Fe V, and Fe VI is not so immediately obvious. The wind-contaminated lines from Fe IV to Fe VI are found mostly in the range of 1250 to 1700 Å: the wavelength range from 1250 to 1400 Å is dominated by lines from Fe V and Fe VI, whereas Fe V lines dominate in the range from 1400 to 1550 Å, and Fe IV and Fe V lines are the most prominent features in the range from 1550 to 1650 Å. Along with the P-Cygni profiles of the light elements referred to above, a comparison of the relative strengths of the lines in these wavelength ranges in the observed and the synthetic UV spectrum allows the effective temperature to be constrained to within ±1000 K. (Using the observed and synthetic UV spectra of the O supergiant α Cam this was demonstrated by Pauldrach et al. 2001, whereas Pauldrach et al. 2004 have demonstrated this by means of quantitative predictions for post-AGB[7] evolutionary calculations.)

**Table 3.** Consistently calculated wind parameters based on standard stellar parameters of ζ Puppis, however, here a value at the upper end has been chosen for $T_{\text{eff}}$ (cf. Fig. 3).

| parameter | | model D$^-$ |
|---|---|---|
| $T_{\text{eff}}$ | (K) | 42000 |
| $\log g$ | (cgs) | 3.60 |
| $R_*$ | ($R_\odot$) | 19.0 |
| $Y_{\text{He}}$ | ($Y_{\text{He},\odot}$) | 2 |
| $v_\infty$ | (km/s) | 2040 |
| $\dot{M}$ | ($10^{-6}\,M_\odot$/yr) | 7.5 |

If we now compare these spectral ranges for the two models shown in Fig. 3, it is readily seen that the iron lines of the higher ionization stages start to become too weak in the spectrum of the upper model (model D$^+$) relative to the lower one (model D$^-$); this is especially the case for the spectral range from 1250 to 1400 Å which is dominated by lines from Fe V and Fe VI. As the spectrum of the model shown in the upper part is based on an effective temperature of $T_{\text{eff}}$ = 39000 K, we regard this value to reflect a lower limit to the effective temperature of ζ Puppis. On the other hand, looking at the spectrum of the model shown in the lower part of Fig. 3 (an effective temperature of $T_{\text{eff}}$ = 42000 K has been assumed for this model, cf. Table 3), it is obvious that the C III and N III lines become much too weak, whereas the N IV and O V lines appear already as much too strong features. The model therefore shows too weak lower ionization stages and too strong higher ionization stages, and this means that the effective temperature of this model is definitely too high for the O star ζ Puppis. *We therefore conclude from our analysis that the effective temperature of ζ Puppis is $T_{\text{eff}}$ = 40000 K.*

### 3.1.3. A complete set of stellar parameters determined from UV diagnostics

In order to present fully consistent models for O-type stars that reproduce the observed UV spectra simultaneously with the observed terminal velocities and mass loss rates, we first have to demonstrate how, as a consequence, the models can be used to determine the stellar parameters and abundances just from the observed UV spectra.

As O stars are surrounded by a stellar wind whose extent is much larger than the photospheres of the stars (on the order of hundred stellar radii), the spectral lines formed in this environment contain geometrical information about the physical size of the object, in contrast to objects whose atmospheres are geometrically so thin that they can be approximated by planeparallel models. Moreover, since these spectral lines are the driving mechanism of the wind they are ultimately responsible for the expansion of the atmospheres. Thus, the information contained in the spectral lines can in principle be extracted by means of a consistent theory that describes the dynamical stratification as a function of the stellar parameters. On basis of this fundamental idea the terminal velocity can be expressed as a function which is primarily proportional to the escape velocity $v_{\text{esc}}$

$$v_\infty = f(T_{\text{eff}}, Z, R_*, M_*)\, v_{\text{esc}}, \qquad (2)$$

where the metallicity is denoted by $Z$. We stress that $f$ does not represent a simple linear function; as was shown by Pauldrach et al. (1990) the ratio of $v_\infty$ to $v_{\text{esc}}$ varies a lot from star to star producing a significant scatter (the possible values of $f$ cover the range of 2.7...5.1); this scatter is caused by the back-reaction of changes of the level populations on the line force and therefore on the wind dynamics (cf. Pauldrach et al. 1990 and Pauldrach 1987).

Despite this scatter, the proportionality of $v_\infty$ to the photospheric escape velocity

$$v_{\text{esc}} = \left(2 g_* R_* (1 - \Gamma)\right)^{1/2} \qquad (3)$$

remains one of the most important parameters in the theory of radiative driven winds ($g_*$ is the photospheric gravity, and $\Gamma$ represents the ratio of radiative Thomson to gravitational acceleration (cf. Eq. (7))), since it is already clear from its definition that the determination of the stellar escape velocity contains knowledge of a combination of the stellar mass and the radius; and it is primarily the dependence of $v_{\text{esc}}$ on the mass $M_*$ of the object which offers the possibility to determine the mass very accurately by the predicted value, and due to the strong relation between $\dot{M}$ and the luminosity $L_*$, the latter value can also be determined precisely (cf. Pauldrach et al. 1986). However, the mass and the luminosity do not have to be determined separately, since as each spectral line profile which is formed in the expanding part of the atmosphere depends crucially on $v(r)$ and $\dot{M}$, so that $M_*$ and $L_*$ are automatically determined along with a fit of the synthetic spectrum. As long as the UV spectrum is not reproduced, the mass and luminosity cannot be expected to be reliable (see Fig. 3 and Table 2).

---

[6] We note that the cooling zones of the shocked gas in the wind (see Sect. 3.1.3) have not been considered in this model calculation, because we don't think that the standard model represents already a realistic description of the expanding atmosphere of ζ Puppis (cf. Sect. 3.1.3), and therefore fine tuning of this kind is not required at this stage.

[7] Asymptotic giant branch

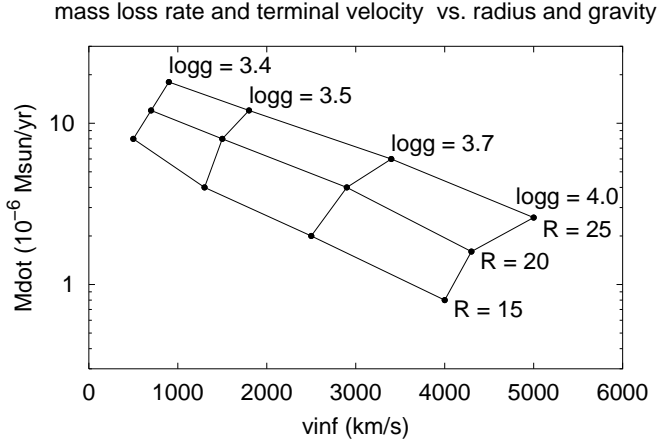

**Fig. 4.** Mass loss rate $\dot{M}$ and terminal velocity $v_\infty$ as a function of radius $R_*$ and surface gravity $\log g$ (as labeled) from a set of dynamically consistent models with $T_{\rm eff} = 40000$ K (i.e., $\zeta$ Puppis-like).

**Table 4.** Radial distribution of the maximum local temperature $T^{\rm max}_{\rm shock}(r)$ in the cooling zones of the shock-heated matter component in the wind of $\zeta$ Puppis (cf. Fig. 7).

| $r$ ($R_*$) | $T^{\rm max}_{\rm shock}(r)$ ($10^6$ K) model A$^+$ | $T^{\rm max}_{\rm shock}(r)$ ($10^6$ K) model A$^-$ |
|---|---|---|
| 1.053 | 0.120 | — |
| 1.058 | 0.120 | — |
| 1.062 | 0.120 | — |
| 1.067 | 0.120 | — |
| 1.084 | 0.120 | — |
| 1.119 | 0.120 | — |
| 1.190 | 0.120 | 0.500 |
| 1.330 | 0.120 | 0.500 |
| 1.590 | 0.600 | 0.500 |
| 1.850 | 0.600 | 0.500 |
| 2.414 | 0.600 | 0.500 |
| 2.849 | 0.600 | 0.500 |
| 3.125 | 0.600 | 0.500 |
| 3.282 | 0.600 | 0.500 |
| 4.793 | 0.600 | 1.420 |
| 6.304 | 0.600 | 1.631 |
| 9.327 | 1.882 | 1.875 |
| 15.37 | 2.118 | 2.098 |
| 27.46 | 2.289 | 2.262 |
| 51.64 | 2.395 | 2.364 |
| 75.82 | 2.434 | 2.402 |
| 95.00 | 2.450 | 2.419 |
| 100.0 | 2.454 | 2.422 |

**The method to derive stellar and wind parameters of massive stars.** As was demonstrated above, it is the consistent hydrodynamics which provides the link between the stellar parameters ($T_{\rm eff}$, $\log g$, $R_*$, and the chemical composition) and the wind parameters ($v_\infty$, $\dot{M}$). Thus, it is primarily the interplay of the non-LTE model and the hydrodynamics that determines the appearance of the UV spectrum.

Computing the wind dynamics consistently therefore permits not only the determination of the wind parameters and the UV spectrum from given stellar parameters, but via this dependence also makes it possible to obtain the stellar parameters from the observed UV spectrum alone. Although the principle behind this idea is not new (cf. Pauldrach et al. 1988, and Kudritzki et al. 1992), it is quite obvious that only the recent generation of models has reached a degree of sophistication that makes such a procedure workable instead of being purely an academic option. How the method works in principle is shown in Fig. 4, where the behavior of the mass loss rate $\dot{M}$ and the terminal velocity $v_\infty$ is presented from consistent models for a systematic variation of the radius $R_*$ and the surface gravity $\log g$ (keeping $T_{\rm eff}$ and the abundances fixed).

To illustrate the crucial point – *the effect of a change in radius and gravity on the UV spectra* –, we have calculated a grid of models with solar abundances and $T_{\rm eff} = 40000$ K and consistent wind dynamics, using radii from 15 to 25 $R_\odot$ and $\log g$ from 3.4 to 4.0. The resulting mass loss rates and terminal velocities are shown in Fig. 4, and the corresponding UV spectra are shown in Fig. 5. In principle, one can now immediately read off the stellar parameters simply by comparing an observed UV spectrum to such a grid at the appropriate effective temperature. However, there is a point which makes our life somewhat harder, and this point regards the abundances. That the properties of O star winds must depend on the abundances is with respect to the driving mechanism obvious: The winds of hot stars are driven by photon momentum transfer through line absorption, and hence the wind momentum rate – the mass loss rate and the terminal velocity – must be a function of the abundances; and this behavior influences the signatures of the spectra in the same way as the abundances do that directly.

If we look at the grid of synthetic UV spectra in some detail, we recognize first that some of the figures shown in Fig. 5 reproduce quite well the overall spectral characteristics of $\zeta$ Puppis, whereas most of the models can definitely be excluded on the basis of the spectra they produce. The decisive point, however, is that none of the spectra presented represents the observed UV spectrum of $\zeta$ Puppis shown in Fig. 3 in detail. And this means that there exists a strict requirement of what has to be done in the next step: *the stellar parameters must be determined along with and simultaneously with the individual values of the abundances!* Thus our final task effectively requires a much finer and more sophisticated grid than the one shown.

The operational procedure of the ultimate method begins with estimates of $T_{\rm eff}$, $R_*$, and $M_*$ (obtained as portrayed above) and a set of reasonable abundances. Concerning our guess of the abundances we have to realize at this point that any change of the chemical composition leads not only to a change of the dynamical parameters, but also a change in the EUV spectral energy distribution (Fig. 1 illustrates the situation clearly) which in turn influences the observable UV spectrum; this behavior regards even elements which show no lines in the observable UV. With these parameters, the model atmosphere is solved in the next step and the velocity field, the mass loss rate $\dot M$, and the synthetic spectrum is calculated. The parameters are then adjusted and the process is repeated until a good fit to all features in the observed UV spectrum is obtained.

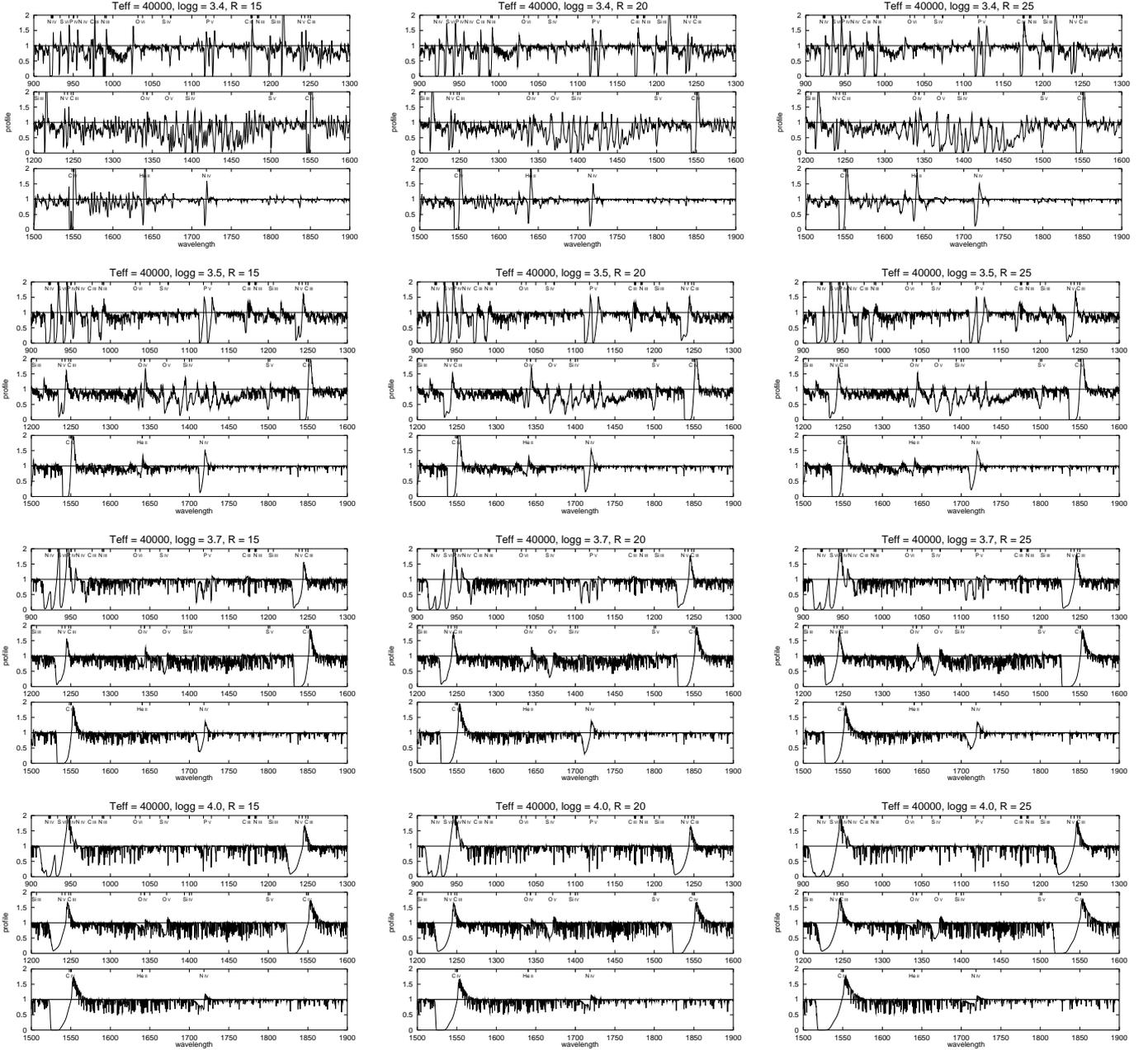

**Fig. 5.** Appearance of the visible UV spectrum as a function of radius $R_*$ and surface gravity $\log g$ for a set of dynamically consistent $\zeta$ Puppis-like models ($T_{\rm eff} = 40000$ K and solar abundances have been assumed for all models). $\log g$ increases toward the bottom, $R_*$ increases to the right.

The derived stellar and wind parameters of $\zeta$ Puppis.
We have carried out this procedure for $\zeta$ Puppis. Comparing the synthetic UV spectrum rendered by our final models to the observations is unquestionably the highlight. As we show in Fig. 6, the observed spectra are reproduced quite well apart from minor differences. The calculated synthetic spectra can therefore be regarded to be in almost perfect agreement with the spectra observed by IUE and Copernicus. A careful look is required to notice the remaining small differences, which are for instance observed for the Si IV and the N IV lines in the upper panel (model A$^+$). However, these small differences just reflect a sensitive behavior of the corresponding lines on the parameters used to describe the shock distribution. This is verified in the spectrum shown in the lower panel of Fig. 6 which offers another comparison (model A$^-$), where only the shock-distribution as input for the model calculation has been changed slightly within the range of uncertainty of the corresponding parameters (cf. Table 4).

The fact that the Si IV resonance line is considerably affected by shock emission has its origin in the soft X-ray radiation of the shock cooling zones which enhances the ionization of Si V, and thus decreases the fraction of Si IV (cf. Pauldrach et al. 1994 and Pauldrach et al. 2001). The significant improvement of the fit of the Si IV resonance line inferred in the lower panel when compared to the upper panel of Fig. 6, is thus a consequence of the change of the soft X-ray radiation in the intermediate part of the wind (Table 4). It is interesting to note that the appreciable differences of the two spectra shown are due to only small differences of the deduced shock distributions, which one would expect to be omnipresent in view of the intrinsic instability of the line-driving force (Owocki et al. 1988), whereas such strong variations in the spectra of O stars are rarely observed (cf. Kaper & Fullerton 1998), implying that possible irregularities in the smoothness of the outflow must be small and evenly distributed.

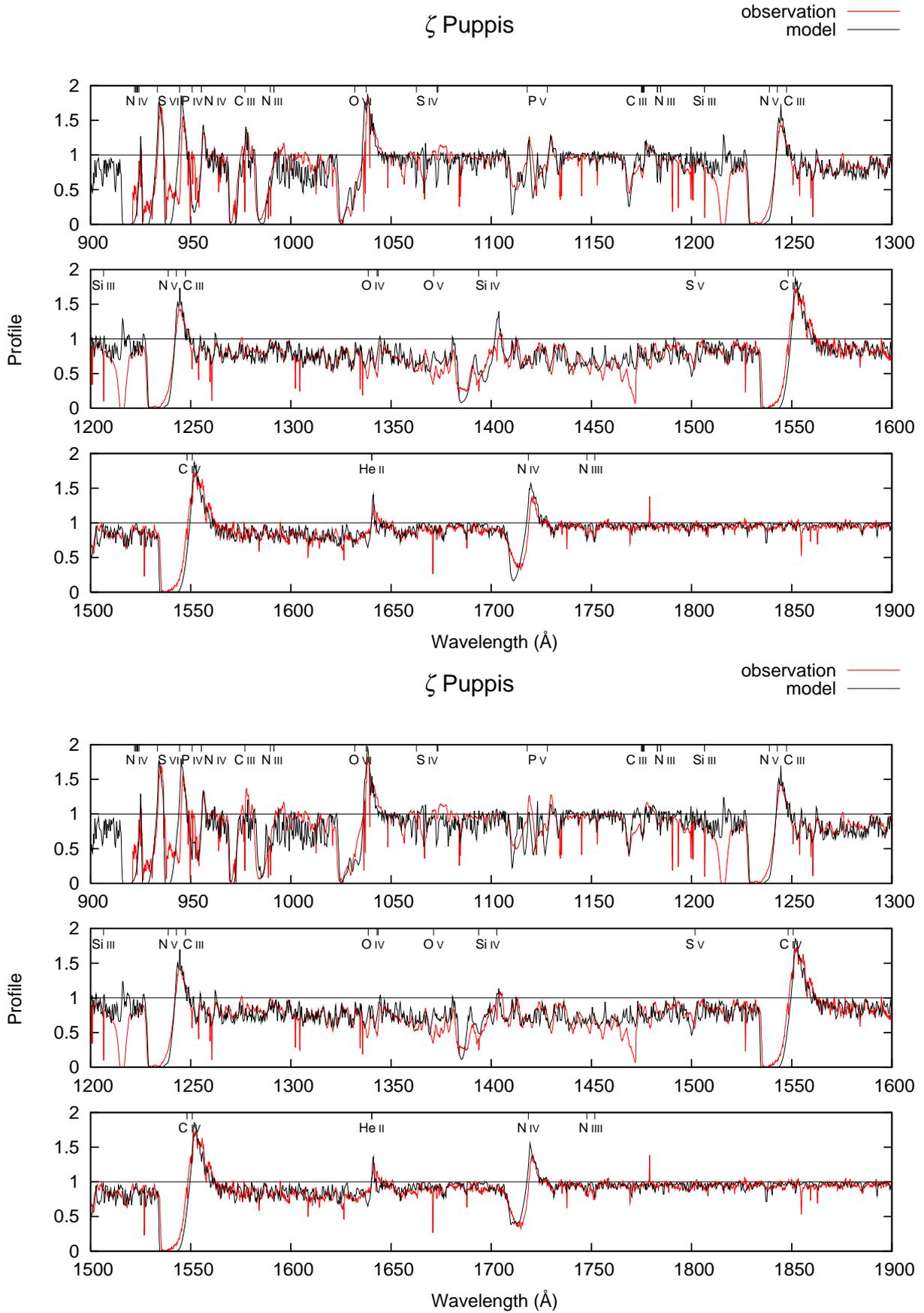

**Fig. 6.** Calculated and observed UV spectra for ζ Puppis. The calculated state-of-the-art spectra represent our final models (upper panel – model $A^+$, lower panel – model $A^-$).

**Table 5.** Stellar and wind parameters determined from UV diagnostics along with consistently calculated atmospheric models of ζ Puppis (cf. Fig. 6).

| parameter | | model A$^+$ | model A$^-$ |
|---|---|---|---|
| $T_{\text{eff}}$ | (K) | 40000 | 40000 |
| $\log g$ | (cgs) | 3.40 | 3.40 |
| $R_*$ | ($R_\odot$) | 28.0 | 28.0 |
| $\log L_*$ | ($L_\odot$) | 6.26 | 6.26 |
| $M_*$ | ($M_\odot$) | 71.9 | 71.9 |
| $Y_{\text{He}}$ | ($Y_{\text{He},\odot}$) | 1.6 | 1.6 |
| $v_{\text{rot}}$[a] | (km/s) | 220 | 220 |
| $v_{\text{turb}}$[b] | (km/s) | 240 | 245 |
| $v_\infty$ | (km/s) | 2075 | 2100 |
| $\dot{M}$ | ($10^{-6}\,M_\odot$/yr) | 13.8 | 13.7 |

[a] $v_{\text{rot}}$ represents the rotational velocity which is deduced from the effect of the rotational broadening. Rotational broadening is important for weak spectral lines which are formed in the velocity region around the sonic point (cf. Puls et al. 1996 and Pauldrach et al. 1994; for instance, the He II line at 1640 Å is considerably influenced by rotational broadening, see Fig. 6).
[b] Hamann (1980) was the first who showed that the assumption of a velocity dispersion $v_{\text{turb}}$ is needed to fit the observed UV P-Cygni profiles of strong lines (see also Puls 1987 and Puls et al. 1993). The physical background of this parameter is well explained by the shock instabilities in the wind flow (cf. Pauldrach et al. 1994) – $v_{\text{turb}}$ can be related to the shock jump velocities. As a consequence of this behavior, the calculated synthetic spectra must account for this velocity dispersion by adopting a depth dependent $v_{\text{turb}}$ law, where the minimum value is usually regarded to be of the order of the sound speed, whereas the maximum value has to be determined from the shape of the strong saturated P-Cygni lines (cf. Pauldrach et al. 1994).

As a general consequence of our procedure and the UV line diagnostics connected the predicted values of $v_\infty$ and $\dot{M}$ are naturally in agreement with the observations; and this regards also the stellar parameters which have been determined precisely via the comparison shown. Since the fit to the UV spectrum provides information about all the basic stellar parameters – effective temperature $T_{\text{eff}}$, stellar radius $R_*$ (or equivalently, stellar luminosity $L_*$), stellar mass $M_*$, terminal wind velocity ($v_\infty$), mass loss rate ($\dot{M}$), and abundances – we therefore have a purely spectroscopic method at hand to obtain an almost complete set of stellar parameters. Table 5 shows the primary stellar and wind parameters obtained by our procedure.

Based on this highly overdetermined system of observables and parameters it turned out that the effective temperature can be determined to within a range of ±1000 K, the abundances to at least within a factor of 2, and the error in $M_*$ is extremely small (< 10 %) due to the sensitive dependence on $v_\infty$ and the small error in this value. Thus *the systematic error in $M_*$ – and hence with respect to $T_{\text{eff}}$ and $\dot{M}$ in $L_*$ – is almost negligible, if the observed UV high resolution spectrum is fully reproduced by the synthetic spectrum.*

The determination of selected important element abundances of ζ Puppis. Not only have the stellar and wind parameters of this object been determined by the model on which the synthetic spectrum is based, but this is also the case for the abundances of the elements C, N, O, Ne, Si, P, S, Cl, Ar, Fe, and Ni (cf. Table 6); and as the model is constrained by the observed UV spectra, the abundances are directly derived from observations.

**Table 6.** Abundances of the elements C, N, O, Ne, Si, P, S, Cl, Ar, Fe, and Ni in units of the corresponding solar values (cf. Asplund et al. 2009) determined from UV diagnostics along with consistently calculated atmospheric models of ζ Puppis (cf. Fig. 6).

| element | model A$^+$ | model A$^-$ |
|---|---|---|
| C | 2.7 | 2.0 |
| N | 8.3 | 6.9 |
| O | 0.15 | 0.15 |
| Ne | 1.7 | 1.7 |
| Si | 8.8 | 7.7 |
| P | 1.0 | 0.62 |
| S | 0.78 | 0.78 |
| Cl | 0.5 | 1.0 |
| Ar | 1.7 | 1.7 |
| Fe | 2.2 | 2.2 |
| Ni | 2.09 | 2.09 |

If we look first at the abundances of the lighter elements, namely C, N, O, we recognize that the sum of the number ratios of the elements is 60 % (model A$^+$) and 30 % (model A$^-$) larger than the solar value obtained by Asplund et al. (2009). Compared to the fact that the corresponding solar value has changed by almost 30 % within the last decade, this is not an explicitly remarkable result. This however is not the case for oxygen: the value we derived for this element from observations contains conspicuous information, since it is reduced strongly compared to the solar value.

Regarding O stars this is however not a brand new result. Pauldrach et al. (2001) found from their diagnostic investigation performed on basis of a model grid that for the most massive stars of their sample (namely HD 93129 A and HD 93250) the oxygen abundance is considerably reduced compared to the solar value. They argued that it is conceivable that these stars are extremely massive precisely because the cooling behavior of the protostellar clouds from which they formed is correlated with a lower oxygen abundance (and possible merging processes of massive stars are also not expected to lower the surface abundance of oxygen). This would mean, however, that a lower oxygen abundance should be observed in all massive and young stars of spectral type O4…O2. Rough estimates of the oxygen abundance support this interpretation, since it is a fact that no very strong O IV and O V lines are observed for this kind of objects (cf. Walborn et al. 1985).

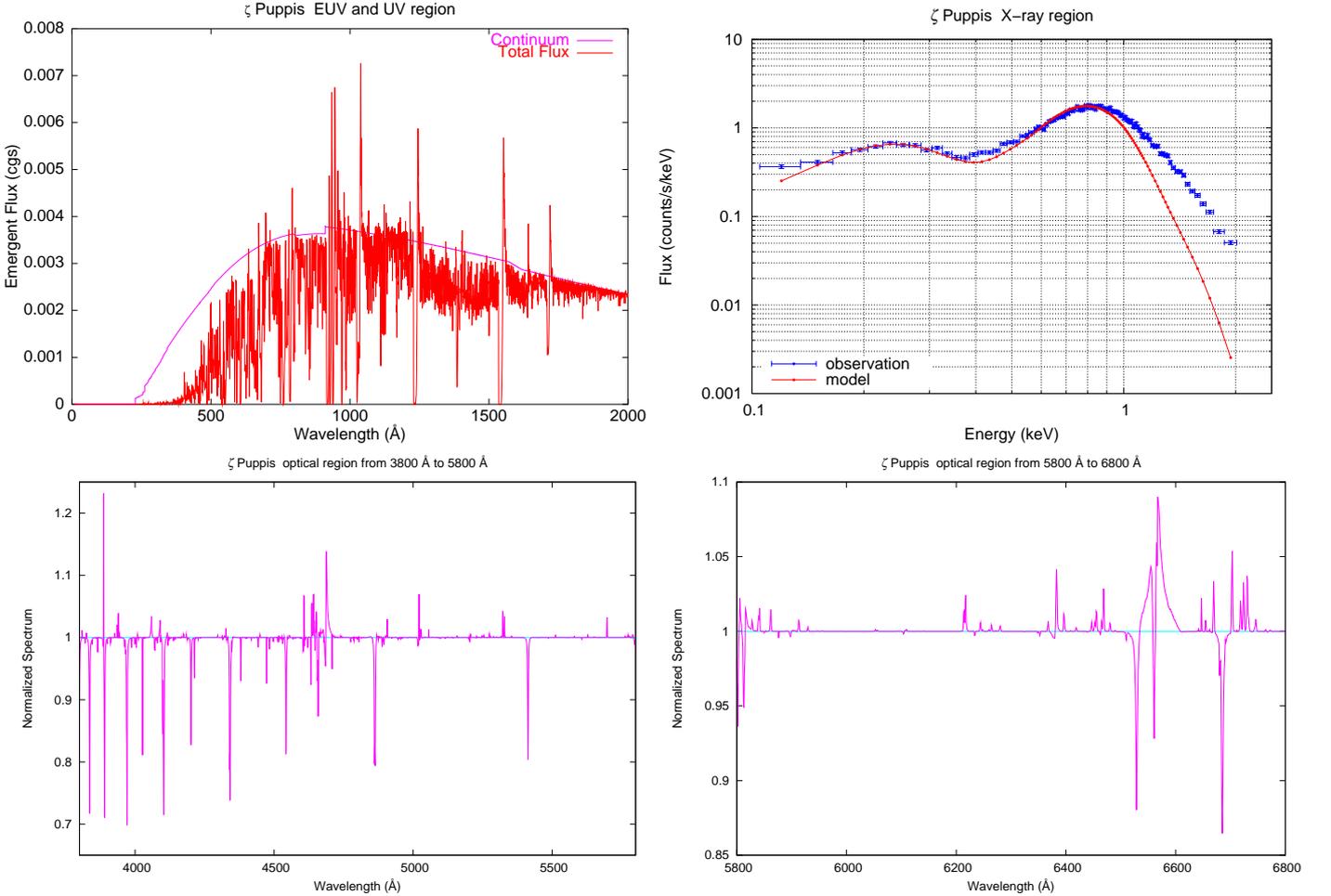

**Fig. 7.** Further calculated spectra for ζ Puppis. *Upper left*: Calculated state-of-the-art spectral energy distribution of the final model of ζ Puppis. *Upper right*: Comparison of the ROSAT-observations (error bars) with the X-ray spectrum of the final model multiplied by the ROSAT PSPC response matrix (thick line)[8]. The calculated state-of-the-art optical spectrum of the final model of ζ Puppis is shown in the lower part. *Lower left*: Spectral range from 3800 to 5800 Å. *Lower right*: Spectral range from 5800 to 6800 Å.

Additionally, the low oxygen abundance is also consistent with the corresponding value found from modelling the X-ray spectrum of ζ Puppis. Cohen et al. (2010) recently obtained a value of $(Z/Z_\odot)_O = 0.2$, and this value is just slightly higher than the value we derived. These authors also found a higher nitrogen abundance of $(Z/Z_\odot)_N = 5.0$, and this value is again only slightly different from the value we obtained here.

We notice moreover that the abundance of oxygen had to be reduced strongly, compared to the solar value, not just with respect to the lines of a single ionization stage, but with respect to the lines of O IV, O V, and O VI. Thus, three succeeding ionization stages of an element are involved in the diagnostics. It is hard to imagine an effect which influences the spectral lines of trace ions and main ionization stages, regardless of whether the lines are subordinate or resonance lines, in the same way.

---

[8] Note that this is not a fit to the ROSAT observations at all, but simply a by-product of determining the shock parameters via modelling the UV spectrum; as such, the excellent agreement must be considered a huge success for UV diagnostics. (The deviation in the hard X-ray regime is explained by the fact that this energy range does not significantly influence the ionization structure in the wind, and thus there are no observables to analyze it via the UV spectrum.)

The only quantity that has the required power of influence is the abundance itself.

Regarding the abundances of the heavier elements, namely Si, Fe, and Ni, we also did not encounter solar-like values. An abundance eight times the solar value is required for Si to account for the opacities and emissivities necessary to produce the corresponding spectral lines, whereas the abundances of iron and nickel must be roughly twice the solar value to reproduce the spectrum of the numerous Fe IV, Fe V, Fe VI, and Ni lines. This result makes clear that the number and mixture of supernovae of type II and Ia that have contributed to the chemical enrichment of the Gum nebula (see next paragraph) must have been considerably different from what has been proposed for the solar neighborhood. Given that star formation is believed to be triggered by superbubbles from stellar winds and supernovae (e.g., Oey et al. 2005), it is not unreasonable to assume that different development histories in the evolution of the interstellar matter leads to regional disparities in the level of the abundances of the heavy elements.

Phosphorus on the other hand show no signs of an underabundance, as is the case for sulfur.

With respect to the structure of the shock cooling zones carbon and nitrogen display a further diagnostic utility. To produce the ionization balance observed in the lighter elements C and N, it turned out that the influence of shock radiation can not start very near the photosphere, because the ionization fractions of the low ionization stages of these elements would then become so small that their resonance lines (C III at 977 Å and N III at 991.6 Å) and excited lines (C III at 1175.7 Å and N III at 1750 Å) would be reduced to just their photospheric components (cf. Pauldrach 1987 and Pauldrach et al. 1994). The way the X-ray spectral region selectively affects the ionization balance of different elements, observable through the lines in the spectrum, therefore provides hard constraints on the lower shock temperatures. Thus, the N III and C III lines have been found to be invaluable diagnostic utilities for this purpose.

From our detailed analysis we conclude that our spectrum synthesis technique allows the determination of abundances precisely because they have such a strong influence on the models and the spectra, so that even abundance adjustments on the order of only 10 % can produce significant changes of the spectra. Thus, the abundances assume a cardinal role among the model parameters. In fact, with respect to any systematic inadequacies that still affect our models, we don't think that the abundances can already be determined to that level of precision, but from an independent objective point of view a determination of the abundances to within a factor of two is certainly reasonable.

**The distance and the mass of $\zeta$ Puppis.** Along with the complete set of stellar parameters we now also obtain the value of the distance almost directly (see discussion above). In contrast to the earlier uncomfortable situation our circumstances have now become quite comfortable, because *the distance is, as usual, as accurate as the mass and the luminosity is. But since the latter values are now highly certain, we now also have an accurate value of the distance at our disposal!* It is therefore quite interesting to look at the value we have obtained (cf. Table 7) and to compare it with the various interpretations offered by the observations.

**Table 7.** Radius and distance of $\zeta$ Puppis – standard values (cf. Puls et al. 2006) versus our values.

| parameter | | standard value | our value |
|---|---|---|---|
| radius | ($R_\odot$) | 18.6 | 28.0 |
| distance | (pc) | 460 | 692 |

In this context one of the most important observations of $\zeta$ Puppis regards the Hipparcos parallax which had been measured to a value of $p = (2.33 \pm 0.51)$ mas (e.g., Schröder et al. 2004). But, as was shown by Schröder et al. (2004)[9], simply deriving a distance $d$ from the Hipparcos parallax $p$ by calculating $d = 1/p$ yields unreliable results for all O stars except for the O9.5V star $\zeta$ Oph and an O8III star. One of the reasons for this finding relies on the well-known Lutz-Kelker bias[10] (Lutz & Kelker 1973). The existence of this bias and the necessity of correcting for it is connected to the conversion from the parallax to its inverse in order to obtain the distance, and a considerable error in the observed value. Because of the fact that there is a larger volume of space behind the star than in front of it, this in general leads to a positive Lutz-Kelker correction term $lk$, such that $d = 1/p + lk$. For the case of $\zeta$ Puppis a representative value of $lk = 115$ pc has been estimated, leading to a range of 419 to 788 pc for the distance to this star; but a different analysis deduced that a plausible value of 330 pc is also in accordance with the Hipparcos measurement (Maíz Apellániz et al. 2008), based on a different reduction of the Hipparcos data by van Leeuwen (2007). However, this modified reduction scheme is also controversially discussed, since it still does not lead to a confirmation of the known distance to the Pleiades[11]. Thus, at the moment the different interpretations of the Hipparcos measurements for $\zeta$ Puppis result in conflicting distance determinations. It is therefore not astonishing that based on their findings Schröder et al. (2004) came to the conclusion that deriving the distance to $\zeta$ Puppis from the Hipparcos parallax seems to be highly risky and that we will have to await future astrometric missions like GAIA, which is expected to achieve a median parallax error of 4 $\mu$as (cf. Perryman 2002), in order to measure a parallax for $\zeta$ Puppis of sufficient accuracy.

On the other hand, our result of 692 pc is astonishingly consistent with an interpretation of distance measurements of $\zeta$ Puppis by Sahu & Blaauw (1993). These authors pointed out that with a space velocity of 70 km/s, $\zeta$ Puppis belongs to the class of runaway OB stars, and that the past projected paths of the star originate from a region close to the young Vela R2 association which has a distance of 800 pc to the sun. Following the past paths Sahu & Blaauw (1993) computed a time scale of $1.5 \cdot 10^6$ years since the star left the Vela R2 association. With respect to this result *the star should thus have an approximate distance of $\sim$ 700 pc to the sun*. The difference to our result is therefore just in the order of 8 pc or 1 %. It should be further noticed that the estimated time scale turned out to be highly consistent with the kinematic ages of massive runaway stars (Blaauw 1993), and it is of course also consistent with the ages of O stars. Moreover, it is also in agreement with the age of the young Vela R2 association (Herbst 1975). With respect to this convincing interpretation of the movement and the past projected paths of the star, $\zeta$ Puppis is obviously not a member of the Vela OB2 association which has a distance of just 450 pc to the sun.

The fact that $\zeta$ Puppis turned out to be a runaway star is also consistent with the mass we determined for this star. With a mass of 71.9 $M_\odot$ $\zeta$ Puppis is possibly the result of stellar merging processes. With respect to this scenario Vanbeveren et al. (2009) presented arguments which favor the dynamical ejection scenario in order to explain runaway stars with a mass in excess of 40 $M_\odot$. Along with these arguments it is very plausible that most of the massive runaway stars, such as $\zeta$ Puppis, are formed during dynamical encounters of massive single stars (or massive close binaries) with massive close binaries. In these cases the runaway star is a merger product of at least two massive stars. That runaway stars can be formed by a binary scenario was already discussed by Blaauw (1961). He pointed out that an exploding massive primary star may disrupt the binary system, leaving a neutron star remnant and a runaway secondary star. Vanbeveren et al. (1998) have investigated this scenario further and considered in some detail its application to the case of ζ Puppis. The mature runaway scenario for ζ Puppis was finally presented by Vanbeveren et al. (2009). By means of this scenario they where able to explain the suspicious surface helium enrichment, the rapid rotation, and the runaway velocity of the star. Their scenario was based on star-binary and binary-binary scattering experiments, where the effects of different masses and different binary periods have been explored. In order to reproduce the observed properties of ζ Puppis by means of the dynamical ejection mechanism with a runaway velocity as observed, it could be shown that the binaries participating in the scattering process always have to be very close. Based on that finding it could be illustrated in numerous numerical experiments that ζ Puppis is obviously a merger of two or three massive stars (cf. Fig. 8).

---

[9] Based on several types of bias reported in the literature in connection with parallax samples of O stars they discussed that straightforwardly deriving a distance from the Hipparcos parallax yields reliable results for two O stars of the sample only.

[10] The Lutz-Kelker bias causes measured parallaxes to be larger than their actual values, and this in turn causes the deduced luminosities and distances to be too small.

[11] Combining precise photometry and radial velocities, firm orbital solutions and accurate physical parameters have been derived for the eclipsing binary HD 23642 by several independent research groups. This object is of importance in determining the distance to the Pleiades, because it is the only known eclipsing binary in this cluster. The resulting distance to the binary is $(132 \pm 2)$ pc (Munari et al. 2004) and $(138 \pm 1.5)$ pc (Groenewegen et al. 2007), whereas the original Hipparcos catalogue gives a distance of 100 to 124 pc and the modified Hipparcos catalogue a distance of 98 to 115 pc to the binary HD 23642.

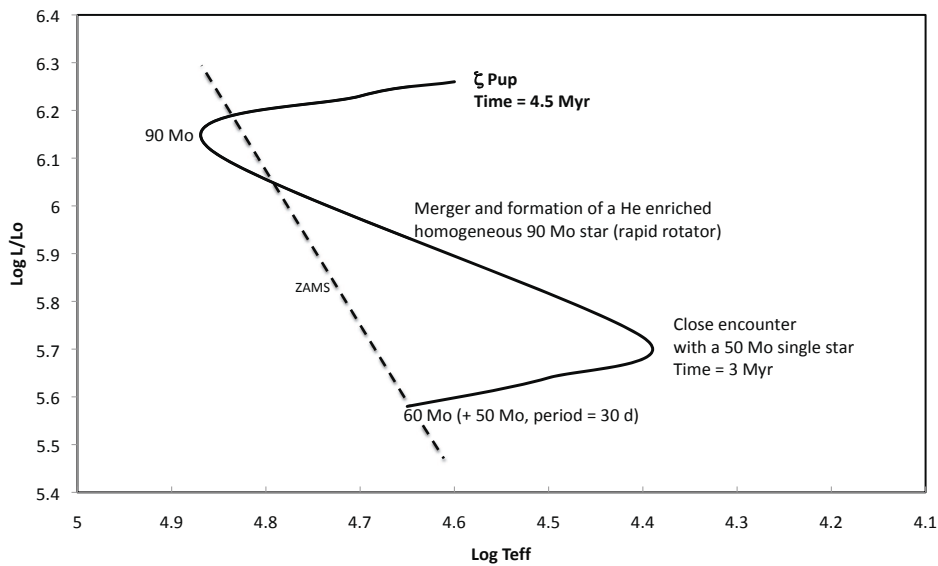

**Fig. 8.** Vanbeveren et al. (2009) discussed a dynamical scenario for ζ Puppis. Here we extend the discussion in view of the stellar parameters derived in the present paper[12]. The figure illustrates a very plausible evolutionary scenario for ζ Puppis, a scenario that also reproduces the stellar parameters discussed in the present paper. A 60 + 50 $M_\odot$ binary with a period of 30 days formed initially in the Vela R2 cluster sinks to the cluster center due to mass segregation. After about 3 Myr, when both stars have transformed already a significant fraction of hydrogen in helium in the core, the binary encounters a 50 $M_\odot$ massive single star[13]. As a consequence of the dynamical interaction, the two binary components merge and the merger acquires a runaway velocity of 60-80 km/s and will leave Vela R2[14]. Since before the encounter, the two cores of the binary components were already significantly He enriched, the mixing results in a homogeneous star with large He abundance (a large He abundance is observed in the case of ζ Puppis). The merger process of two binary components obviously results in a rapid rotator, and the observations reveal that ζ Puppis is a rapid rotator. About 1.5 Myr after the merger process, the resulting star has all the properties of ζ Puppis.

---

[12] In order to discuss the dynamical scenario for ζ Puppis we performed about 1 million binary-single or binary-binary scattering experiments with different binary and single star masses, with binary periods ranging between 5-1000 days. With regard to these calculations the binary eccentricities were taken from a thermal distribution. We notice that many combinations of the parameters produce the required mass and runaway velocity of ζ Puppis, but a most probable one involves at least one massive binary with a period of 30 days.

[13] Because of its high mass, the 50 $M_\odot$ massive single star also sunk to the cluster center due to mass segregation.

[14] As shown by Suzuki et al. (2007) the merger of two massive stars results in a new star with a mass approximately equal to the sum of masses of the two stars (note that during the merger process some 10 $M_\odot$ may be lost) and the merger is efficiently mixed, which means that the merger is a largely homogenized stellar object.

*The X-ray spectrum of the final model for ζ Puppis.* The *EUV and X-ray radiation* produced by cooling zones which originate from the simulation of *shock heated matter* arising from the non-stationary, unstable behavior of radiation driven winds (cf. Lucy & Solomon 1970, Lucy & White 1980, Lucy 1982, Owocki et al. 1988) is, together with K-shell absorption, included in our non-LTE treatment and the radiative transfer calculations (see Pauldrach et al. 2001 and Feldmeier et al. 1997). This renders the possibility to compare our automatically calculated X-ray spectrum with corresponding observations. As we want to inspect the behavior here just qualitatively we have chosen the ROSAT PSPC observations of ζ Puppis for this comparison (cf. Fig. 7).

It is important to emphasize that the calculation shown in Fig. 7 does not present a best fit to the ROSAT-observations, but instead just shows the emergent X-ray spectrum of the final model whose parameters were determined from UV diagnostics. Thus, this comparison is only a by-product of our model calculations, but it nevertheless demonstrates the quality of our parameter determination.

The fact that our treatment accounting for the structured cooling zones behind the shocks reproduces consistently the ROSAT PSPC spectrum *as well as* the resonance lines of N v and O vi gives us great confidence in our present approach.

*The optical spectrum of the final model for ζ Puppis.*
Fig. 7 displays the consistently calculated spectrum for our final model of ζ Puppis in the wavelength range of 3800 Å to 6800 Å, showing that photospheric as well as optical wind lines are also computed by our models. We note at this point that Stark broadening, which plays an important role in shaping the line profiles of several of these lines, has now been included as a new feature in our procedure (Kaschinski et al. 2011). Given that our parameters of ζ Puppis have been derived solely from an analysis of the UV spectrum, it is certainly of interest to see how well these predicted optical lines compare to the observed profiles.

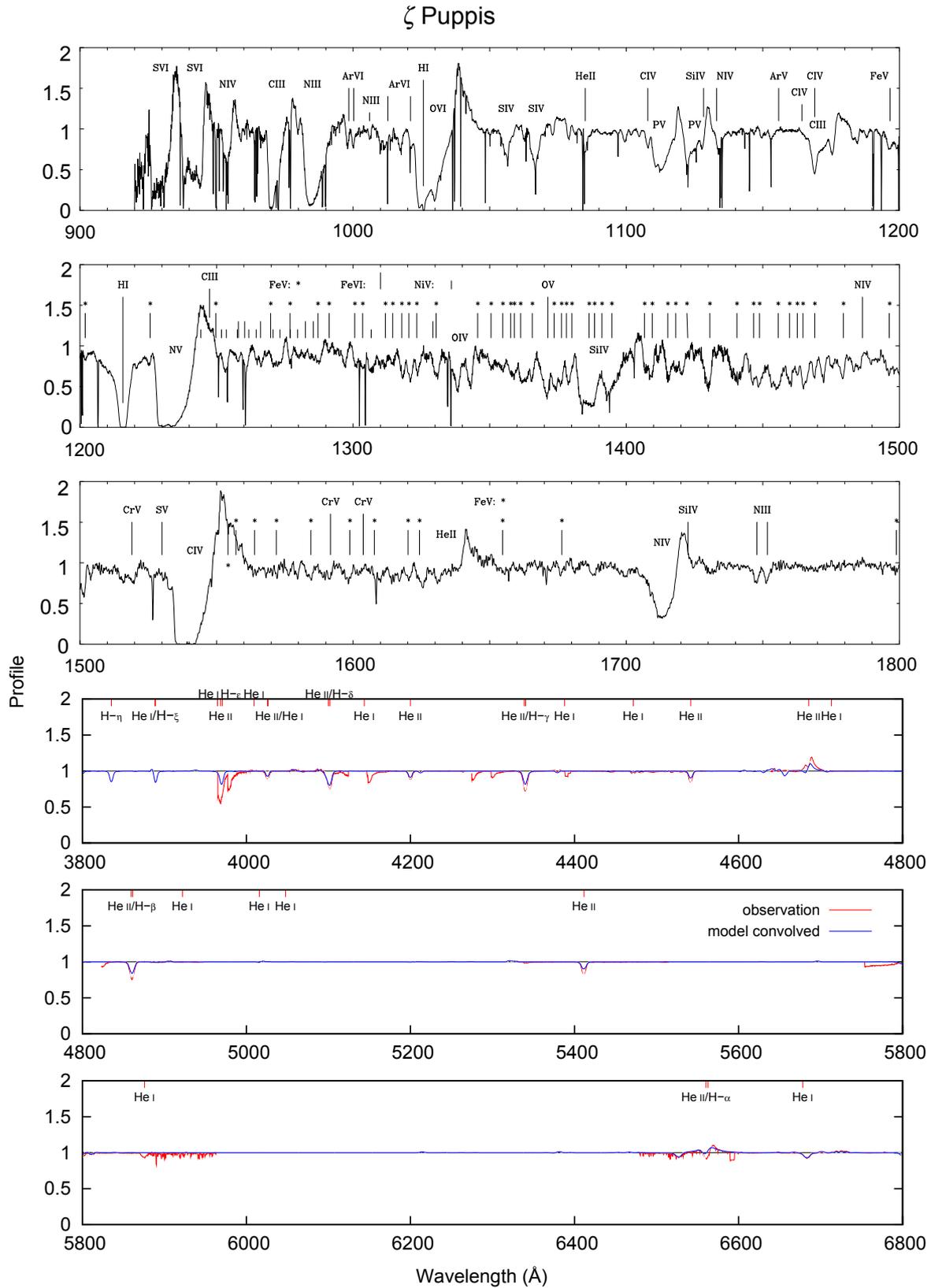

**Fig. 9.** Calculated and observed optical spectrum for $\zeta$ Puppis (lower three panels; the model spectrum has been convolved with a rotational broadening profile and an instrumental gaussian). Although the spectrum in this wavelength range is also computed by our code, it is not our primary means of spectral diagnostics, because compared to the UV spectral range (upper three panels) it is obviously just "the tail of the beast", since it contains only few and only rather weak lines. When compared on the same vertical scale (as shown here) it becomes immediately obvious that the UV spectral range contains much more information, and due to the much stronger lines in the UV part of the spectrum, individual features are far less sensitive to minor uncertainties in the modelling than in the optical range.

This comparison is shown in Fig. 9. Like the X-ray spectrum, the agreement of the optical lines must be considered excellent, considering that these lines have not been used in determining the model parameters. On the other hand, we note minor deviations in the strengths of several of these lines, but we must caution against reading too much into these deviations. Not only does the optical range contain only few lines, but these are weak, subordinate lines, and as such much more susceptible to uncertainties in the modelling than the lines in the UV. In fact, there is currently an ongoing debate regarding in particular inconsistencies in the strengths of optical emission lines, the currently favored explanation being small inhomogeneities ("clumps") in the density (cf. Repolust et al. 2004 and Puls et al. 2006; see also next paragraph), and an ad-hoc "clumping factor" is often employed to bring the line profiles of the model into agreement with the observed line profiles. But supposing that clumping plays a role, then – as long as we have no consistent physical description relating the behavior of the clumping to the fundamental stellar parameters – arbitrarily adapting the clumping factor to fit the optical emission lines will not supply us with any useful information about the stellar parameters, and furthermore involves the risk of covering up intrinsic weaknesses in other parts of the model description. Whatever the real reason for the deviations in the optical spectra will finally turn out to be, *at the present stage it is highly questionable whether the optical hydrogen and helium lines of O stars should really be regarded as superior observables from which reliable information about the stellar parameters can be deduced*. In particular, the wind emission in H$\alpha$ as a quantitative indicator of mass loss rate must be currently considered untrustworthy.

Non-stationarity and clumpiness.  Although the reliability of our models of expanding atmospheres has been proven via the strictest of criteria, namely the comparison to observations (at this point we must stress again that the system of observables and parameters is highly overdetermined and that it is therefore quite impossible for a non-realistic method to find a solution to the problem), a comment is required regarding inhomogeneous features which might be embedded in the wind. It is certainly true that evidence for non-stationarity and clumpiness has been found in the atmospheres of hot stars (e.g., Moffat & Robert 1994, Kaper & Fullerton 1998), but it should also be noted that the amplitudes of the deviations from a smooth stationary model are not very large in general (cf. Kudritzki 1999, as well as Fig. 6 and Table 4 in this paper). Thus, it is not really astonishing that our UV-analysis yields reliable models for time-averaged stellar winds.

Summary.     When using the same stellar parameter for $\zeta$ Puppis as in the analyses of the optical lines, the X-ray regime, and the radio data, we obtain a mass loss rate which is the same as that determined by the other groups (Tables 1 and 2). On the other hand, we obtain a higher value of the mass loss rate as a consequence of the changes in the stellar parameters and abundances necessary to reproduce the observed UV spectrum with the synthetic UV spectrum resulting from a consistent hydrodynamic model. As shown in Table 5 and Fig. 4, this mass loss rate increase is primarily due to the increased radius of the star – note that this is a parameter that cannot be deduced from an analysis of the optical, the X-ray, and the radio spectral parts. Note also that the mass loss rate deduced from an analysis of optical emission lines scales with $R_*^{3/2}$, and therefore the difference to our new value is significantly smaller than it appears to be.

From the discussion above we conclude at this stage that the present method of quantitative spectral UV analysis of hot stars leads to models which can be regarded as being realistic, and the reasoning is that *realistic models are characterized by at least a quantitative spectral UV analysis calculated together with consistent dynamics*. We have demonstrated here that the modifications to the models concerning the energy distributions, ionizing continua, and line spectra lead to much better agreement with the observed UV spectra, and this has important repercussions for the quantitative analysis of the spectra of hot stars. Consequently, we consider this kind of quantitative spectral UV analysis the ultimate test for the accuracy and the quality of the new generation of stellar model atmospheres for hot stars, and based on the improvements discussed in Sect. 2 it defines the status quo.

### 3.2. Comparison of the results of a model grid of massive O stars with observations

With this new method in hand we will present in the following a basic model grid of massive O stars of solar and subsolar metallicity in order to further test the quality of our models. We have also computed selfconsistent models using the stellar parameters of a well-analyzed sample of observed Galactic O stars, namely those of Repolust et al. (2004) (to a large part a re-analysis of the earlier Puls et al. (1996) sample with improved models), to verify that our models do reproduce all main features of the observed winds of individual massive O stars.

Figure 10 compares the observed terminal velocities and the mass loss rates determined by Repolust et al. (2004) to those predicted by our models for the same stellar parameters (always assuming solar abundances). The mass loss rates agree mostly to within a factor of 2 to 3, and the terminal velocities show the observed trend, albeit with a number of differences between observed and predicted values for some of the stars. In general, the reason for these differences can individually only be judged via a detailed comparison of observed and predicted UV spectra, but in most cases they are simply due to a discrepancy in the surface gravity, the main factor which decides between a faster, thinner wind and a slower, denser wind. (Note that differences between predicted and observed values usually appear anticorrelated in the plots of $\dot{M}$ and $v_\infty$.) In some cases (fast rotators) differences are due to having used the "effective" $\log g$ (including centrifugal correction) given by Repolust et al. (2004), which is a good choice near the photosphere but a poorer approximation at farther distances from the stars.

Fig. 11 shows the predicted mass loss rates compared to similar earlier predictions. As can be seen, within the luminosity range of the sample the mass loss rates agree well with earlier predictions, but there is a noticeable scatter (again, for a given stellar radius and luminosity, it is the surface gravity (or equivalently, the stellar mass) which determines the balance between a thin, fast wind and a dense, slow wind, cf. Fig. 4 and Fig. 5), and the slope of a fit to the predicted mass loss rates as function of luminosity will obviously depend on the distribution of the stars with different stellar parameters within the sample. Thus, extrapolating the predictions to higher luminosities may lead to significant errors.

Although from a theoretical standpoint the mass loss rates are the quantity of primary interest regarding the evolutionary development of the stars, it is nonetheless useful to test the predictions of our atmospheric models using a quantity which can be observationally determined with less scatter than the mass loss rates themselves, namely the (modified) wind momentum rates.

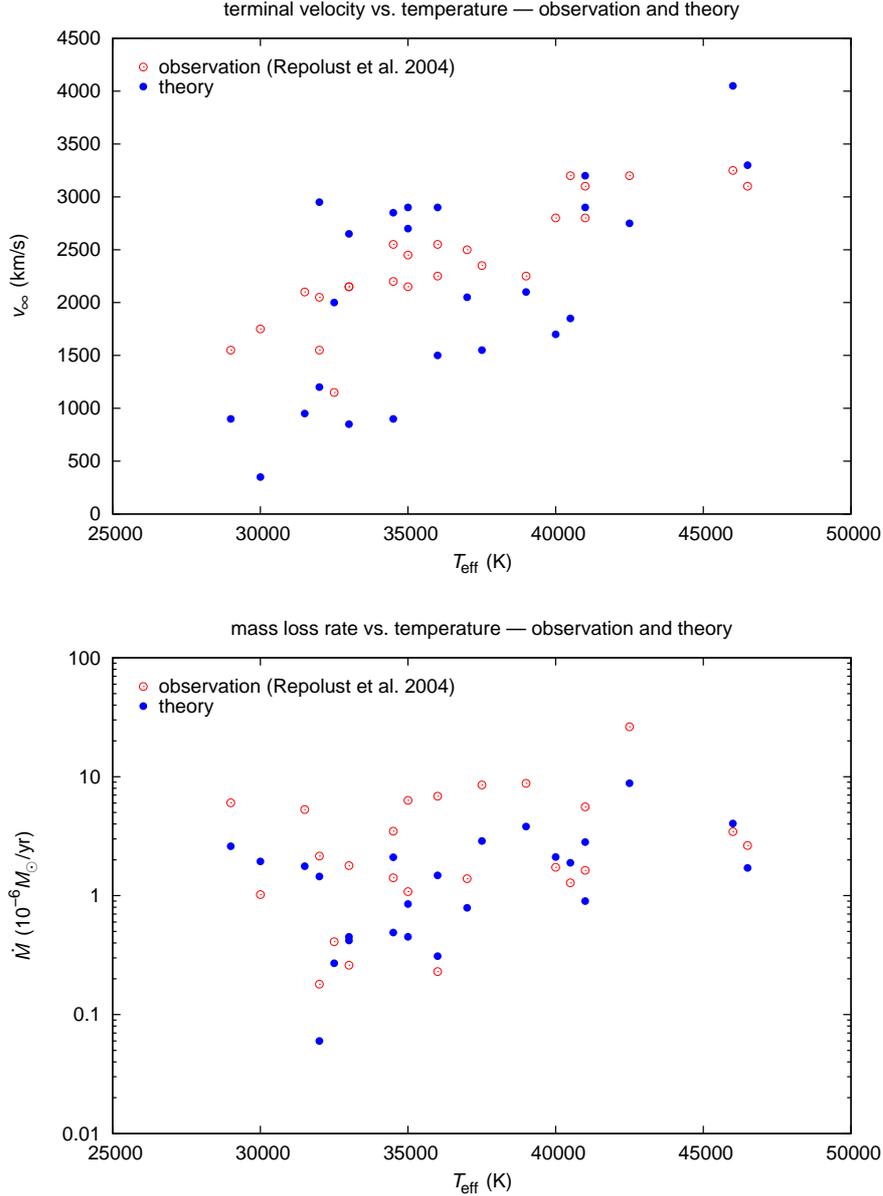

**Fig. 10.** Terminal velocities and mass loss rates computed for models using the stellar parameters of Repolust et al. (2004). Filled symbols denote the values predicted by our model calculations, open symbols represent the observed terminal velocities (top) and the mass loss rates determined by Repolust et al. (2004) from an analysis of the optical spectra (bottom).

One of the fundamental results of the spectroscopic analysis of radiation-driven winds is the existence of a wind-momentum–luminosity relation (WLR) for massive stars (e.g., Lamers & Leitherer 1993, Kudritzki et al. 1995, Puls et al. 1996). Due to the driving mechanism of the wind – transfer of momentum from the radiation field to the gas via photon absorption in metal lines – the mechanical momentum of the wind flow ($\dot{M}v_\infty$) is mostly a function of photon momentum ($L/c$) and is therefore related to the luminosity. Thus, the theory of radiation-driven winds predicts, for fixed abundances, a simple relation between the quantity $\dot{M}v_\infty$, which has the dimensions of a momentum loss rate, and the stellar luminosity.

$$\dot{M}v_\infty \sim R_*^{-1/2} L_*^{1/\alpha'} \qquad (4)$$

where $\alpha'$, related to the power law exponent of the line strength distribution function, is $\approx 2/3$ ($\alpha' = \alpha - \delta$ where $\alpha$ and $\delta$ are parameters of the the force multiplier concept deduced from a correct calculation of the radiative line acceleration which includes line-overlap and multiple scattering – see A). It is practical to plot the log of $D = \dot{M}v_\infty R_*^{1/2}$ (known as the modified wind momentum rate) as a function of $\log L_*$. In this kind of plot the theory predicts, in first approximation, a linear relation, which is indeed followed by all kinds of massive hot stars (cf. Fig. 14).

One of the puzzling results in the study of the WLR of O stars up to recently had been the fact that supergiants seem to follow a distinct WLR different from that of giants and dwarfs (Puls et al. 1996), a finding which cannot be explained by theory under the usual assumptions (i.e., that the winds represent a smooth, stationary, spherically symmetric flow). Although the separation between the supergiants and the other luminosity classes seemed to have decreased with the new parameters of Repolust et al. (2004) compared to the earlier values of Puls et al. (1996), a plot of the wind momenta (Fig. 12) still shows a number of distinct outliers, which are associated with very high wind performance numbers. In contrast, the wind momenta from the hydrodynamic models do not show a division between the luminosity classes. This is the same result as had already been found by Puls et al. (1996) using comparable hydrodynamic calculations; a similar theoretical result was found by Vink et al. (2000) using a completely different method to compute mass loss rates of a grid of models.

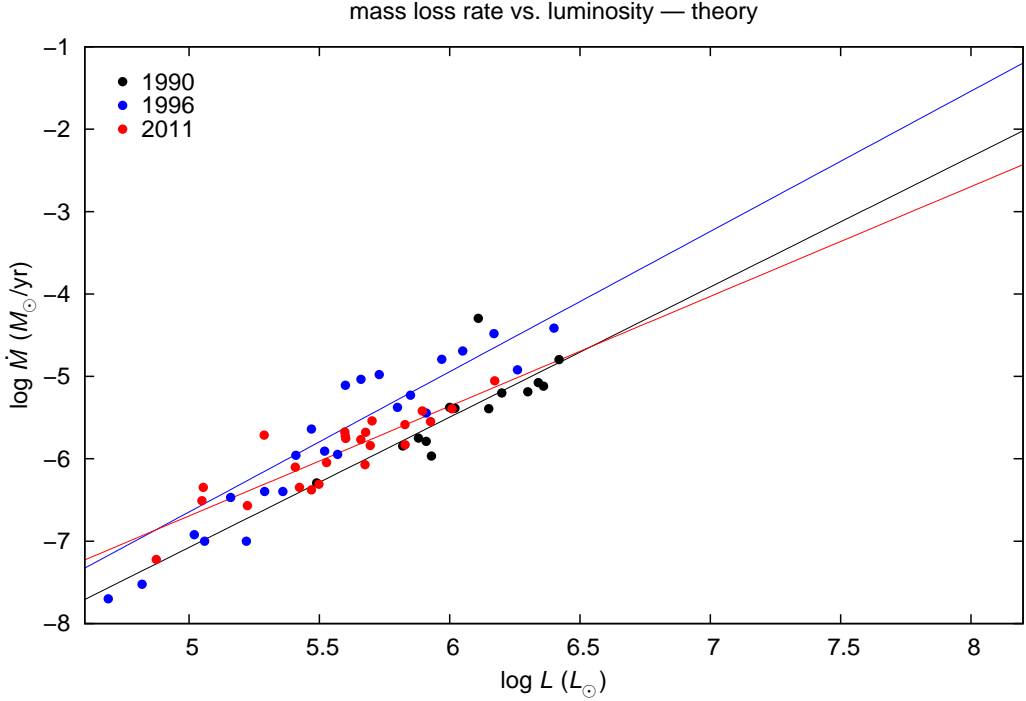

**Fig. 11.** Theoretically predicted mass loss rates versus luminosity for different stages in the development of our model atmosphere code. Although the description of the physics in the model has been continually refined, the predictions regarding the mass loss rates have remained remarkably stable. (The mass loss rates were computed using different sets of stellar parameters, but assuming solar metallicity, from: 1990: theoretical evolutionary tracks (cf. Pauldrach et al. 1990); 1996: observationally determined parameters from Puls et al. 1996; 2011 (current version of the code): observationally determined parameters from Repolust et al. 2004.) The solid lines represent simple linear fits to the calculated values. (For the 1996 sample, the five stars with the lowest luminosities have not been included in the fit.) Although the predicted mass loss rates agree quite well within the luminosity range of the three parameter sets, the slope of the fit nevertheless depends on the distribution of the stellar parameters in each parameter set. Thus, blindly extrapolating the predictions into the regime of very massive stars (with higher luminosities) may lead to significant errors.

Based on the fact that a characteristic property of the outliers was a strong H$\alpha$ emission, Puls et al. (2003) and again Repolust et al. (2004) had proposed that the winds of these objects may be clumped in the H$\alpha$ forming region, with subsequent stronger emission in H$\alpha$ than in the case of an unclumped wind (since H$\alpha$ is a recombination line in the wind, the emission is proportional to the square of the density). Indeed, the latest mass loss determinations by Puls et al. (2006), accounting for wind clumping and the associated enhanced emission in their analysis of the H$\alpha$ profiles, now show no significant differences between the wind momenta of dwarfs and supergiants (Fig. 13), and agree well with our hydrodynamic predictions based on the same stellar parameters.

The significance of the WLR lies in the fact that with observed wind-momentum rates of supergiant winds it allows for a determination of distances (Kudritzki et al. 1995). Thus, this relation makes it possible, in principle, to determine the absolute luminosity of these stars from the observed spectra alone. In combination with measured apparent magnitudes and knowledge of the interstellar extinction this allows determination of the distances to these objects even out to the Virgo and Fornax clusters, beyond the local group. However, a calibration of the relation by means of stars with accurately known parameters is essential for its reliable application to distance measurements. In particular the dependence on metallicity is of fundamental relevance.

Since the winds are driven by metal lines, it is obvious that the wind strengths will also depend on metallicity. This plays a significant role, for example, in applying the WLR to distance measurements of extragalactic objects with a predominantly different metallicity. To avoid a systematic error in the derived distance the metallicity dependence of the wind momentum must be taken into account. Theory predicts, based on line-strength distribution statistics (e.g. Kudritzki et al. 1987, Puls et al. 2000), that the wind momentum $D$ should scale with metallicity $Z$ as $D \sim Z^{(1-\alpha)/\alpha'}$. To verify whether this dependence is reproduced by our detailed numerical simulations, we have computed the wind momenta for a grid of models – a supergiant and a dwarf sequence with temperatures from 30000 K to 50000 K – at two different metallicities, solar and one-fifth solar metallicity. The calculations, plotted in Fig. 15, show that wind momenta of the lower metallicity grid agree well with the theoretically predicted scaling relation with an exponent of 1/2.

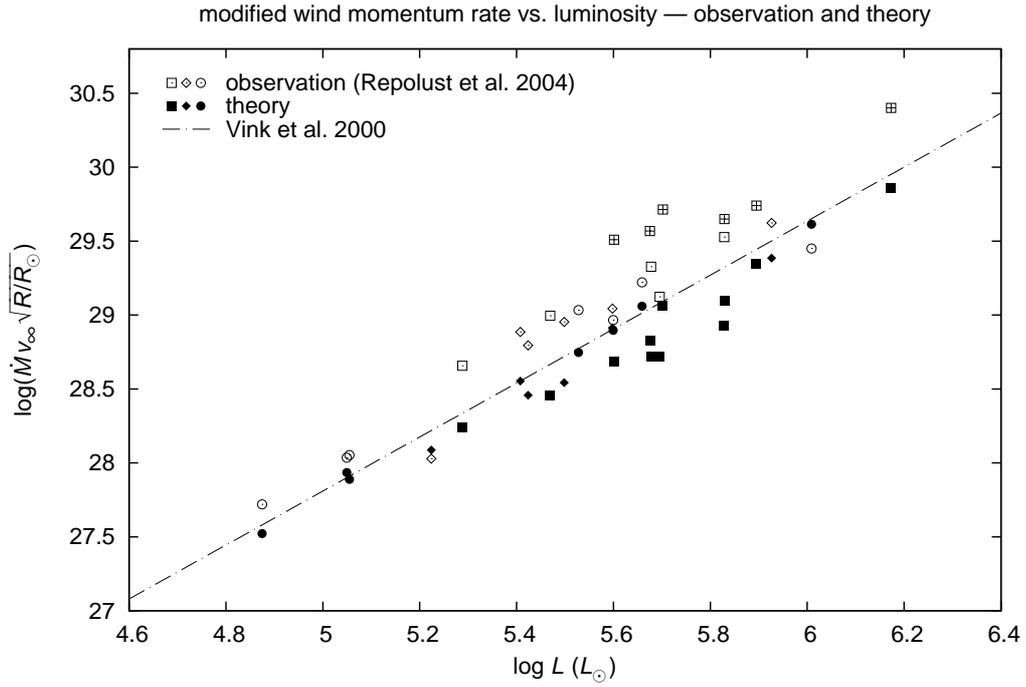

**Fig. 12.** Wind momenta based on the parameters given by Repolust et al. (2004) (open symbols) compared to those resulting from consistent wind dynamics based on the same stellar parameters (filled symbols). The objects for which the observations seemed to indicate anomalously high wind performances > 1 are marked with a "+". (For comparison we have also shown the relation derived theoretically by Vink et al. (2000) using an entirely different approach.)

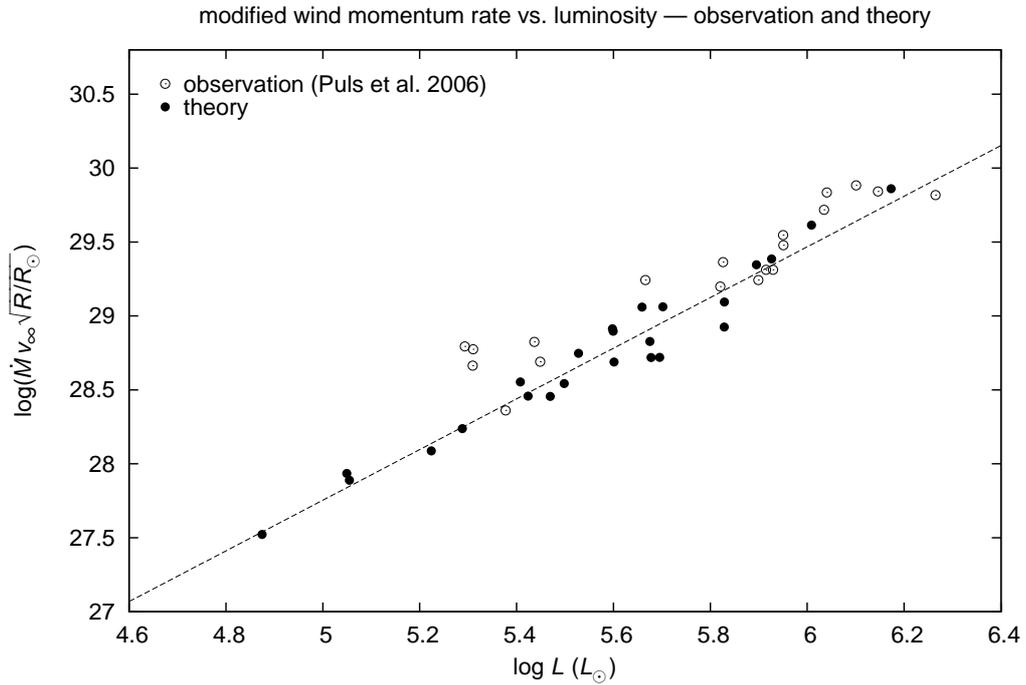

**Fig. 13.** Wind momenta from the analysis by Puls et al. (2006) (open symbols) compared to our consistent wind dynamics (filled symbols). These latest mass loss determinations, accounting for enhanced H$\alpha$ emission from wind clumping, do not indicate systematically higher mass loss rates for supergiants anymore. The dashed line is a simple linear fit to our predicted values, and has a shallower slope than the relation predicted by Vink et al. (2000) (cf. Fig. 12).

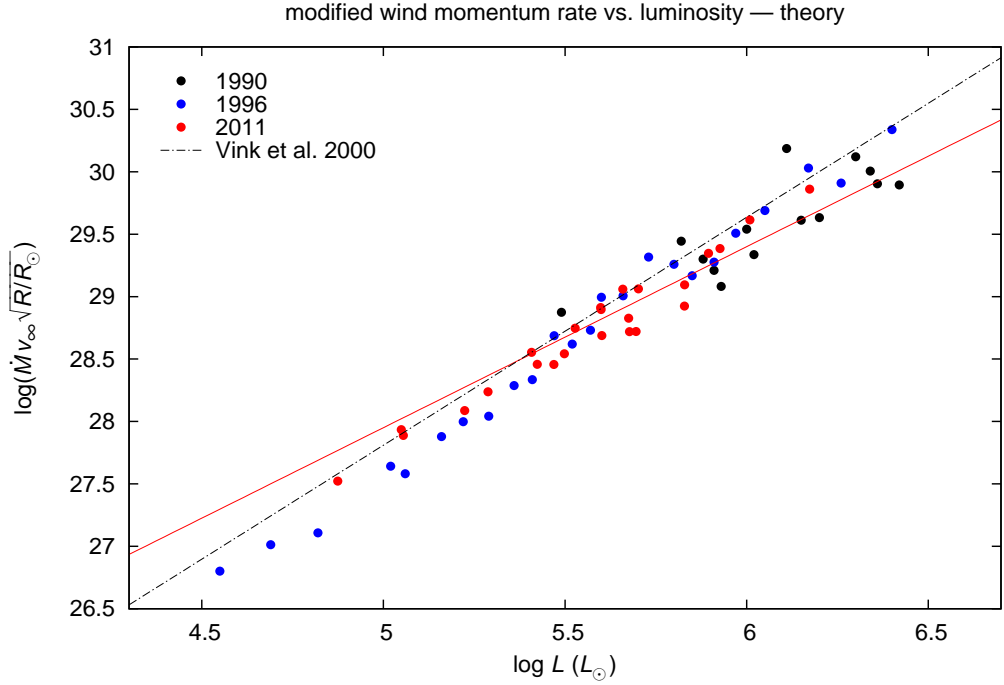

**Fig. 14.** Theoretically predicted wind momentum rates for the same objects as in Fig. 11. As the (modified) wind momentum rates are to first order independent of the masses, a tighter correlation is evident than for the mass loss rates themselves. The dash-dotted line is the corresponding prediction of Vink et al. (2000). Although this predicted relation appears to follow our computed wind momenta in the shown luminosity range quite well, extrapolating it to higher luminosities will lead to significant errors, as the real luminosity-dependence of the wind momenta is actually represented by the solid line, which – anticipating results from further on in this paper – incorporates computed wind momenta for O-type stars with much lower (central stars of planetary nebulae) and much higher luminosities (very massive stars).

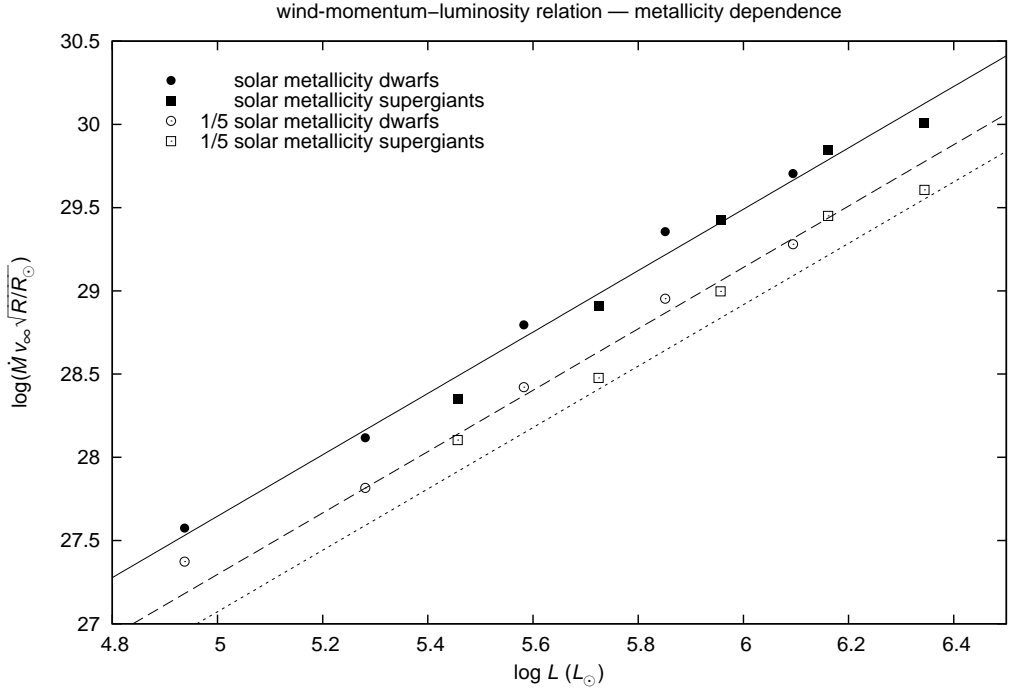

**Fig. 15.** The theoretical wind-momentum–luminosity relation computed for solar metallicity stars (filled symbols) and one-fifth solar metallicity stars (open symbols). The solid line is a linear fit to the calculated wind momentum rates for the solar metallicity stars. The dashed line is the same relation scaled by $(Z/Z_\odot)^{1/2}$ with $Z/Z_\odot = 0.2$, and is in excellent agreement with the actual computed wind momentum rates at one-fifth solar metallicity (open symbols). For the dotted line we have scaled the wind momenta according to the relations predicted by Vink et al. (2001), i.e., $\dot{M}v_\infty \sim (Z/Z_\odot)^{m+pq+q} = (Z/Z_\odot)^{0.82}$ with $m = 0.85$, $p = -1.226$, and $q = 0.13$. (According to Vink et al., $\dot{M}$ scales as $Z^m v_\infty^p$ and $v_\infty$ scales as $Z^q$.) This larger exponent gives a much more pronounced dependence on metallicity, and as can be clearly seen from the figure, for subsolar metallicities this scaling relation predicts wind momenta that are obviously much smaller than those predicted by our models.

# 4. Evolutionary properties of very massive stars (VMS) and their mass loss history

In the introduction we discussed indirect and direct evidence for the existence of VMS, during the early metal poor evolutionary phase of the universe and at present. In this section we combine stellar interior structure calculations and the hydrodynamical atmosphere model described in the previous sections in order to study the wind mass loss properties and the evolution of VMS.

## 4.1. The structure, the stellar parameters, and the evolution of VMS

Very massive stars develop a structure where most of the mass is concentrated in the convective core, whilst only a small fraction of the mass is located in the extended radiative envelope (i.e., VMS have a core-halo configuration). In all stellar evolutionary codes this envelope is treated in hydrostatic equilibrium. However, using recent OPAL[15] opacities, models of VMS with solar metallicity and with an initial mass larger than 1000 $M_\odot$ reach a phase where the radiative force in the outermost layers of the envelope becomes larger than gravity, and this means that hydrostatic equilibrium is not a valid assumption anymore. Fortunately, due to the core containing most of the mass being convective, they evolve quasi-homogeneously and their overall evolution (the temporal evolution of the luminosity, the chemical abundances, the mass) is only marginally affected by the details of the adopted physics of the envelope (Yungelson et al. 2008).

This quasi-homogeneous evolution of the core allows us to use the formalism outlined by Belkus et al. (2007) in the evolutionary calculations of very massive stars, and which is suitable for implementation within a dynamic $N$-body population code (cf. Sect. 6). To illustrate the validity of the recipe, in Fig. 16 we compare detailed evolutionary results of a Galactic 500 $M_\odot$ star calculated with the Eggleton code (cf. Yungelson et al. 2008) with the evolution predicted by the Belkus et al. (2007) method. As shown, both methods agree well in their predictions of the temporal development of the mass, the key parameter with regard to the evolution of a possible merger object in dense cluster simulations.

In the present work, we consider VMS with initial masses between 150 $M_\odot$ and 3000 $M_\odot$ (corresponding to luminosities between $\log(L_*/L_\odot) = 6.57$ and $\log(L_*/L_\odot) = 8.0$), effective temperatures $T_{\text{eff}}$ between 65000 K and 20000 K, and metallicities $Z$ between 0.001 and 0.02 (i.e., between 0.05 $Z_\odot$ and 1 $Z_\odot$).

## 4.2. The expanding atmospheres of VMS

As we have shown that our current models produce realistic results for massive Population I stars, it is now a natural step to also make quantitative predictions regarding the expanding atmospheres of VMS. With respect to this objective we have used our models to calculate $v_\infty$ and $\dot{M}$ for a grid of very massive stars with parameters based on theoretical evolutionary models.

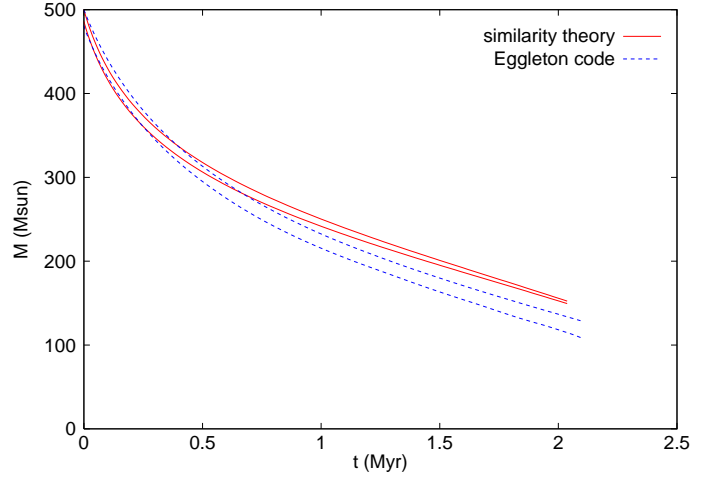

**Fig. 16.** The temporal evolution during core hydrogen burning of the total mass (top line of each pair) and of the core mass (bottom line of each pair) of a 500 $M_\odot$ star calculated with the Eggleton stellar evolutionary code (solid lines) (cf. Yungelson et al. 2008) and with the approximate evolutionary method for very massive star discussed by Belkus et al. (2007) (dashed lines). Since the latter gives results which are in sufficiently good agreement with the results of the detailed calculations, but is computationally much faster, it is an excellent candidate for the use in cluster evolutionary $N$-body codes dealing with a large number of individual stars.

**Spectral energy distributions of VMS.** As is shown in Fig. 17, the numerical model grid produces spectral energy distributions whose characteristics are not much different from those of massive Population I stars (cf. Fig. 1). However, it is also evident from the figure that the metallicity $Z$ as well as the atmospheric hydrogen abundance $X$ have an enormous influence on the behavior of the spectral energy distributions. This influence on the spectra is not directly caused by changes of the abundances themselves, but rather primarily by corresponding changes of the hydrodynamical structures that result as a consequence of the changes of the metallicity and the atmospheric hydrogen abundance.

**Dynamical parameters of VMS.** Regarding the quantitative predictions of our dynamical parameters $v_\infty$ and $\dot{M}$ it is in a next step certainly a good strategy to compare previous estimates of VMS mass loss rates with those calculated with our "WM-basic" procedure[16]. We proceeded as follows:

1. Since our present code differs significantly from the one used to calculate the mass loss rates of Kudritzki (2002), we have recalculated a sub-grid of the one considered by Kudritzki (stars with initial masses $\leq 300 M_\odot$) and compared our values to his.
2. We then computed the mass loss rates of stars with initial masses of up to 3000 $M_\odot$ and compared them with extrapo-

---

[15] The OPAL code was developed at the Lawrence Livermore National Laboratory to compute opacities of low- to mid-Z elements. The calculations are based on an approach that carries out a many-body expansion of the grand canonical partition function. The method includes electron degeneracy and the leading quantum diffraction term. The atomic data are obtained from a parametric potential method and the calculations use detailed term accounting – bound-bound transitions for example are treated in full intermediate or pure LS coupling depending on the element. Line broadening is also treated – based on a Voigt profile that accounts for Doppler, natural width, electron impacts, and for neutral and singly ionized metals broadening by H and He atoms – Iglesias & Rogers 1996.

[16] WM-basic is an interface to a program package which models the atmospheres of hot stars, supernovae Ia, and gaseous nebulae (the latter in 1-dimensional or 3-dimensional radiative transfer) on a Windows platform. The Release comprises all programs required to calculate model atmospheres which especially yield ionizing fluxes and synthetic spectra – cf Pauldrach et al. 2001.

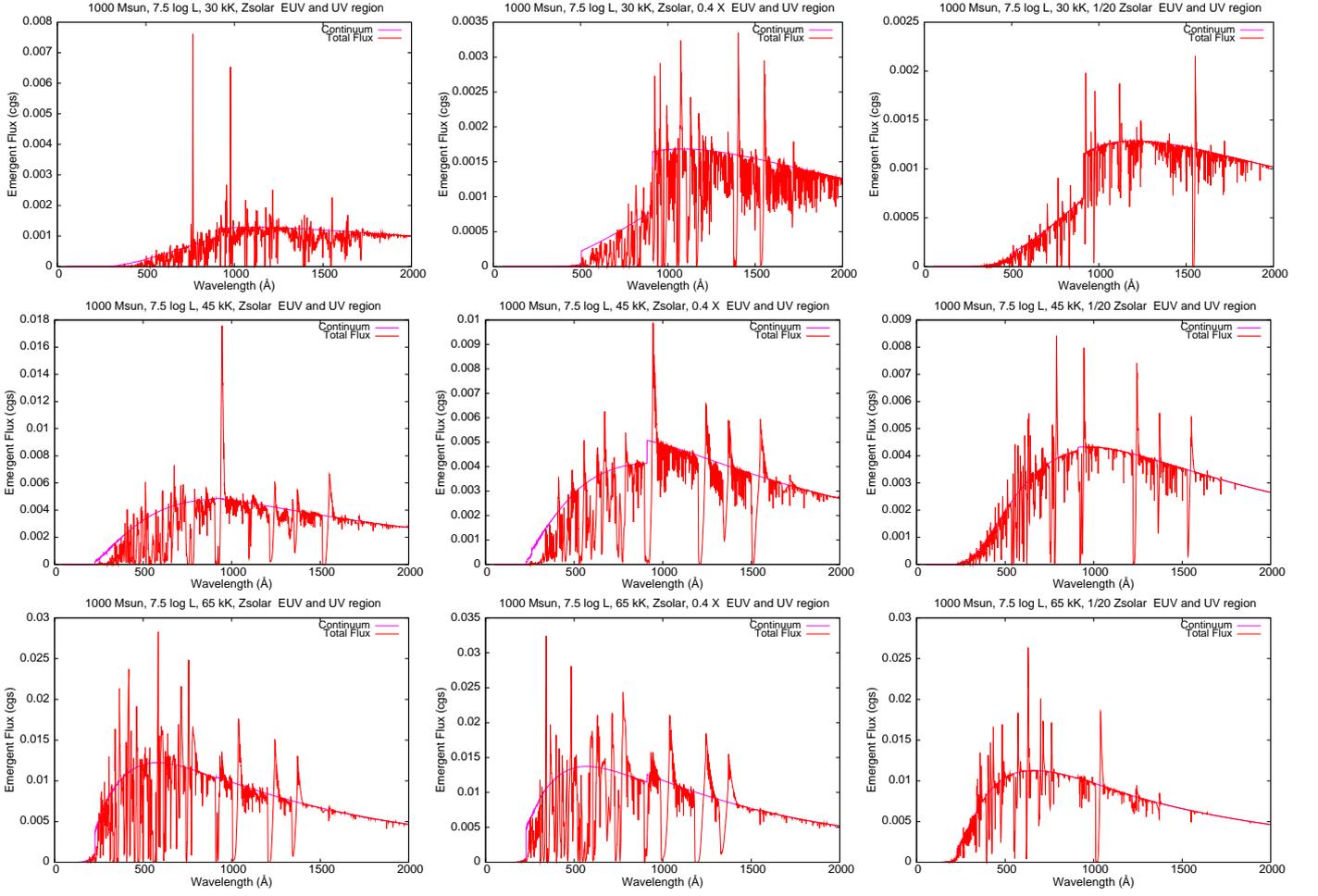

**Fig. 17.** Calculated spectral energy distributions (Eddington flux in cgs units) versus wavelength of the 1000 $M_\odot$ sequence of the VMS model grid stars. The effective temperature of the models in the first row is equal to $T_{\rm eff}$ = 30000 K, in the second row to $T_{\rm eff}$ = 45000 K, and in the third row to $T_{\rm eff}$ = 65000 K. For the models shown in the left column solar abundances have been assumed. This is also the case for the models in the middle column, but for which the atmospheric hydrogen abundance has been additionally reduced to $X = 0.4$. For the models shown in the right column, finally, the metallicity has been scaled down to 1/20 of the solar value. As the figures convincingly show, the metallicity $Z$ and the atmospheric hydrogen abundance $X$ have an enormous influence on the behavior of the spectral energy distributions (mainly due to the corresponding changes of the hydrodynamic structures).

lations using the formalisms of Kudritzki ($\dot{M}_{\rm K02}$), Vink et al. 2001 ($\dot{M}_{\rm V01}$), Puls et al. 1996 ($\dot{M}_{\rm P96}$), and Pauldrach et al. 1990 ($\dot{M}_{\rm P90}$).

3. We calculated the stellar wind mass loss rates of some VMS test cases with a reduced atmospheric hydrogen abundance of $X_{\rm atm} = 0.4$, corresponding to objects in an advanced stage of core hydrogen burning, and compared our values with the previously applied Nugis & Lamers 2000 rates ($\dot{M}_{\rm N00}$).

### 4.2.1. The CHB[17] mass loss rates of VMS with $M_* \leq 300\,M_\odot$

In this subsection we compare our mass loss rates for VMS with those calculated by Kudritzki (2002), which were the first mass loss rate calculations for stellar masses above the usually accepted limit of 120 $M_\odot$. Results of the calculated sub-grid are shown in Table 8. For VMS with a mass up to 300 $M_\odot$, our new rates are only about a factor of 2 to 3 smaller, the differences increasing slightly towards the higher-mass models. Since the core hydrogen burning lifetime of the stars in Table 8 closely approaches 2 Myr it is clear that radiation driven mass loss during CHB is rather unimportant for the overall evolution. Although the effect of rotation is not the scope of the present paper, rotation may of course alter the foregoing conclusion provided that rotation alters the stellar wind mass loss rate of a massive star (since VMS during CHB develop very large hydrogen burning cores and thus the largest part of the star is convective, rotational mixing of the stellar interior is not very important for the evolution of VMS). To illustrate, the rotation-wind relation proposed by Maeder & Meynet (2000) may increase the rates of the VMS considerably since VMS have very high $\Gamma$-factors. However, the relation of Maeder and Meynet has not yet been confirmed either by observations or by detailed hydro-dynamical simulations. Moreover, it can readily be checked that even at critical rotation, the rotation energy is at most 50 % of the escape energy so that the effect of rotation on stellar wind mass loss remains questionable.

---

[17] Core hydrogen burning

### 4.2.2. The CHB mass loss rates of VMS with $M_* > 300\,M_\odot$

Kudritzki (2002) developed an analytical fit formula to his numerical results of the mass loss rates of VMS. Marigo et al. (2003) and Belkus et al. (2007) used this formula in their studies

**Table 8.** Lower part of the VMS model grid: mass loss rates predicted by Kudritzki 2002 ($\dot{M}_{K02}$) versus our values ($\dot{M}_{ov}$).

| $\log L_*$ ($L_\odot$) | $T_{\text{eff}}$ (K) | $\log g$ (cgs) | $M_*$ ($M_\odot$) | $Z_*$ ($Z_\odot$) | $\dot{M}_{K02}$ ($10^{-6}\,M_\odot/\text{yr}$) | $\dot{M}_{ov}$ ($10^{-6}\,M_\odot/\text{yr}$) |
|---|---|---|---|---|---|---|
| 7.03 | 40000 | 3.25 | 300 | 1 | 528 | 220 |
|  |  |  |  | 0.05 | 121 | 59 |
| 6.93 | 40000 | 3.35 | 300 | 1 | 300 | 82 |
|  |  |  |  | 0.05 | 69 | 13 |
| 6.57 | 50000 | 3.79 | 150 | 1 | 48.3 | 23 |
|  |  |  |  | 0.05 | 11.1 | 5.0 |
| 6.57 | 40000 | 3.41 | 150 | 1 | 36.3 | 26 |
|  |  |  |  | 0.05 | 8.3 | 4.7 |

**Table 9.** Upper part of the VMS model grid: mass loss rates extrapolated from the predictions by Kudritzki 2002 ($\dot{M}_{K02}$) versus our values ($\dot{M}_{ov}$).

| $\log L_*$ ($L_\odot$) | $T_{\text{eff}}$ (K) | $M_*$ ($M_\odot$) | $Z_*$ ($Z_\odot$) | $\dot{M}_{K02}$ ($10^{-6}\,M_\odot/\text{yr}$) | $\dot{M}_{ov}$ ($10^{-6}\,M_\odot/\text{yr}$) |
|---|---|---|---|---|---|
| 8.0 | 65000 | 3000 | 1 | 95050 | 385 |
|  | 45000 | 3000 | 1 | 95050 | 923 |
|  | 30000 | 3000 | 1 | 95050 | 1025 |
| 8.0 | 65000 | 3000 | 0.05 | 95050 | 132 |
|  | 45000 | 3000 | 0.05 | 95050 | 285 |
|  | 30000 | 3000 | 0.05 | 24103 | 179 |
| 7.5 | 65000 | 1000 | 1 | 76740 | 98 |
|  | 45000 | 1000 | 1 | 14130 | 190 |
|  | 30000 | 1000 | 1 | 12810 | 250 |
| 7.5 | 65000 | 1000 | 0.05 | 17650 | 35 |
|  | 45000 | 1000 | 0.05 | 3250 | 53 |
|  | 30000 | 1000 | 0.05 | 2946 | 34 |
| 7.03 | 50000 | 600 | 1 | 530 | 34 |
| 7.03 | 50000 | 600 | 0.05 | 120 | 6 |

As in the work of Belkus et al. (2007), for those models for which this extrapolation predicts mass loss rates in excess of the upper limit given by Owocki et al. (2004), we have used the upper limit instead. Note that the true mass loss rates ($\dot{M}_{ov}$) are significantly lower than the extrapolated values.

of the evolution of VMS with masses up to 1000 $M_\odot$. We have checked the validity of this extrapolation by calculating the mass loss rates of stars from 600 to 3000 $M_\odot$ for a number of different values of $T_{\text{eff}}$ and $Z$. Table 9 gives our computed mass loss rates ($\dot{M}_{ov}$) and compares them with the results of extrapolating the Kudritzki fit ($\dot{M}_{K02}$). As can be seen, the extrapolation of the Kudritzki fit largely overestimates the mass loss rates and therefore the VMS evolutionary computations of Marigo et al. and of Belkus et al. may possibly need revision.

### 4.2.3. The CHB mass loss rates of VMS compared to previous predictions

In this paragraph we compare our computed stellar wind mass loss rates of VMS with the rates obtained by extrapolating earlier mass loss rate predictions for O stars, from Pauldrach et al. 1990 ($\dot{M}_{P90}$), Puls et al. 1996 ($\dot{M}_{P96}$), and Vink et al. 2001 ($\dot{M}_{V01}$). Table 10 displays the results. For stars with $M_* \leq 300\,M_\odot$ the predictions differ by about a factor of 2 to 4 (the $\dot{M}_{P90}$ values are surprisingly good in this range). For the higher-mass stars ($M_* \geq 600\,M_\odot$) these extrapolations reproduce the actual computed mass loss rates somewhat better than the extrapolation of the Kudritzki fits, but the same caveats regarding extrapolation beyond the intended range apply. Note that these deviations do not necessarily mean that the original computations are flawed, since the mass loss rates do not differ that much in the luminosity range for which they have been calculated (see Fig. 11), but rather reflect the fact that, as a result of the intrinsic scatter as function of the stellar parameters, the extrapolations will strongly depend on the selection of objects chosen for the original sample.

We also compare the mass loss rates of a VMS with $M_* = 800\,M_\odot$ and with $T_{\text{eff}} = 20000$ K. Here the difference is very large, at least a factor of 10, and we attribute this to the bi-stability jump around 25000 K (Pauldrach & Puls 1990) defended by Vink et al. but not confirmed by our calculations for this model. Note also that our mass loss rates roughly follow the canonical $Z^{1/2}$ relationship.

**Table 10.** Our computed mass loss rates for the VMS model grid ($\dot{M}_{\rm ov}$) compared to previous predicted values: Vink et al. 2001 ($\dot{M}_{\rm V01}$), Puls et al. 1996 ($\dot{M}_{\rm P96}$), Pauldrach et al. 1990 ($\dot{M}_{\rm P90}$).

| $\log L_*$ ($L_\odot$) | $T_{\rm eff}$ (K) | $M_*$ ($M_\odot$) | $Z_*$ ($Z_\odot$) | $\dot{M}_{\rm V01}$ ($10^{-6}\,M_\odot/{\rm yr}$) | $\dot{M}_{\rm P96}$ ($10^{-6}\,M_\odot/{\rm yr}$) | $\dot{M}_{\rm P90}$ ($10^{-6}\,M_\odot/{\rm yr}$) | $\dot{M}_{\rm ov}$ ($10^{-6}\,M_\odot/{\rm yr}$) |
|---|---|---|---|---|---|---|---|
| 8.00 | 65000 | 3000 | 1 | 680 | 29000 | 4600 | 385 |
|  | 45000 | 3000 | 1 | 1380 | 29000 | 4600 | 923 |
|  | 30000 | 3000 | 1 | 680 | 29000 | 4600 | 1025 |
| 8.00 | 65000 | 3000 | 0.05 | 79 | 6500 | 1000 | 132 |
|  | 45000 | 3000 | 0.05 | 160 | 6500 | 1000 | 285 |
|  | 30000 | 3000 | 0.05 | 79 | 6500 | 1000 | 179 |
| 7.50 | 65000 | 1000 | 1 | 230 | 4100 | 750 | 98 |
|  | 45000 | 1000 | 1 | 470 | 4100 | 750 | 190 |
|  | 30000 | 1000 | 1 | 230 | 4100 | 750 | 250 |
| 7.50 | 65000 | 1000 | 0.05 | 28.3 | 910 | 170 | 35 |
|  | 45000 | 1000 | 0.05 | 57.8 | 910 | 170 | 53 |
|  | 30000 | 1000 | 0.05 | 28.3 | 910 | 170 | 34 |
| 7.40 | 20000 | 800 | 1 | 1000 | 2800 | 520 | 98 |
| 7.03 | 50000 | 600 | 1 | 79 | 650 | 140 | 34 |
|  |  |  | 0.05 | 9.7 | 140 | 30 | 6.0 |
| 7.03 | 40000 | 300 | 1 | 200 | 650 | 140 | 220 |
|  |  |  | 0.05 | 25 | 140 | 30 | 59 |
| 6.93 | 40000 | 300 | 1 | 122 | 440 | 90 | 82 |
|  |  |  | 0.05 | 9.5 | 100 | 20 | 13 |
| 6.57 | 50000 | 150 | 1 | 48 | 110 | 25 | 23 |
|  |  |  | 0.05 | 5.9 | 24 | 6.0 | 5.0 |
|  | 40000 | 150 | 1 | 49 | 110 | 25 | 26 |
|  |  |  | 0.05 | 6.0 | 24 | 6.0 | 4.7 |

The metallicity-scaling of the Vink et al. mass loss rates has been carried out according to their recipe, whereas the Puls et al. and Pauldrach et al. mass loss rates have been scaled with $(Z/Z_\odot)^{1/2}$.

**Table 11.** Hydrogen-deficient objects ($X = 0.4$) of our VMS model grid: mass loss rates predicted by Nugis & Lamers 2000 ($\dot{M}_{\rm N00}$) versus our values ($\dot{M}_{\rm ov}$).

| $\log L_*$ ($L_\odot$) | $T_{\rm eff}$ (K) | $M_*$ ($M_\odot$) | $Z_*$ ($Z_\odot$) | $\dot{M}_{\rm N00}$ ($10^{-6}\,M_\odot/{\rm yr}$) | $\dot{M}_{\rm ov}$ ($10^{-6}\,M_\odot/{\rm yr}$) |
|---|---|---|---|---|---|
| 8.0 | 65000 | 3000 | 1 | 11700 | 260 |
|  | 45000 | 3000 | 1 | 11700 | 459 |
|  | 30000 | 3000 | 1 | 11700 | 394 |
| 7.5 | 65000 | 1000 | 1 | 2650 | 102 |
|  | 45000 | 1000 | 1 | 2650 | 142 |
|  | 30000 | 1000 | 1 | 2650 | 115 |
| 7.4 | 20000 | 800 | 1 | 1970 | 43 |

### 4.2.4. The core hydrogen burning mass loss rates of hydrogen-deficient VMS

When VMS evolve and lose mass by stellar winds, there may be a phase where the star is still CHB, but has an atmospherical hydrogen abundance $X_{\rm atm} \leq 0.4$ and a $T_{\rm eff} \geq 10000$ K (cf. Crowther et al. 2010). In many stellar evolutionary calculation the star is then defined as a WNL-type[18] Wolf-Rayet star and the further evolution is calculated by adopting the stellar wind mass loss rate formalism presented by Nugis & Lamers (2000). When this is done for VMS it can readily be checked that also here we are dealing with an extrapolation. We therefore calculated the stellar wind mass loss rates of some VMS test cases where $X_{atm} = 0.4$ (Table 11). The same conclusion applies as in the previous subsections, namely the extrapolation predicts mass loss rates for CHB WNL stars that are much too large.

---

[18] Wolf-Rayet stars are divided in several sub-types reflecting the emission lines in their spectra. WNL-type objects are the youngest of the Wolf-Rayet stars, and they still have substantial amounts of hydrogen in their envelope. Moreover, they tend to be the largest and most massive objects, having yet to loose substantial amounts of mass via their stellar winds. Their emission lines show abundant nitrogen with some carbon present.

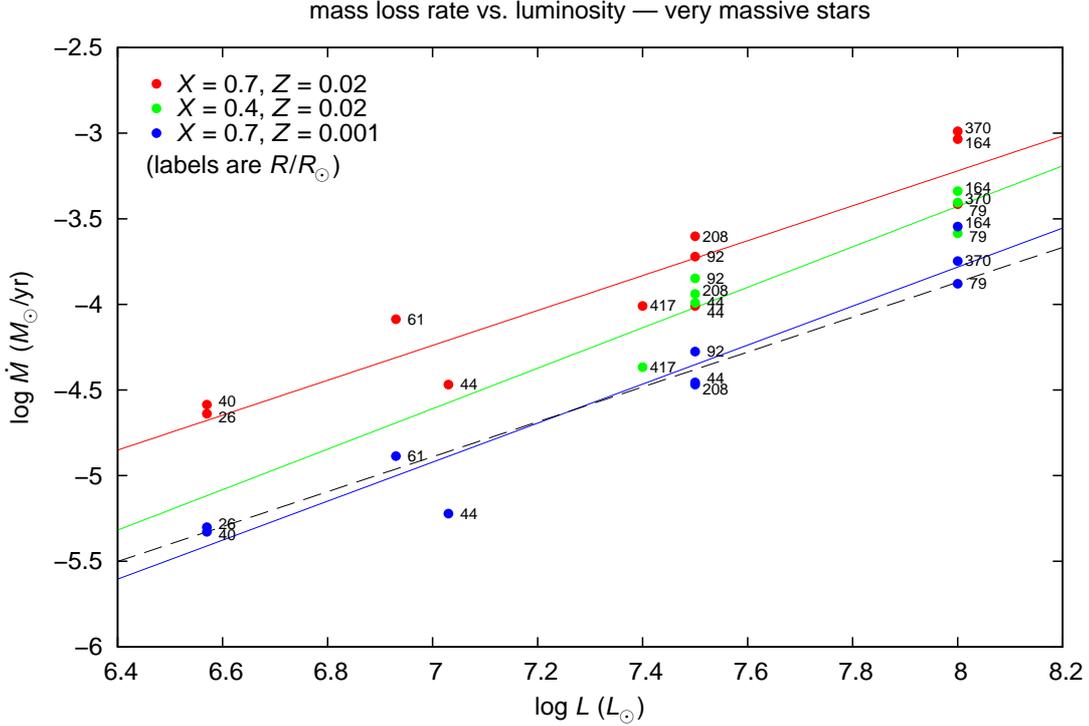

**Fig. 18.** Our theoretically predicted mass loss rates of VMS versus luminosity for solar metallicity, low metallicity ($Z_*/Z_\odot = 0.05$), and for hydrogen-deficient objects ($X = 0.4$) at solar metallicity. The solid lines are linear fits to the calculated values. The dashed line represents the mass loss rates of the solar-metallicity stars (red line) scaled by $(Z/Z_\odot)^{1/2}$ with $Z/Z_\odot = 0.05$.

4.2.5. Overall

Fig. 18 illustrates our mass loss rate calculations and we show rough linear relations of the mass loss rates as a function of luminosity. The results holding for $X = 0.7$ but for different values of $Z_*/Z_\odot$ further illustrate the $Z^{1/2}$ metallicity dependence. Note the significant spread of the individual values around the regression lines, which is a result of stars with different surface temperatures (and thus winds driven by different ionization stages with different line strength distributions) sharing the same luminosity. The $X = 0.4$ results deserve some attention. Since VMS evolve quasi-homogeneously (this is true a fortiori for the most massive ones), the core hydrogen burning stars with $X = 0.4$ obey almost the same mass-luminosity relation as those with $X = 0.7$ and this explains why our VMS with lower atmospheric hydrogen have a smaller stellar wind mass loss rate, i.e., it looks as if very massive core hydrogen burning "Wolf-Rayet" stars have a smaller stellar wind mass loss than their O-type star counterparts. This seems to be at odds with the results of the WNL star WR22 of Gräfener & Hamann (2008) and with the results of the WNL stars in 30 Dor (LMC) of Crowther et al. (2010). We will discuss this in more detail in Sect. 5.1.

## 5. Mass loss rates and wind momenta across five decades in luminosity

From Sect. 3.2 it became clear that it is straightforward to discuss the strengths of stellar winds in terms of their wind-momentum–luminosity relationship (WLR)

$$\log\left(\dot{M}v_\infty(R_*/R_\odot)^{1/2}\right) \sim \log(L_*). \qquad (5)$$

In Fig. 19 we compare computed and observed wind momentum rates of different classes of stars with O-type atmospheres, but widely differing luminosities. From this figure it is evident that there exists a clear common relationship that extends to five orders of magnitude in luminosity.

The fact that the VMS and the CSPNs[19], as objects of significantly higher and lower luminosities and of completely different evolutionary status, have wind momenta corresponding to the extrapolations of the WLR of massive O-stars has to be regarded as an enormous success of the interpretation of their winds being radiatively driven, and that this common relationship reflects the similar physical conditions in their atmospheres that also lead to all these stars being classified as O stars. But we note also that within the parameter space of observed "normal" O stars there is a slight bias towards hotter stars at the high-luminosity end and towards cooler stars at the low-luminosity end, which would unintentionally mislead into predicting a steeper wind-momentum–luminosity relation, and thus higher predicted mass loss rates for the very massive stars at higher luminosities, than is actually the case, since despite their high luminosities our very massive stars in fact have "normal" O-star temperatures.

---
[19] Central stars of Planetary Nebulae

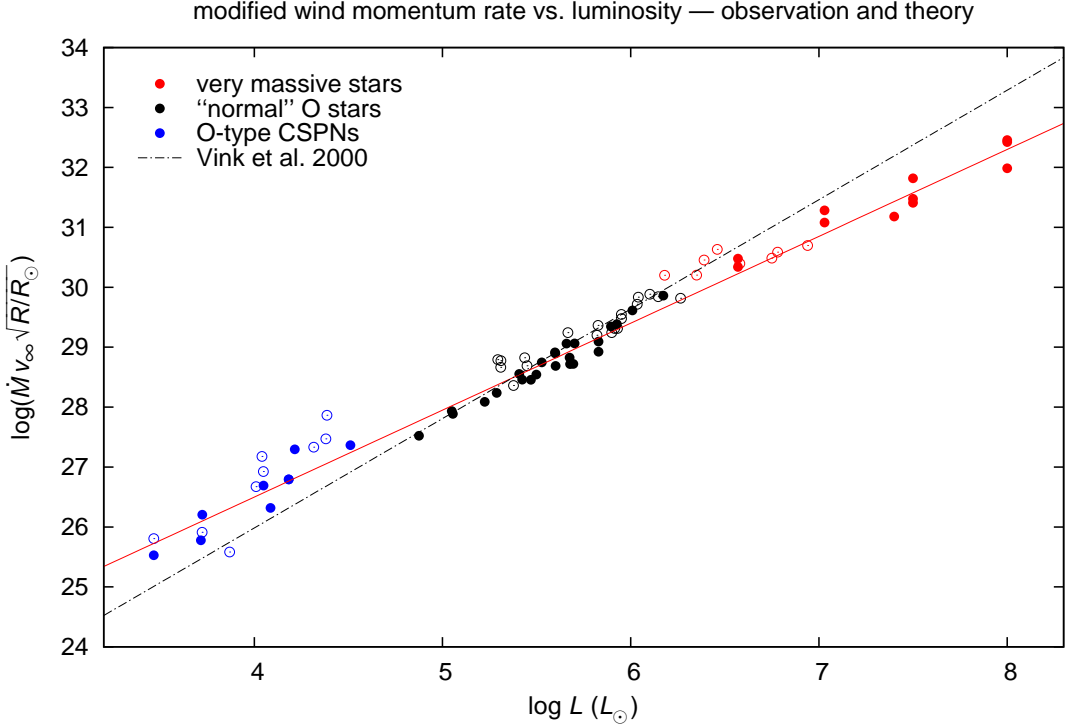

**Fig. 19.** Our computed wind momentum rates (filled symbols) for hot stars spanning almost 5 orders of magnitude in luminosity, compared to observationally derived wind momenta (open symbols; O-type CSPNs: Pauldrach et al. 2004; "normal" O stars: Repolust et al. 2004; very massive stars: Crowther et al. 2010). When viewed in this larger context, the slope of the wind-momentum–luminosity relation is shallower than predicted for "normal" O stars alone (e.g., Vink et al. 2000).

To discuss the strengths of stellar winds in terms of the WLR turned out to be a successful strategy, because the mechanical momentum of the stellar wind flow has to be a function of the photon momentum. This behavior is expressed by the theory of radiation driven winds through the prediction of a dependence of the stellar wind momentum on the luminosity (cf. Castor et al. 1975, Pauldrach et al. 1986, Pauldrach et al. 1990)

$$\dot{M} \propto L_*^{\frac{1}{\alpha'}} \; (M_*(1-\Gamma))^{1-\frac{1}{\alpha'}} \; . \tag{6}$$

Figure 20 demonstrates that the theory of line driven winds does indeed also lead to a proportionality of $\dot{M}$ to $L_*$, as is indicated by the comparison to observations.

However, it also has to be pointed out that the physical nature of the line driving mechanism leads, for example in terms of the chemical composition and the corresponding ionization structure, to an intrinsic scatter in the correlations shown in the Fig. 19 and Fig. 20. A reduction of the metallicity produces for instance much more weaker spectral lines (see Fig. 17) which in turn produce slower winds (cf. Pauldrach et al. 1988), whereas enhanced abundance values produce more strong spectral lines which lead to faster winds.

The mass-loss rate on the other hand is not only influenced by both line strength effects, but also the terminal velocity itself – via a combination of the equation of continuity (cf. Eq. (1)) and the WLR. This means that for a given photon momentum the mass loss rate has to compensate a smaller value obtained for $v_\infty$. As the terminal velocity can for instance be considerably reduced in cases where the star is close to the Eddington-limit, since the escape velocity

$$v_{\rm esc} = \left(2GM_*(1-\Gamma)/R_*\right)^{1/2} \tag{7}$$

(where $\Gamma = L_*/(4\pi cGM_*/(\chi_{\rm Th}/\rho))$ is the ratio of stellar luminosity to Eddington luminosity, $G$ is the gravitational constant, $\chi_{\rm Th}$ is the Thomson absorption coefficient, and $\rho$ is the density) is significantly reduced in these cases, and since the terminal velocity is correlated with the escape velocity (cf. Eq. (2)), and since the photon momentum is not decreased by this effect, this behavior must lead to an enhanced $\dot{M}$ – producing of course an important inherent scatter in the correlations shown.

Figure 21 furthermore demonstrates that the basics of the theory of line-driven winds – represented here in form of the approximative formula given in Eq. (6) for the mass loss rate as a function of $\Gamma$, the ratio of stellar luminosity to Eddington luminosity, and the stellar mass – can still be regarded as a quite reasonable description even for the results of our detailed model calculations, since the data points are principally in accordance with the theoretically derived series of curves – note that the inherent scatter mentioned above distorts of course also this relation somewhat.

As multiple photon scattering, which is inherently considered in our procedure, does not significantly change the behavior of $\alpha'$ (the representative power law exponent of the line strength distribution function appearing in Eq. (6)) for O-stars, the figure mainly reflects the increasing strengths of radiation-driven winds with decreasing distance from the Eddington limit. And as a consequence of this behavior the relation curves upward for all stellar masses as one approaches $\Gamma = 1$ – this was pointed out first by Castor, Abbott, and Klein (Castor et al. 1975).

Moreover, the mass–luminosity parameters of our VMS models are based on evolutionary tracks representing the early phases of stellar evolution. It is therefore not surprising that the objects of our sample do not show so-called optically thick winds (such as those which define, for instance, the envelopes of Wolf-Rayet stars). That the winds of our particular objects are not optically thick is – for the given radii of these objects – already indicated by the $\Gamma$-values obtained, which are not too close to 1 (cf. Fig. 21), and it can easily be verified by means of the spectral energy distributions shown in Fig. 17 for our sequence of 1000 $M_\odot$ models. As the plots show, in addition to the typical P-Cygni profiles of the strong resonance lines there are numerous absorption and emission lines above and below the continuum curve of each model, which clearly illustrates that we are looking down into an extended photosphere.

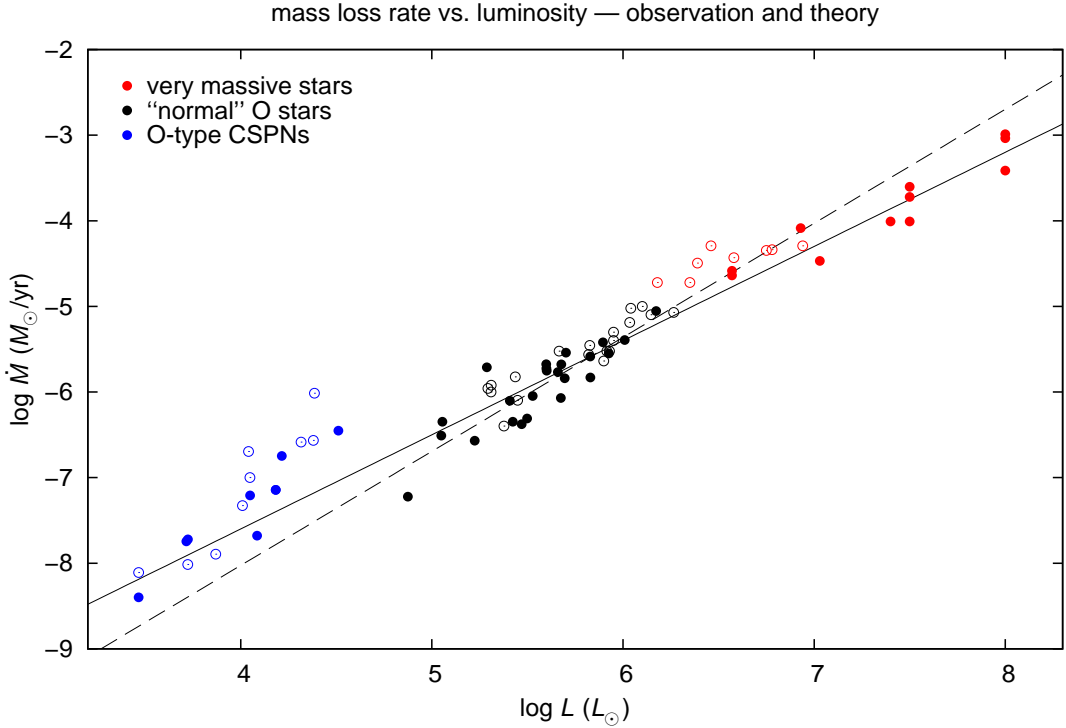

**Fig. 20.** Predicted and observed mass loss rates for for the same objects as in Fig. 19. The dashed line represents a linear fit to the predicted mass loss rates of the "normal" O star sample alone, whereas the solid line represents the "true" relation which extends from O-type CSPNs up to very massive O stars.

As was shown first by Pauldrach et al. (1985) the winds will not be optically thin if not only the $\Gamma$-values become larger than 0.8, but at the time the stellar radii compared to O-stars become much smaller. Thus, the winds of such objects become optically thick in the continuum already in highly supersonic regions (e.g., in Wolf-Rayet stars). It is therefore important to realize that $\Gamma$ is not the only parameter that determines whether a wind is optically thin or optically thick, as Vink et al. (2011) appear to suggest. As was shown by Puls & Pauldrach (1991), for a given mass loss rate and terminal velocity (and assuming a similar velocity profile), the (normalized) radius at which the electron-scattering optical depth ($\tau_e$) reaches unity is inversely proportional to the stellar radius,

$$\tau_e \propto \frac{\dot{M}}{v_\infty R_*}, \qquad (8)$$

and thus for the radii of normal O stars the winds will be optically thin, whereas for much smaller radii (WR stars) the winds can become optically thick (Pauldrach et al. 1985); note that in these cases the wind efficiency – $\dot{M}v_\infty c/L_*$ – also increases considerably and certainly exceeds values of 2 (cf. Pauldrach et al. 1985 and Vink et al. 2011); note also that the finding of Pauldrach et al. (1985), who showed that the proximity to the Eddington limit is the primary parameter to produce the huge mass loss rates of Wolf-Rayet stars (for suitable stellar parameters), has recently been confirmed by Gräfener & Hamann (2008).

Thus, the spread in mass loss rates at a given $\Gamma$ can be naturally understood as a consequence of the variation in stellar parameters of our sample of objects. If we had included in our sample objects whose stellar parameters gave rise to optically thick winds, the scatter in Fig. 21 would become even larger. This is somewhat in contradiction with the result of Vink et al. (2011) who found a strong correlation between $\dot{M}$ and $\Gamma$ for high $\Gamma$-values. However, as only stellar parameters leading to optically thick winds had been chosen by these authors, our finding is an extension of their results, clarifying that a simple correlation between $\dot{M}$ and $\Gamma$ cannot exist.

As the $\Gamma$-values of our present calculated models are not very close to 1 and the radii are large (cf. Fig. 18), it is therefore not surprising that we are exclusively dealing with O-type stars in this paper, even at the high-mass end, and all of our objects have optically thin winds in the optical and observable UV continuum.

### 5.1. The WN5h star R136a3

The LMC cluster R136 in the Tarantula nebula hosts a number of very luminous WR stars. From spectroscopic analysis Crowther et al. (2010) concluded that some of these stars are descendants of VMS with an initial mass larger than 100 $M_\odot$ (perhaps up to 300 $M_\odot$) thus occupying the lower end of the VMS. Since most of these VMS WR stars belong to the WNL sequence, i.e., they have a significant amount of hydrogen in their atmosphere, they may be core hydrogen burning stars and a comparison between their stellar wind properties and our results may therefore be very instructive.

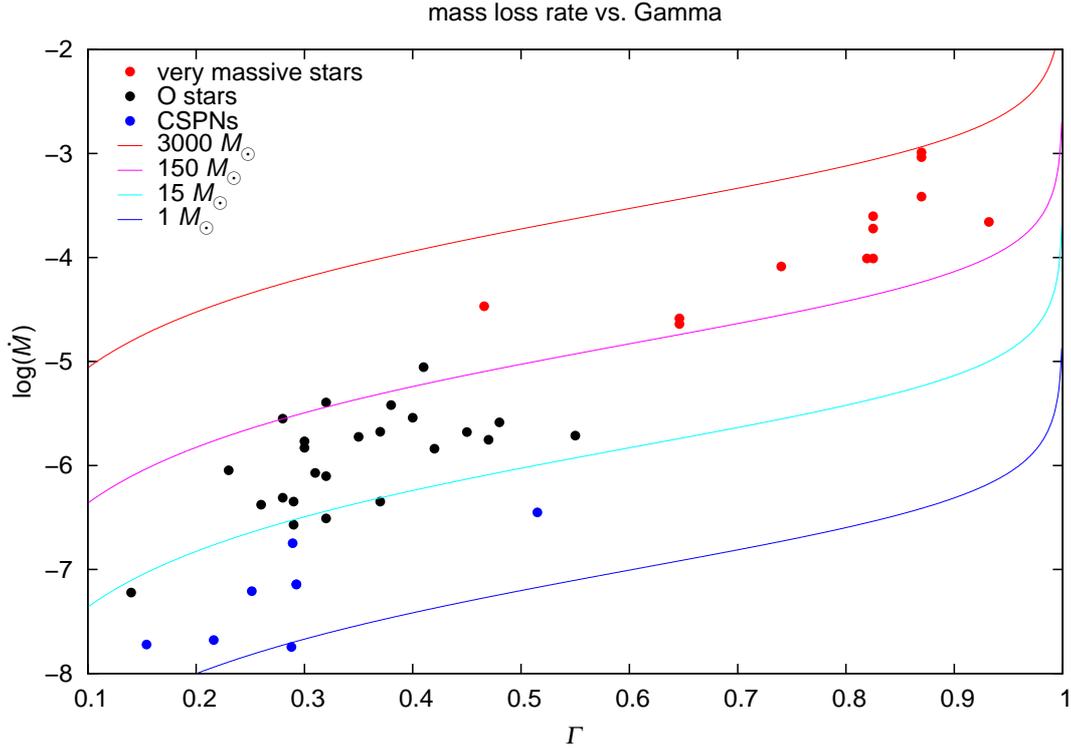

**Fig. 21.** Predicted mass loss rates for the same objects as in Fig. 19 versus $\Gamma$, the ratio of stellar luminosity to Eddington luminosity. The curves represent the mass loss rates according to the approximative formula of Eq. (6) for various stellar masses as labelled. (Note that $\Gamma$ is defined with respect to electron scattering, as it is used to define a reference scale only. But the actual model computations of course contain all relevant line and continuum forces (cf. Appendix A); still, for an optically thin wind, the free-free and bound-free forces are marginal compared to the Thomson force, in contrast to an optically thick wind where they become appreciable.)

In Fig. 19 and Fig. 20 we also show the corresponding results of the analysis by Crowther et al. (2010) (red open circles in our figures). They lie in good agreement along the line defined by the regular massive O stars and the solar-metallicity objects of our VMS model grid. As these results are not only extremely interesting but also quite spectacular, we will, as an example, concentrate in this section on one of these objects, in order to investigate whether the characteristics of this object are indeed in line with our results. We have chosen the WN5h star R136a3 for this purpose, because its parameters are close to those of one of our $150\,M_\odot$ grid stars (cf. Table 10 and Table 12), and because a model UV spectrum has been shown for this object by Crowther et al. (2010).

By inspecting our $150\,M_\odot$ model sequence – shown in Table 10 – in comparison with the model of Crowther et al. (2010) – presented in Table 12 – one might not immediately be convinced that the results of the R136 WNL stars of Crowther et al. (2010) are really in line with our results. The skepticism is based on the fact that the R136 stars are objects with a reduced hydrogen mass fraction of $X_H = 0.4$, and since these objects belong to the LMC their metallicity is furthermore lower than the solar value. We would thus have to compare the values of Table 12 with the values displayed by the green curve of Fig. 18 and these values additionally have to be downscaled with respect to the metallicity relationship of the mass loss rate. One might therefore be inclined to conclude that the Crowther et al. (2010) mass loss rates are much higher.

However, the basis of the discrepancy we are concerned with is much more complicated, and as we will see, global arguments, as the above chain of reasoning involves, might easily lead to wrong conclusions.

Nevertheless, from our grid shown in Table 10 we have to note for the moment that we observe indeed what we just suspected: for lower metallicities our $\dot{M}$ value will become smaller than the corresponding value of Crowther et al. (2010), but it is on the other hand important to realize that our $v_\infty$ value will also become much higher than the observed $v_\infty$ of R136a3; and the compensating effect of the latter fact is the reason why our results shown in Fig. 19 are in agreement with the wind-momentum–luminosity relationship. Thus, with respect to the stellar parameters we used it seems that our calculated $\dot{M}$ values are too small, and our calculated $v_\infty$ values are too large in order to reproduce the properties of the R136 WNL stars.

However, in the first part of this paper we have shown in detail that our models yield correct values for the dynamical parameters. We therefore conclude that the disagreement of our $\dot{M}$ values with those of Crowther et al. (2010) strongly indicates that the input parameters of R136a3 need revision.

In order to figure out which of the parameters might be affected we calculated a consistent model for R136a3 based on the same parameter set as has been used by Crowther et al. (2010) and compared the dynamical parameters and the UV spectrum to the corresponding values of these authors and to the observations, respectively. Model "A$^-$" in Table 12 shows our consistently calculated dynamical values, and as expected a too low mass loss rate $\dot{M}$ and a too high terminal velocity $v_\infty$ is obtained. This result is also recognized by a comparison of the calculated

**Table 12.** Stellar and wind parameters for the WN5h star R136a3 – a member of the central ionizing cluster R136 in the Tarantula nebula (30 Doradus) located within the Large Magellanic Cloud (LMC).

| parameter | | model CSH | model A$^-$ |
|---|---|---|---|
| $T_{\rm eff}$ | (K) | 53000 | 52000 |
| $R_*$ | ($R_\odot$) | 23.4 | 24.2 |
| $\log L_*$ | ($L_\odot$) | 6.58 | 6.59 |
| $M_*$ | ($M_\odot$) | 135 | 135 |
| $X_{\rm H}$[a] | (%) | 40 | 40 |
| $X_{\rm C}$ | (%) | 0.004 | 0.004 |
| $X_{\rm N}$ | (%) | 0.35 | 0.35 |
| $X_{\rm O}$ | (%) | 0.004 | 0.004 |
| $X_{\rm Fe}$[b] | (%) | ssa | ssa |
| $X_{\rm Ni}$ | (%) | ssa | ssa |
| $v_\infty$ | (km/s) | 2200 | 3300 |
| $\dot M$ | ($10^{-6}\,M_\odot$/yr) | 37.0 | 14.3 |

The parameters denoted by "CSH" have been taken from Crowther et al. (2010), whereas the parameters of model "A$^-$" show our consistently calculated dynamical values for almost the same set of stellar parameters. The result is a too low mass loss rate $\dot M$ and a too high terminal velocity $v_\infty$ of our model, indicating strongly that the mass $M_*$ of the object, which has been assumed by Crowther et al. (2010) on basis of main-sequence evolutionary models (cf. Hirschi et al. 2004), is too high for the given radius (cf. Fig. 22 and Table 13). Scaled solar abundances with the usual scaling factor of 0.5 for the LMC have been adopted for all metals other than those given in the table. (Regarding the values of the solar abundances we refer to Asplund et al. 2009.)
[a] The $X_{element}$ values represent the mass fractions of individual elements.
[b] "ssa" is an abbreviation for "scaled solar abundances".

**Table 13.** Stellar and wind parameters determined from consistently calculated atmospheric models along with UV diagnostics of the WN5h star R136a3 (cf. Fig. 22).

| parameter | | model A$^+$ | model A |
|---|---|---|---|
| $T_{\rm eff}$ | (K) | 52000 | 52000 |
| $R_*$ | ($R_\odot$) | 24.2 | 24.2 |
| $\log L_*$ | ($L_\odot$) | 6.59 | 6.59 |
| $M_*$ | ($M_\odot$) | 95 | 110 |
| $X_{\rm H}$ | (%) | 40 | 40 |
| $X_{\rm C}$ | (%) | 0.028 | 0.014 |
| $X_{\rm N}$ | (%) | 0.43 | 0.43 |
| $X_{\rm O}$ | (%) | 0.004 | 0.004 |
| $X_{\rm Fe}$[a] | (%) | ssa | 0.19 |
| $X_{\rm Ni}$[a] | (%) | ssa | 0.011 |
| $v_\infty$ | (km/s) | 2250 | 2900 |
| $\dot M$ | ($10^{-6}\,M_\odot$/yr) | 33.0 | 32.3 |

With respect to the stellar and wind parameters determined by Crowther et al. (2010) (cf. Table 12), the main difference to the parameters of our prototype model – model "A$^+$" – is the mass of the object $M_*$. From our analysis it turned out that the mass must be lowered by almost 30 % in order to reproduce the dynamical parameters and the UV spectrum of the star. By means of the parameters determined by model "A" we show further, that the problem of a too high terminal velocity $v_\infty$ (cf. Table 12) can not been solved just by a moderate reduction of $M_*$ and an additional enhancement of the abundances of the elements which increase $\dot M$ (cf. Pauldrach 1987) – namely iron and nickel –, because although $\dot M$ increases in this case to the required value, $v_\infty$ remains at a value which is still too high.
[a] For model "A" the iron and nickel abundance has been increased by a factor of three relative to the adopted scaled solar abundance values.

synthetic spectrum to the observed one – Fig. 22 shows that the synthetic spectrum from this model is radically different from the observed spectrum.

With respect to the procedure required we first need to consider what we know of this object. Item one concerns the uncertainty of the stellar mass. As our result is a consequence of the assumed compactness of the star, indicated by an extremely high surface gravity, a correct determination of this parameter is essential. However, the mass of the object has been assumed by Crowther et al. (2010) on the basis of main-sequence evolutionary models and not been determined, and the assumed value appears too high for the given radius, since the computed terminal velocity from a consistent model with these parameters is higher than the observed one (cf. Fig. 19 and Fig. 20). Thus, from our point of view $M_*$ has to be lowered, because this step would decrease $v_\infty$ and enhance $\dot M$. Item two concerns the uncertainty of the abundances – we note that these important values of the elements have also been assumed and have not been consistently determined. As we have shown is the first part of this paper, a consistent analysis is the only reliable way to determine the abundances along with dynamical parameters. This analysis requires however the consideration of the individual values of the abundances and their back-reaction on the dynamics of the wind, and the crucial point is that the abundances of just a few elements, which might not conform to a simple reduction of the metallicity from solar values, can dominate the quantity $\dot M$. The mass loss rate can therefore increase just by means of small increases of, for instance, Fe and Ni (cf. Pauldrach 1987). The problem that we do not yet know the individual values of the abundances involves the reason why global arguments, as those given above, might lead to wrong conclusions, and it is the reason why we are allowed to compare our results of the dynamical parameters obtained for solar metallicity even with corresponding values of objects of the LMC. Such a comparison does not even indicate an upper limit, because a small enhancement of Fe above the solar values (produced in a starburst region by more supernovae than suspected) would yield even larger consistently obtained mass loss rates. Thus, regarding the need of refinement of the input values used, the question which of the two possibilities described covers the real situation can certainly not be answered by global arguments, but it can be answered by a comparison of predicted spectra with the observations.

In order to facilitate such a comparison we have calculated two additional consistent models for R136a3 based on different stellar parameter sets (cf. Table 13) and compared the dynamical parameters and the UV spectrum to the corresponding values of Crowther et al. (2010) and to the observations. Model "A" clearly shows that the problem of a too high terminal velocity $v_\infty$ can not been solved just by a moderate reduction of $M_*$ and an additional enhancement of the abundances of Fe and Ni – the elements which primarily determine $\dot M$ –, because although $\dot M$ increases in this case to the required value, $v_\infty$ remains at a value which is still too high. This result is additionally emphasized by a comparison of the calculated synthetic spectrum to the observed one (cf. Fig. 22). The corresponding synthetic spectrum clearly shows that the Fe lines start to become too strong in the emission parts of the N V and C IV resonance lines. This however is not observed in the synthetic spectrum of our prototype model (model "A⁺"). Although the He II line at 1640 Å turns out to be somewhat too weak, this spectrum is in quite good agreement with the observations. With respect to the stellar and wind parameters determined by Crowther et al. (2010) (cf. Table 12) the main difference to the parameters of this model is the mass of the object. From our analysis it turned out that the mass of the star must be reduced by almost 30 % in order to reproduce the dynamical stellar parameters in a consistent manner.

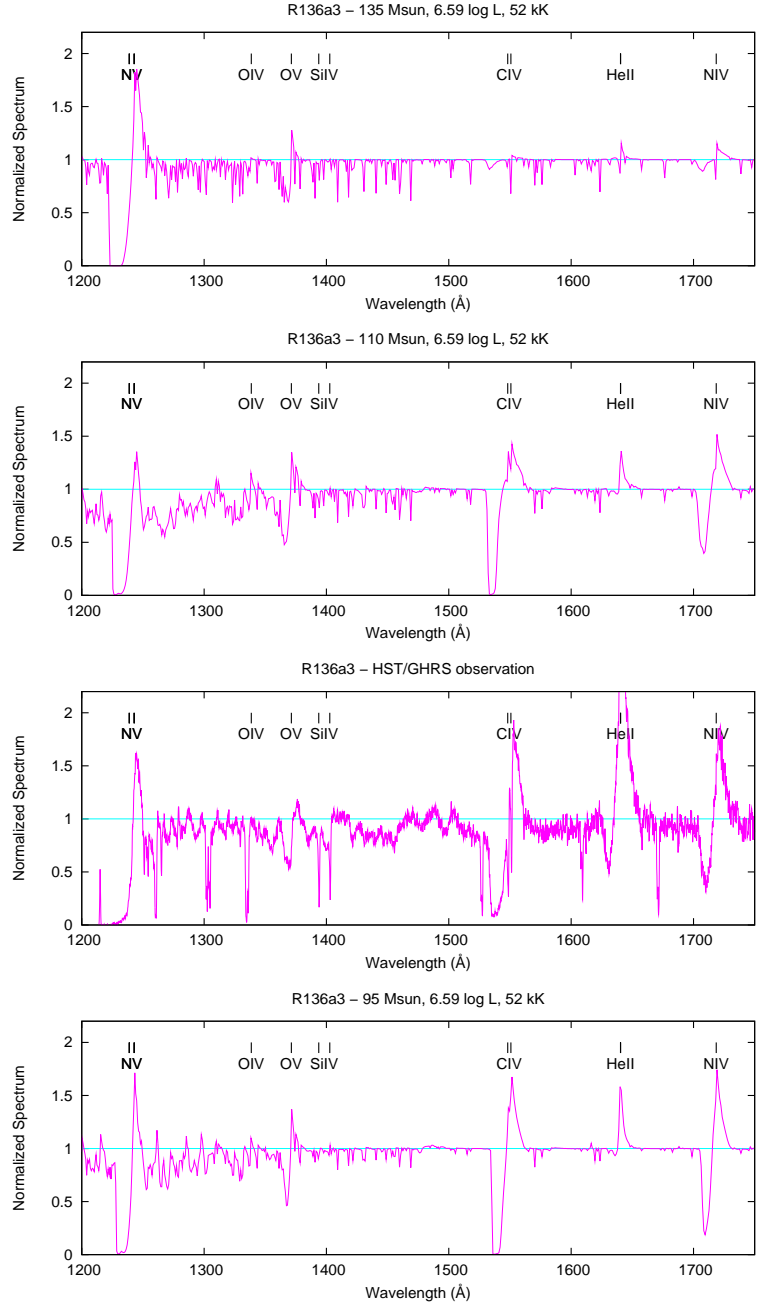

**Fig. 22.** Calculated and observed UV spectra for the WN5h star R136a3. The calculated spectra represent our prototype model (model A$^+$ – lowest panel), whereas the model in the upper panel (model A$^-$) shows the result of our consistently calculated dynamical values based on the stellar parameter set of Crowther et al. (2010). As a consequence of the assumed compactness of the object[20] the result of a too low mass loss rate $\dot{M}$ and a too high terminal velocity $v_\infty$ of this model (cf. Table 12) is easily recognized by a comparison to the observed spectrum shown in the third panel[21]. The spectrum presented in the second panel shows clearly that a moderate reduction of the stellar mass along with an enhancement of the abundances of iron and nickel – which leads to an enhancement of $\dot{M}$ – does not solve the problem ("model A" in Table 13), because the Fe lines start to become too strong in the emission parts of the N v and C iv resonance lines[22]. On the other hand, model A$^+$ (lowest panel) is not affected by these weak points. Although the He ii line at 1640 Å is somewhat too weak, the spectrum is together with the dynamical parameters (cf. Table 13) in quite good agreement with the observations[23].

---

[20] The compactness of the object is indicated by an extremely high surface gravity.

[21] The result therefore indicates that the mass $M_*$ of the object adopted by Crowther et al. (2010) is certainly too high for the given radius.

[22] With respect to the width of the absorption parts of the P-Cygni lines the spectrum shows further that the problem of a too high terminal velocity $v_\infty$ remains.

[23] The main difference of the parameters Crowther et al. (2010) and we have determined regards therefore the mass of the star. With respect to our analysis $M_*$ has to be lowered by almost 30 % in order to reproduce the dynamical parameters quantitatively and the UV spectrum of the star at least qualitatively.

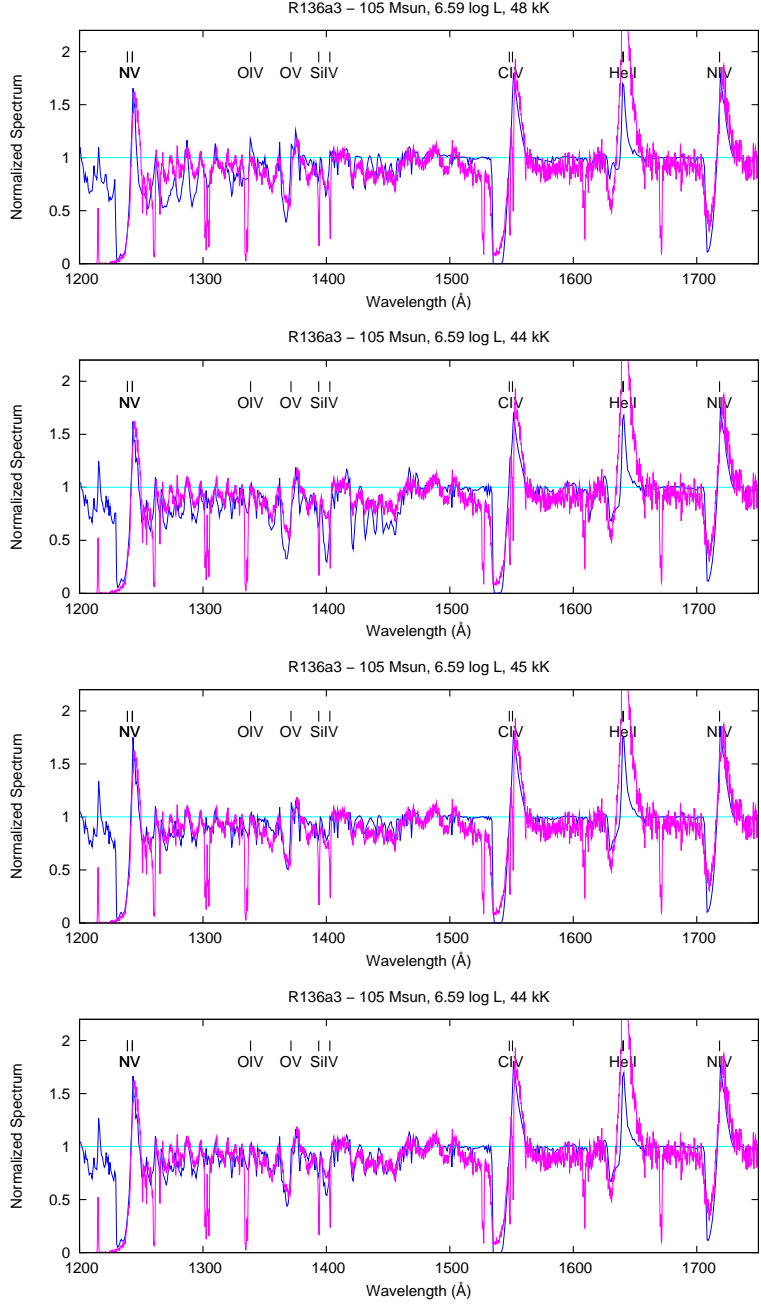

**Fig. 23.** Synthetic UV spectra of a consistently calculated model sequence in comparison with the observed UV spectrum of the WN5h star R136a3. The calculated spectra represent our current best model (A*) in the *lowest panel* (cf. Table 14), as well as model calculations in the panels above whose changes in the parameters can be ruled out. The *uppermost panel* shows clearly the result of a too high effective temperature ($T_{\rm eff}$ = 48 kK), as it reflects too strong Fe VI wind-contaminated lines in the wavelength range from 1250 to 1400 Å whereas the Fe V lines in the wavelength range from 1400 to 1550 Å are much too weak. In the *second panel* it is easily recognized by a comparison to the observed spectrum shown that this problem is solved by adopting a lower value of the effective temperature ($T_{\rm eff}$ = 44 kK). However, now the spectral signatures of both ionization stages – Fe V and Fe VI – are too strong. This problem can be solved by a reduction of the abundances of iron and nickel by a factor of 2.0 and a small increase of the effective temperature ($T_{\rm eff}$ = 45 kK). Now, as is shown in the *third panel* the Fe VI lines are quite well in agreement with the observed structures, but the Fe V lines in the range from 1400 to 1550 Å are again somewhat too weak. In the last step – *lowest panel* – the effective temperature has been reduced slightly again to the value of $T_{\rm eff}$ = 44 kK and the abundances of iron and nickel have been decreased just by a factor of 1.5. Apart from the He II line at 1640 Å (see text) the spectrum is now quite well in agreement with the observations.

As is shown in Fig. 23 we obtained a similar result for the effective temperature of this star. From an inspection of our proto-type model – the lowest panel in Fig. 22 – it is quite obvious that the value of the effective temperature adopted from Crowther et al. (2010) is considerably too large, because the spectrum does not reproduce the Fe v lines in the wavelength range from 1400 to 1550 Å, whereas the Fe vi wind-contaminated lines in the wavelength range from 1250 to 1400 Å are already too strong. As we know from Sect. 3.1.2 that the effective temperature can be accurately determined from the ionization balance in the wind, which in this case is reflected in the strengths of the spectral lines of the successive ionization stages of Fe v and Fe vi, this disagreement has important consequences. To overcome the described problem we calculated a consistent model with a lower effective temperature of $T_{\rm eff}$ = 48 kK and the result of the synthetic UV spectrum is shown in the uppermost panel of Fig. 23. As can be seen, this change of the parameter $T_{\rm eff}$ results in only a small improvement, because the Fe vi lines in the shorter wavelength region are still too strong and the Fe v lines in the longer wavelength region are still too weak. We therefore calculated an additional model with an even lower effective temperature of $T_{\rm eff}$ = 44 kK and the synthetic UV spectrum from this model is shown in the second panel of Fig. 23, from which it is easily seen that this problem is solved. However, this synthetic spectrum shows also clearly that the spectral signatures of both ionization stages of iron are now too strong. This unexpected new problem can of course be solved by a reduction of the abundance of iron and accordingly a small increase of the effective temperature ($T_{\rm eff}$ = 45 kK has been assumed for the effective temperature and a factor of 2 relative to the adopted scaled solar abundance values has been applied for the iron abundance). From this procedure, which is illustrated by means of the obtained synthetic spectrum in the third panel, it turned out that the Fe vi lines are now for the first time quite well in agreement with the observed structures; however, the Fe v lines in the range from 1400 to 1550 Å are again persistently somewhat too weak. Thus, a last step is required where the effective temperature has to be reduced again slightly – back to the value of $T_{\rm eff}$ = 44 kK – and the abundance of iron has to be decreased less drastically – this time just by a factor of 1.5. With respect to this fine-tuning procedure the calculated synthetic spectrum of this last step represents our current best model (A*), shown in the lowest panel of Fig. 23. Although the He ii line at 1640 Å is still somewhat too weak, the spectrum is now quite well in agreement with the observations (from our point of view it is extremely astonishing that obviously just H and He lines are affected by a process which causes stronger emission parts of some profiles – we notice that in Sect. 3.1.3 we obtained a comparable result for the case of the normal O star $\zeta$ Puppis; the usual interpretation to assume a clumped wind for the cause of this process – cf. Moffat & Robert 1994 – should therefore certainly be revised in order to be able to explain this very special selection effect). As a final conclusion of this digression our results indicate strongly that the value of the effective temperature obtained by Crowther et al. (2010) is certainly too high for the luminosity given of this star. Table 14 summarizes the final results of our current best model A* with respect to the stellar and wind parameters of R136a3. We have obtained a lower effective temperature, a larger radius, a lower iron and nickel abundance, a larger mass loss rate, and most interestingly, a smaller mass of the object, whereas our values of the luminosity, the terminal velocity, and the hydrogen mass fraction are almost identical when compared to the values of Crowther et al. (2010) (cf. Table 12).

With respect to this case study we have three additional conclusions: The first one is that the scientific content of our Fig. 19

**Table 14.** Stellar and wind parameters of our current best model – model "A*" – of the WN5h star R136a3 (cf. Fig. 23).

| parameter | | model A* |
|---|---|---|
| $T_{\rm eff}$ | (K) | 44000 |
| $R_*$ | ($R_\odot$) | 33.8 |
| $\log L_*$ | ($L_\odot$) | 6.59 |
| $M_*$ | ($M_\odot$) | 105 |
| $v_{\rm rot}$ | (km/s) | 200 |
| $X_{\rm H}$ | (%) | 40 |
| $X_{\rm C}$ | (%) | 0.0083 |
| $X_{\rm N}$ | (%) | 0.41 |
| $X_{\rm O}$ | (%) | 0.004 |
| $X_{\rm Fe}$[a] | (%) | 0.043 |
| $X_{\rm Ni}$[a] | (%) | 0.0026 |
| $v_\infty$ | (km/s) | 2240 |
| $\dot{M}$ | ($10^{-6}\,M_\odot$/yr) | 43.4 |

The presented parameters have been determined from UV diagnostics based on our consistently calculated atmospheric models. Compared to the values of Crowther et al. (2010) (cf. Table 12) we obtained a lower effective temperature ($T_{\rm eff}$), a larger radius ($R_*$), a lower iron and nickel abundance, a larger mass loss rate ($\dot{M}$), and a smaller mass ($M_*$) of the object, whereas our values of the luminosity ($L_*$), the rotational velocity ($v_{\rm rot}$), the terminal velocity ($v_\infty$), and the hydrogen mass fraction ($X_{\rm H}$) are almost identical.
[a] The iron and nickel abundance has been decreased by a factor of 1.5 relative to the adopted scaled solar abundance value.

and Fig. 20 has, as anticipated in the previous section, indeed a convincing character regarding the significance of the dynamical parameters. The second one is that the results of the R136 WNL stars of Crowther et al. (2010) are in quite good agreement with our results, if their assumed masses are reduced by up to 20 % (that we obtained a lower effective temperature for R136a3 is not considered here as a general conclusion, because it might be based on a selection bias). The third one is that, with the parameters we derived from a comparison with the observed UV spectrum, the wind of this particular star is still optically thin in the optical and observable UV continuum (cf. Fig. 24; the star is classified as a WNL star but its UV spectrum still shows many similarities to that of an O star). However, the issue of optically thin vs. optically thick winds can not be generalized to all WNL stars, as the R136 stars are hydrogen-rich high-mass stars with high terminal velocities, whereas many WNL stars have lower terminal velocities (Hamann et al. 2006) and, according to Gräfener & Hamann (2008), optically thick winds (note that the influence of the terminal velocity on the optical thickness of these winds is indicated in Eq. (8)). Thus it appears that the WNL stars do not represent a homogeneous class of objects, and depending on their stellar parameters and chemical composition the whole range of optically thin to optically thick winds seems to be realized.

Regarding the second point we reencounter from our point of view the persisting problem of the mass discrepancy (cf. Groenewegen et al. 1989), this time for WNL stars if these stars really are core hydrogen burning stars. However, that might not be true. A close inspection of the mass and the luminosity of R136a3 leads to the conclusion that it is likely that the "WR star" is a core helium burning object rather than core hydrogen burning. Interestingly, the mass and the luminosity of the WNL star WR22 proposed by Gräfener & Hamann 2008 are also much more consistent with the statement that the WNL star is core helium burning.

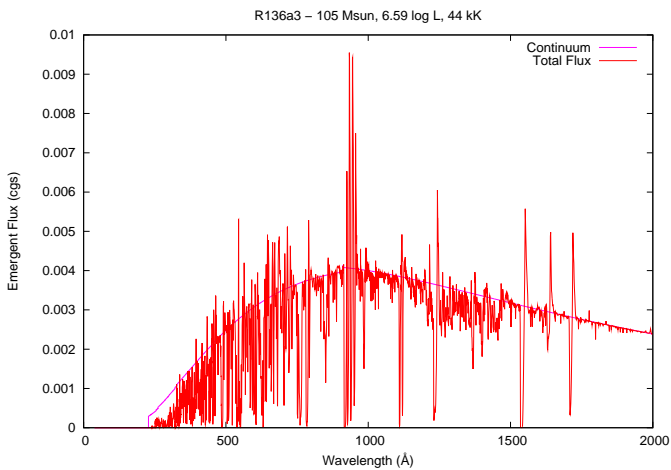

**Fig. 24.** Calculated spectral energy distributions (Eddington flux $H_\nu$ in cgs units) versus wavelength of our current best model ($A^*$) of R136a3. The spectrum contains, in addition to the typical P-Cygni profiles of the strong resonance lines, numerous absorption and emission lines above and below the continuum curve, illustrating that we are looking down into an extended photosphere similar to that of an O star. Thus, the behavior of the spectral energy distribution clearly shows that this model has an optically thin wind in the optical and observable UV continuum[24].

## 6. The evolution of very massive stars (VMS) in dense stellar systems

In this section we reconsider the evolution of very massive stars and study the effect of our VMS mass loss rates. As the evolutionary program used here has been described by Belkus et al. (2007) we will here only give a brief overview of the physics to be treated.

*The general concept of the evolutionary algorithm of VMS in dense clusters.* The dynamical evolution of a dense cluster is characterized by the process called mass segregation where the most massive stars sink to the cluster core. Depending on the initial cluster size and cluster mass the spatial density of massive stars in the core may become so large that the gravitational encounter of two massive objects becomes very probable and the stars collide and merge. This collision merger eventually initiates a chain reaction where the same object merges with other massive stars: the term "runaway merger" is used for this process. In many cases, this runaway merger may become as massive as $1000\,M_\odot$ or even larger (Portegies Zwart et al. 2006 and references therein).

To investigate the consequences of this process it is indispensable to know the properties of such an object and how it evolves further. Using 3d smoothed-particle hydrodynamics (SPH), Suzuki et al. (2007) have simulated the collision and merging of two massive stars. The evolution during merging depends on the mass ratio of the two colliding stars, but after thermal adjustment (Kelvin-Helmholtz contraction) the merger is nearly homogenized. Suzuki et al. (2007) simulated the collision of a star of $88\,M_\odot$ with a star of the same mass, and with one with a mass of $28\,M_\odot$. The merging process is very short (on the order of days) and, interestingly, during the merging the collision object loses $\sim 10\,M_\odot$. It is tempting to link such a collision and merging process to the eta Carinae outburst in the nineteenth century.

In our $N$-body simulations we therefore assume that massive collision objects are homogenized and become zero-age main-sequence stars (with the appropriate chemical composition) on a time-scale which is short compared to the stellar evolutionary time-scale. Since very massive stars remain quasi-homogeneous, the further evolution of the merger object is conveniently described by the fast and robust algorithm of Belkus et al. (2007), based on the similarity theory of stellar structure (Nadyozhin & Razinkova 2005). This evolution depends critically on the adopted stellar wind mass loss rate formalism (outlined in the next paragraph) and which relies on the previously described results of the present paper.

*The influence of the mass loss rate of VMS on the evolution of dense clusters.* For solar-type VMS, when the atmospheric hydrogen abundance $X_H$ is larger than 0.4 we calculate the mass loss rate with the uppermost linear fit (red curve) of Fig. 18 (note that similar results are obtained when the fit of Fig. 20 is used). When $X_H \leq 0.4$ the mass loss is calculated with the middle fit (green line) holding for WNL stars. Since our mass loss rate computations confirm the canonical $Z^{1/2}$ dependence, it is adopted here. Slightly different Z-formulae have been proposed in the recent past, but it can easily be understood that they will not alter our main conclusions. Our mass loss rates differ considerably from previous studies and we therefore focus on the total mass that is lost during core hydrogen burning. Table 15 summarizes evolutionary properties during the CHB phase. As can be seen the total ejected mass predicted by the earlier studies is significantly larger than those calculated with our stellar wind mass loss rate estimates and this has a few interesting consequences related to (i) the primordial chemical enrichment of Population III very massive stars and the age determination of globular clusters; and (ii) the importance of stellar wind mass loss on the formation of intermediate mass black holes (IMBH). This will be discussed in the next subsections.

**Table 15.** The evolution of VMS with initial masses of 3000 $M_\odot$, 1000 $M_\odot$, and 750 $M_\odot$ and metallicities 1 $Z_\odot$, 0.05 $Z_\odot$, and 0.005 $Z_\odot$. Shown are the values of the total ejected mass predicted by earlier studies, $\Delta M_{BM}^{ej}$ (Belkus et al. 2007, Marigo et al. 2003) versus our new values, $\Delta M_{ov}^{ej}$.

| $T_{BM}^{CHB}$ (Myr) | $\Delta M_{BM}^{ej}$ ($M_\odot$) | $Z_*$ ($Z_\odot$) | $\Delta M_{ov}^{ej}$ ($M_\odot$) | $T_{ov}^{CHB}$ (Myr) |
|---|---|---|---|---|
| | | 3000 $M_\odot$ | | |
| 1.7 | 2680 | 0.005 | 120 | 1.6 |
| 1.7 | 2730 | 0.05 | 304 | 1.6 |
| 1.8 | 2820 | 1 | 1105 | 1.6 |
| | | 1000 $M_\odot$ | | |
| 1.9 | 581 | 0.005 | 20 | 1.8 |
| 1.9 | 779 | 0.05 | 72 | 1.8 |
| 2.0 | 847 | 1 | 195 | 1.9 |
| | | 750 $M_\odot$ | | |
| 1.9 | 146 | 0.005 | 6.5 | 1.8 |
| 1.9 | 528 | 0.05 | 18 | 1.8 |
| 2.0 | 597 | 1 | 74 | 1.9 |

---

[24] Note that this is the same spectrum as shown in the comparison to the observed UV spectrum in the lowest panel of Fig. 23 and that the stellar and wind parameters of our final results for this WN5h star model are summarized in Table 14.

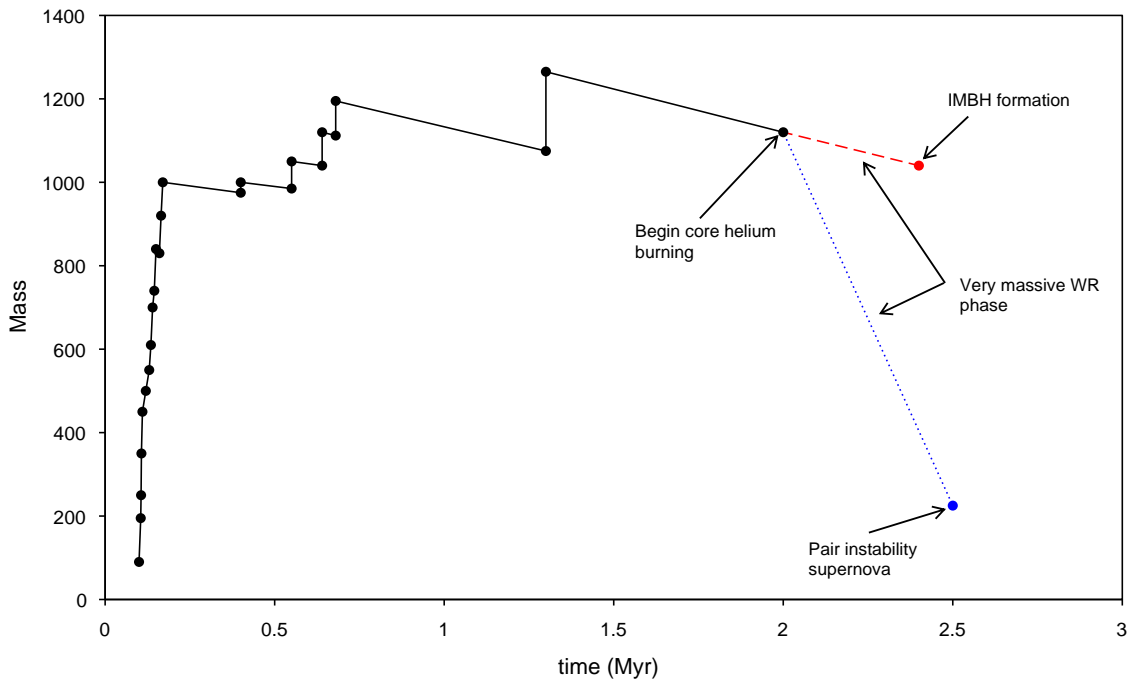

**Fig. 25.** Mass evolution of a collision runaway. The figure illustrates the influence of the mass loss rate on the evolution of a massive merger object. The dashed line represents the mass evolution during the core helium burning phase (CHeB) if a constant mass loss rate of hydrogen-deficient WR stars is adopted that closely matches our computed rates. In this case the VMS will end its life with a mass greater than 1000 $M_\odot$ and the formation of an intermediate-mass black hole (IMBH) takes place. In contrast, we also followed the CHeB evolution by using an extrapolation of the mass loss rates presented by Nugis & Lamers (2000) for hydrogen-deficient WR stars (dotted line). In this case the star loses much more mass and therefore the occurrence of a pair-instability supernova is more likely than the formation of an IMBH.

### 6.1. The primordial chemical enrichment of Population III very massive stars

Since VMS evolve quasi-homogeneously, mass lost by stellar wind enriches the intergalactic medium in helium ($Y$) and CNO elements. Bond et al. (1983) and Carr (1994) argued that a first generation of stars where the IMF is more skewed towards massive and very massive stars, can mask the true primordial abundance of helium if stellar wind mass loss is sufficiently large. By using Kudritzki (2002) rates, Marigo et al. (2003) obtained a (maximum) helium enrichment of $\Delta Y \approx 0.01$ caused by a population of metal-free VMS. Our stellar wind calculations indicate that this enrichment may be at least a factor 10 smaller. We conclude that a significant helium enrichment of an initial population of very massive stars is highly questionable. Note that $\Delta Y \approx 0.001$ is not sufficient to affect the age determination of globular clusters (as was proposed by Shi 1995).

### 6.2. The influence of VMS mass loss on the formation of intermediate mass black holes

At very low to zero metallicities, very massive stars are likely to keep most of their mass, and, depending on their initial mass,

they will thus very probably end their life as an IMBH. This means that a top-heavy IMF in a low-metallicity region predicts the existence of intermediate-mass black holes. A more controversial question was discussed in the introduction and is related to the link between the dynamical formation of IMBHs in young dense clusters with solar-type metallicity, the existence of ULXs and the possibility that these IMBHs are the building blocks of the AGN in galactic centers. Vanbeveren et al. (2009) combined the evolutionary algorithm of VMS proposed by Belkus et al. (2007) with an $N$-body code that follows the dynamical evolution of a starburst, in order to study the possible formation of an IMBH and a ULX in a cluster like MGG-11 (see Sect. 1). They concluded that IMBH-formation is unlikely. However, this conclusion relied almost entirely on the adopted Kudritzki (2002) stellar wind mass loss rates of VMS during core hydrogen burning.

As outlined above, our new mass loss rate predictions are significantly smaller and we have therefore repeated the MGG-11 simulation with the mass loss rates presented here. As in Vanbeveren et al. (2009) we assume a cluster containing 3000 massive single stars (with masses between 10 $M_\odot$ and 120 $M_\odot$, satisfying a Salpeter initial mass function), distributed according to a King model[25] with $W_0 = 9$ and a half-mass radius of 0.5 pc. Mass segregation causes the most massive stars to sink towards the center of the cluster where the spatial density is so large that collisions occur. Fig. 25 shows the rapid growth in mass of the

---

[25] In a self-gravitating $N$-body system constrained by a fixed energy and mass the entropy is extremized by the distribution of an isothermal sphere. A set of models closely related to the isothermal sphere was introduced by King (1966). A common parameter identifying the models is the so-called King-parameter $W_0$. King models are still the standard model for globular clusters, as they fit the observed density-profiles very well.

A. W. A. Pauldrach et al.: Very massive stars and the evolution of dense stellar clusterscollision object. The frequency of collisions gradually decreases and the collision runaway process finally stops with a mass of the merger object of around 1000 to 1200 $M_\odot$, from where the evolution is followed adopting a stellar wind mass loss rate that closely matches the computed rates discussed in Sect. 4.2.4. The figure marks the moment that core hydrogen burning is finished (and core helium burning starts). When we compare the mass at that point with the corresponding mass at the end of the CHB phase in the simulations of Vanbeveren et al. (2009), we see that with the new rates the final mass is about 7 times larger than with the Kudritzki rates, of course in line with the evolutionary results in Table 15. At the beginning of core helium burning (CHeB) the star is a very massive WR star and the further evolution depends entirely on the mass loss rate of such stars. To illustrate this we first computed the CHeB phase by adopting a constant mass loss rate of $2 \cdot 10^{-4}$ $M_\odot$/yr which corresponds to the observed maximum mass loss rate of hydrogen-deficient WR stars (dashed line in Fig. 25). As can be seen the star will end its life with a mass $> 1000$ $M_\odot$ and the formation of an IMBH is very likely. Alternatively, we also followed the CHeB evolution by using extrapolated Nugis & Lamers (2000) rates for hydrogen-deficient WR stars (the dotted line in Fig. 25). In this case the star loses so much mass that it does not form an IMBH but instead a pair instability SN becomes very plausible. We conclude that the final answer is blowing in the wind mass loss rates of very massive core helium burning WR stars, but this is beyond the scope of the present paper.

# 7. Conclusions and outlook

Stellar wind lines can presently be identified in the spectra of O supergiants up to distances of 20 Mpc. Moreover, they also show up in the integrated spectra of starburst galaxies even at redshifts up to $z \sim 4$ (cf. Steidel et al. 1996). Clearly, these spectral features can provide important information about the chemical composition, the stellar populations, and thus the galactic evolution even at extragalactic distances. That this is a feasible task has already been demonstrated by Pettini et al. (2000), who showed that the O star wind line features can indeed be used to constrain star formation processes and the metallicity. We are thus close to the point of making use of complete and completely independent quantitative spectroscopic studies of the most luminous stellar objects, this possibility being offered by the stellar winds of these objects.

However, to realize this objective a diagnostic tool for determining the physical properties of hot stars via quantitative UV spectroscopy is required. We have described here the status of our continuing work to construct an advanced diagnostic tool that includes an assessment of the accuracy of the determination of the parameters involved for this purpose. We have shown that the atmospheric models for hot stars of this *new generation* are realistic. They are *realistic* with regard to a *quantitative spectral UV analysis* calculated *along with consistent dynamics*, which, in principle, allows to determine the stellar parameters by comparing an observed UV spectrum to a suitable synthetic spectrum. The new generation of models has therefore reached a degree of sophistication that makes such a procedure practicable. The astronomical perspectives are enormous, not only for applications of the diagnostic techniques to massive O-type stars and low-mass stars of post-AGB evolution, but also extended to very massive stars (VMS). The role of massive stars as tracers of the chemical composition and the population of starbursting galaxies at high redshift can therefore be efficiently tracked.

While in principle there is of course always the possibility that the model predictions have systematic weaknesses, we tend to have a stronger confidence in these models when the models have been tested against observations and are found to be confirmed by this test. Regarding such a test, we have compared our consistently calculated mass loss rate ($\dot{M} = 3.8 \cdot 10^{-6}$ $M_\odot$/yr) for $\zeta$ Puppis with the corresponding values obtained from an analysis of the optical lines ($\dot{M} = 4.2 \cdot 10^{-6}$ $M_\odot$/yr), the X-ray regime ($\dot{M} = 3.5 \cdot 10^{-6}$ $M_\odot$/yr), and the radio data ($\dot{M} = 3.9 \cdot 10^{-6}$ $M_\odot$/yr), using the same stellar parameters and abundances as adopted in these other analyses. Despite this excellent agreement, a comparison of the observed UV spectrum with the synthetic UV spectrum obtained from the consistent solution between non-LTE-radiative transfer and hydrodynamics has led us to revise the radius of $\zeta$ Puppis and some of the abundances, overall resulting in a higher mass loss rate ($\dot{M} = 13.7 \cdot 10^{-6}$ $M_\odot$/yr). So while it is not the case in general that our mass loss rate predictions are higher than the optically determined mass loss rates (rather, they agree quite well on average), UV analyses of specific stars (such as $\zeta$ Puppis) may yield differences.

Based on the success of this essential step in modelling the hydrodynamics of the stellar outflow, we have used our self-consistent atmospheric models to calculate the mass loss rates for a grid of massive and very massive stars. These computed mass loss rates turned out to be lower than previous mass loss rate predictions for regular massive O stars extrapolated to very massive stars. A plot of the wind momentum rate vs. luminosity shows that there exists a common relation that extends over five decades in luminosity, from O-type CSPNs over "normal" massive O stars (dwarfs and supergiants) up to the very massive stars, and as one might suspect from this result, the mass loss rates show a similar dependence on luminosity over the same full range. Furthermore, our calculations confirm the canonical $Z^{1/2}$ metallicity dependence of the mass loss rates.

We have also calculated stellar wind mass loss rates of very massive stars with hydrogen-deficient atmospheres, assuming that these are core hydrogen burning (such stars are commonly considered to be WNL stars). When we compare our rates with those of Crowther et al. (2010) for the WNL stars in R136 it appears at first glance that our rates are smaller, whereas, for the given stellar parameters, our computed terminal velocities are higher than observed, implying that for those VMS models the assumed masses are too high for their luminosities. (This is not the case for "normal" massive O stars based on observed stellar parameters, for which, as we have shown, our computed mass loss rates and terminal velocities reproduce the observations and thus the underlying mass–luminosity relation.) Our re-analysis of one of the WNL stars (R136a3) of Crowther et al. (2010), for which we have determined a set of stellar and hydrodynamically consistent wind parameters that reproduces the observed UV spectrum with, surprisingly, a wind that is still optically thin in the optical and observable UV continuum despite the star being classified as WNL. We argue that the WNL stars do not represent a homogeneous class of objects, and depending on their stellar parameters and chemical composition the whole range of optically thin to optically thick winds may be realized (the R136 stars are hydrogen-rich high-mass stars with high terminal velocities, but other WNL stars exist that have lower terminal velocities and optically thick winds). Moreover, the mass and luminosity we find for R136a3 indicate that this star is very likely a core helium burning star. These points may re-open the discussion on the evolutionary phase of WNL stars.

From an application of the mass loss rates of VMS obtained in the present paper in an evolutionary code we conclude that VMS with a small metallicity lose a very small amount of their mass. This means that it is unlikely that population-III VMS cause a significant helium enrichment of the interstellar medium. Even more, it cannot be excluded that VMS with a solar-type metallicity, formed by dynamical processes in dense clusters, end their life massive enough to form intermediate mass black holes. The final answer will depend on the stellar wind mass loss rates during the core helium burning phase.

*Acknowledgements.* We thank an anonymous referee for helpful comments which improved the paper. This work was supported by the *Deutsche Forschungsgemeinschaft (DFG)* under grants Pa 477/7-1 and Pa 477/9-1.

## Appendix A:
## A consistent concept for the calculation of the radiative acceleration

The theoretical concept for calculating the radiative acceleration consistently with the non-LTE model has been based on a radiation field which stems from the photosphere of the star and supplies the stellar wind with photons (cf. Castor et al. 1975, Pauldrach et al. 1986, Pauldrach et al. 1994).

The severe modifications of the non-LTE model which now includes a realistic treatment of line blocking and blanketing in the total atmosphere including the effects of line-overlap and multiple scattering (cf. Pauldrach et al. 2001) obviously requires also a consistent treatment of the calculation of the radiative acceleration. Although the corresponding new concept has already been used by Pauldrach (2003), Pauldrach et al. (2004), and Pauldrach (2005), it has up to now not been described. This gap will be closed in the following.

In general the hydrodynamics is solved by iterating the complete continuum acceleration $g_{cont}(r)$ (which includes in our case the force of Thomson scattering and of the continuum opacities $\chi_\nu^{cont}(r)$ – free-free and bound-free – of all important ions(cf. Pauldrach et al. 2001)) together with the line acceleration $g_{lines}(r)$ – obtained from the spherical non-LTE model – and the density $\rho(r)$, the velocity $v(r)$, and the temperature structure $T(r)$.

Starting from the equation of motion for stationary, spherically symmetric line driven winds

$$v(r)\frac{dv(r)}{dr} = -\frac{dp(r)}{dr}\frac{1}{\rho(r)} + g_{cont}(r) + g_{lines}(r) - g(r), \quad (A.1)$$

where $g(r) = GM_*/r^2$ is the local gravitational acceleration, and $p(r) = a^2(r)/\rho(r)$ is the local gas pressure based on the isothermal sound speed $a(r)$, the radiative continuum acceleration is defined as

$$g_{cont}(r) = \frac{4\pi}{c}\frac{1}{\rho(r)} \int_0^\infty \chi_\nu^{cont}(r) H_\nu(r)\, d\nu \quad (A.2)$$

where $c$ is the speed of light, and $H_\nu(r)$ is the local monochromatic Eddington flux at the frequency $\nu$. Allowing for Doppler shifting of the line profiles along the chosen path, the radiative line acceleration is written as

$$g_{lines}(r) = \frac{2\pi}{c}\frac{1}{\rho(r)}\frac{\pi e^2}{m_e c}\sum_{lines} f_{lu}g_l\left(\frac{n_l(r)}{g_l} - \frac{n_u(r)}{g_u}\right) \\
\cdot \int_0^\infty \int_{-1}^{+1} I_\nu(r,\mu)\phi\left(\nu - \nu_0\left(1+\mu\frac{v(r)}{c}\right)\right)\mu\, d\mu\, d\nu, \quad (A.3)$$

which can be rewritten in terms of the local line absorption coefficients $\chi_{line}(r)$ as

$$\chi_{line}(r) = \frac{\pi e^2}{m_e c}f_{lu}g_l\left(\frac{n_l(r)}{g_l} - \frac{n_u(r)}{g_u}\right)$$

$$g_{lines}(r) = \frac{2\pi}{c}\frac{1}{\rho(r)}\sum_{lines}\chi_{line}(r)\int_0^\infty\int_{-1}^{+1} I_\nu(r,\mu)\phi\left(\nu-\nu_0\left(1+\mu\frac{v(r)}{c}\right)\right)\mu\, d\mu\, d\nu, \quad (A.4)$$

where $e$ is the charge and $m_e$ the mass of an electron; $f_{lu}$ is the oscillator strength for a line with upper level $u$, lower level $l$, statistical weights $g_l$ and $g_u$, and local occupation numbers $n_l(r)$ $n_u(r)$; $I_\nu$ is the specific intensity and $\phi(\nu)$ the line broadening function accounting for the Doppler effect; and finally $\mu$ is the usual cosine of the angle between the ray direction and the outward normal on the spherical surface element. As $g_{cont}(r)$ and $g_{lines}(r)$ account for the transfer of momentum from the radiation field to the matter – only absorptive processes and stimulated emission contribute to the radiative acceleration, whereas spontaneous emission is treated as isotropic and does not provide a net transfer of momentum to the material.

Application of the Sobolev-approximation now gives for the radiative line acceleration (cf. Pauldrach et al. 1994 and references therein)

$$g_{lines}(r) = \frac{2\pi}{c}\frac{1}{\rho(r)}\sum_{lines}\chi_{line}(r)\int_{-1}^{+1} I_{\nu_0}(r,\mu)\frac{1-e^{-\tau_s(r,\mu)}}{\tau_s(r,\mu)}\mu\, d\mu, \quad (A.5)$$

where

$$\tau_s(r,\mu) = \chi_{line}(r)\frac{c}{\nu_0}\left[(1-\mu^2)\frac{v(r)}{r} + \mu^2\frac{dv(r)}{dr}\right]^{-1} \quad (A.6)$$

is the Sobolev optical depth, and $\nu_0$ is the frequency at the center of each line.

Eq. (A.5) presents the mathematical procedure the radiative line acceleration has to be calculated. However, a simple tabulation and a corresponding usage of the numbers obtained does not present a proper algorithm for solving the hydrodynamics. Such a method converges extremely slowly – if it does it at all (cf. Pauldrach et al. 1986 and Puls 1987) –, and this is especially the case, if the objects considered are extremely close to the Eddington-limit. A proper parameterization scheme which is based on the numerical values obtained from Eq. (A.5) is therefore required, in order to overcome the problems outlined. Such a suitable scheme was introduced by Castor et al. (1975), modified by Abbott (1982) and further improved by Pauldrach et al. (1986), Pauldrach (1987), Pauldrach et al. (1990), and Pauldrach et al. (1994). In the following this "force multiplier" scheme will be adapted to our present treatment of blocking and blanketing of the radiative transfer.

The primary idea of the force multiplier concept is to find a simple parametrization of the calculated radiative line acceleration values by considering in a first step just "radial streaming" photons – thus, for the construction of the radiative line acceleration just photons which are connected to the approximation $\tau_s(r,\mu=1)$ are considered in this step, and just these photons are used to parameterize the line force in terms of $M(t,r)$, where the dimensionless parameter $t$ corresponds to the local optical thickness of a line for which the integrated line opacity exactly represents the adequate Thomson scattering value ($t = \chi_{Th}v_{th}/(dv/dr)$

– where $v_{\text{th}}$ is the thermal velocity of hydrogen) –, and to apply in a second step a precisely calculated correction factor $CF(r)$, accounting for the non-radial streaming photons in the way that the product of $M(t, r)$, $CF(r)$, and a radius dependent factor, represents exactly Eq. (A.5).

With respect to this idea we first split the integrals in Eq. (A.5) in two parts

$$\sum_{\text{lines}} \chi_{\text{line}}(r) \int_{-1}^{+1} I_{\nu_0}(r, \mu) f(\mu^2) \mu \, \mathrm{d}\mu = \sum_{\text{lines}} \chi_{\text{line}}(r)$$
$$\cdot \left( \int_{\mu_*}^{+1} I_{\nu_0}(r, \mu) f(\mu^2) \mu \, \mathrm{d}\mu - \int_{\mu_*}^{-1} I_{\nu_0}(r, \mu) f(\mu^2) \mu \, \mathrm{d}\mu \right) \quad \text{(A.7)}$$

(here $f(\mu^2)$ describes the $\tau_s(r, \mu)$ dependent part of the kernel of the integral – cf. Eq. (A.5) and Eq. (A.6) – and the cone angle $\mu_*$ is defined by $\mu_* = [1 - (R_*/r)^2]^{1/2}$).

In order to see how the correction factor for the non-radial streaming photons has to be applied, we replace the angle dependent intensities at the line centers temporarily by suitable mean values ($\bar{I}^{\pm}_{\nu_0}(r)$). For the terms in brackets we thus obtain

$$\cdot \left( \bar{I}^{+}_{\nu_0}(r) \int_{\mu_*}^{+1} f(\mu^2) \mu \, \mathrm{d}\mu - \bar{I}^{-}_{\nu_0}(r) \int_{\mu_*}^{-1} f(\mu^2) \mu \, \mathrm{d}\mu \right) \quad \text{(A.8)}$$

As $f(\mu^2)$ is for both cases – radial and non-radial streaming photons – an even function with respect to $\mu$ the integrals in Eq. (A.8) are identical. Hence, we obtain instead of Eq. (A.8)

$$\cdot \left( \left( \bar{I}^{+}_{\nu_0}(r) - \bar{I}^{-}_{\nu_0}(r) \right) \int_{\mu_*}^{+1} f(\mu^2) \mu \, \mathrm{d}\mu \right) \quad \text{(A.9)}$$

Now, for the radial streaming approximation $\tau_s(r, \mu = 1)$ we obtain

$$\cdot \left( \left( \bar{I}^{+}_{\nu_0}(r) - \bar{I}^{-}_{\nu_0}(r) \right) \frac{R_*^2}{2r^2} f(\mu^2 = 1) \right)$$
$$\text{or} \quad \text{(A.10)}$$
$$\cdot \left( \frac{H_{\nu_0}(r)}{H(r)} \frac{v_{\text{th}} \nu_0}{c} \frac{1 - e^{-\tau_s(r, \mu=1)}}{t} \right) \frac{1}{\chi_{\text{line}}(r)} 2 H(r) \chi_{\text{Th}}(r)$$

where $H(r)$ is the local frequency integrated Eddington flux which gives $H(r) = L_*/(16\pi^2 r^2)$.

Summed over all lines the term in brackets defines the force multiplier $M(t, r)$

$$M(t, r) = \sum_{\text{lines}} \left( \frac{H_{\nu_0}(r)}{H(r)} \frac{v_{\text{th}} \nu_0}{c} \frac{1 - e^{-\tau_s(r, \mu=1)}}{t} \right)$$
$$= k t^{-\alpha} \left[ \frac{n_e(r)}{10^{11} [\text{cm}^{-3}]} \frac{2}{1 - \mu_*} \right]^{\delta} \quad \text{(A.11)}$$

where the parameters $k$, $\alpha$, and $\delta$ appearing in the second equation are deduced from the evaluation of the first equation ($n_e(r)$ is the local electron particle density).

The correction factor $CF(r)$ for non-radial streaming photons now follows directly from the integral of Eq. (A.9) (cf. Pauldrach et al. 1986, and Pauldrach et al. 1994)

$$CF(r) = \frac{2r^2}{R_*^2} \int_{\mu_*}^{+1} \left[ \frac{(1-\mu^2)\frac{v(r)}{r} + \mu^2 \frac{\mathrm{d}v(r)}{\mathrm{d}r}}{\frac{\mathrm{d}v(r)}{\mathrm{d}r}} \right]^{\alpha} \mu \, \mathrm{d}\mu \quad \text{(A.12)}$$

With respect to Eq. (A.5), Eq. (A.10), Eq. (A.11), and Eq. (A.12), the radiative line acceleration is now given in a parameterized form by

$$g_{\text{lines}}(r) = \frac{2\pi}{c} \frac{1}{\rho(r)} 2 H(r) \chi_{\text{Th}}(r) M(t, r) CF(r)$$
$$= \frac{\chi_{\text{Th}}(r) L_*}{4\pi r^2 c \rho(r)} M(t, r) CF(r) \quad \text{(A.13)}$$

Thus, although the radiative line acceleration is calculated by Eq. (A.5), Eq. (A.13) represents the algorithm for solving the hydrodynamics.

But Eq. (A.13) is not merely the parameterized form of the radiative line acceleration, it exists in its own right, and contains basic physics which can be readily interpreted: as is shown in Eq. (A.11) the line acceleration is a non-linear function of the velocity gradient, which enters into the equation by means of the parameter $t(r)$

$$t(r) = \chi_{\text{Th}}(r) \frac{v_{\text{th}}}{\mathrm{d}v/\mathrm{d}r}, \quad \text{(A.14)}$$

and with respect to its exponent $\alpha$ two extremes can be realized, namely $\alpha = 0$ and $\alpha = 1$. These two extremes represent the optically thin and the optically thick case of the spectral lines, respectively. The fact that $\alpha \sim 0.7$ is a typical value for massive O-stars shows that optically thick and hence strong lines are of primary importance (note that $v_{\text{th}}$ is just a scaling factor in Eq. (A.14)). (The problem of self-shadowing of strong lines[26] (cf. Pauldrach et al. 2001) can easily be omitted in Eq. (A.5), as well as Eq. (A.10) by using an incident intensity which is not affected by the considered strong line itself in the corresponding frequency interval given by the Doppler shift with respect to the terminal velocity (cf. Pauldrach et al. 1998).)

It is further important to realize that, since the ionization balance changes in the wind flow, the total line-strength changes accordingly, and thus, the value of $\alpha$ would even roughly not be a constant value. However, the biggest part of a possible change of $\alpha$ is compensated by an additional term for which the parameter $\delta$ is the exponent. This term accounts for changes of the ionization structure via changes of the electron density and the geometrically diluted radiation field (cf. Abbott 1982 and Pauldrach et al. 1986) – we note that $\delta \sim 0.1$ is a typical value for massive O-stars.

However, this term just accounts for changes of the ionization structure, it does not account for possible changes of the slope of $\alpha$, thus, for changes of the total line-strength induced by changes of the ionization structure. The values of $\alpha$ and $\delta$ can therefore in general only be treated as constants in specified intervals of the wind (cf. Pauldrach 1987). This regards also the parameter $k$ which can be interpreted as a measure of the total number of lines whose opacities are important for the driving mechanism.

---

[26] The problem of self-shadowing occurs because the incident intensity used for the calculation of a bound-bound transition is already affected by the line transition itself, since the opacity of the line has been used for the computation of the radiative quantities in the previous iteration step. If the lines contained in a frequency interval are of almost similar strength, this is no problem, since the used intensity calculated represents a fair mean value for the true incident radiation of the individual lines in the interval. It is however a problem, if a line has a strong opacity with a dominating influence in the interval. In this case the source function of this line is underestimated and the radiative processes – the scattering part is mostly affected – are not correctly described in the way that the line appears systematically too weak.

Actually, the contributions of all these lines are calculated from the opacities and the radiation field. This regards hundred thousands of lines, and from these calculated contributions new force multiplier parameter sets ($k_i$, $\alpha_i$, $\delta_i$) are deduced (the index $i$ corresponds to specified intervals in $t$ of the wind, within which the parameter sets are treated as constants; depending on the wind structure the number of intervals used may vary from at least one up to seven).

Regarding the iterative solution of the complete system of equations, the force multiplier concept has, compared to other methods, a big advantage, namely its primary proportionality to the variable $1/t$, which in turn is directly proportional to the velocity gradient. The only requirement for finding a solution to the hydrodynamic equations is, therefore, that the structure of the velocity gradient allows the critical point to be passed at some point in the atmosphere (the critical point is given by the eigenvalue of the problem and its corresponding velocity is characterized by the sound velocity (cf. Pauldrach et al. 1986). This requirement can, however, be achieved in all cases where a solution exists, even for stars which are positioned extremely close to the Eddington-limit. As this is the case for several VMS, it is quite obvious that the hydrodynamics of these objects can only be treated in an iterative way with a concept that is analogous to the force multiplier procedure (Pauldrach et al. 1986).

The state of the art iterative solution of the complete system of equations gives us henceforth not only the *synthetic spectra and ionizing fluxes* which can be used to determine stellar parameters and abundances, but also the hydrodynamical structure of the wind – thus, important constraints for the *mass-loss rate* $\dot{M}$ and the *velocity structure* $v(r)$ are derived in a consistent way.